\documentclass[fleqn,usenatbib]{mnras}

\usepackage{mathptmx}
\usepackage[T1]{fontenc}
\usepackage{multirow}

\DeclareRobustCommand{\VAN}[3]{#2}
\let\VANthebibliography\thebibliography
\def\thebibliography{\DeclareRobustCommand{\VAN}[3]{##3}\VANthebibliography}

\usepackage{graphicx}	% Including figure files
\usepackage{amsmath}	% Advanced maths commands
\usepackage{amssymb}	% Extra maths symbols
\usepackage{color}
\usepackage{adjustbox}
\usepackage[table,xcdraw]{xcolor}
\usepackage{diagbox}
\usepackage{pifont}% http://ctan.org/pkg/pifont
\newcommand{\cmark}{\ding{51}}
\newcommand{\xmark}{\ding{55}}
\usepackage{color}
\definecolor{mygreen}{rgb}{0.01, 0.75, 0.24}
\definecolor{myblue}{rgb}{0.0, 0.25, 0.69}

\newcommand{\msun}{\,\rm{M}_{\odot}}

\newcommand{\lgal}{\texttt{L-Galaxies}}

\title[X-ray fields of massive black hole mergers]{Galaxy fields of LISA massive black hole mergers in a simulated Universe}

\author[Lops et al.]{Gaia Lops,$^{1}$\thanks{E-mail: gaia.lops@unimib.it}
David Izquierdo-Villalba,$^{1,2}$\thanks{E-mail: david.izquierdovillalba@unimib.it}, Monica Colpi$^{1,2}$,  Silvia Bonoli$^{3,4}$,
Alberto Sesana$^{1,2}$, \newauthor Alberto Mangiagli$^{5}$
\\
$^{1}$ Dipartimento di Fisica ``G. Occhialini'', Universit\`{a} degli Studi di Milano-Bicocca, Piazza della Scienza 3, I-20126 Milano, Italy\\
$^{2}$ INFN, Sezione di Milano-Bicocca, Piazza della Scienza 3, 20126 Milano, Italy\\
$^{3}$ Donostia International Physics Centre (DIPC), Paseo Manuel de Lardizabal 4, 20018 Donostia-San Sebastian, Spain\\
$^{4}$ IKERBASQUE, Basque Foundation for Science, E-48013, Bilbao, Spain\\
$^{5}$ Université Paris Cité, CNRS, Astroparticule et Cosmologie, F-75013 Paris, France\\}
\date{Accepted XXX. Received YYY; in original form ZZZ}

\pubyear{2022}

\begin{document}
\label{firstpage}
\pagerange{\pageref{firstpage}--\pageref{lastpage}}
\maketitle

\begin{abstract}
LISA will extend the search for gravitational waves (GWs) at $0.1\,{-}\,100$ mHz where loud signals from coalescing binary black holes of $ 10^4 \,{-}\,10^7\,\msun$ are expected. Depending on their mass and luminosity distance, the uncertainty in the LISA sky-localization decreases from hundreds of deg$^2$ during the inspiral phase to fractions of a deg$^2$ after the merger. By using the semi-analytical model \lgal{} applied to the  \texttt{Millennium-I} merger trees, we generate a simulated Universe to identify the hosts of $z\,{\leq}\,3$ coalescing binaries with total mass of $3\,{\times}\,10^{5}$, $3\,{\times}\,10^6$ and $3\,{\times}\,10^7\msun$, and varying mass ratio. We find that, even at the time of merger, the number of galaxies around the LISA sources is too large (${\gtrsim}\,10^2$) to allow direct host identification. However, if an {\it  X-ray counterpart} is associated to the GW sources at $z\,{<}\,1$, all LISA fields at merger are populated by ${\lesssim}\,10$ AGNs emitting above ${\sim}\, 10^{-17} \, \rm erg\,cm^{-2}\,s^{-1}$. For sources at higher redshifts, the poorer sky-localization causes this number to increase up to ${\sim}\, 10^3$. Archival data from eRosita will allow discarding ${\sim}\, 10\%$ of these AGNs, being too shallow to detect the dim  X-ray luminosity of the GW sources. Inspiralling binaries in an active phase with masses ${\lesssim}\,10^6\msun$ at $z\,{\leq}\,0.3$ can be detected, as early as $10$ hours before the merger, by future X-ray observatories in less than a few minutes. For these systems, ${\lesssim}\,10$ AGNs are within the LISA sky-localization area. Finally, the LISA-Taiji network would guarantee the identification of an X-ray counterpart $10$ hours before merger for all binaries at $z\,{\lesssim}\,1$.
%Inspiralling binaries with masses ${\lesssim}\,10^6\msun$ at $z\,{\leq}\,0.3$ can be detected, in less than a few minutes, within an X-ray field of view of ${\sim}\,0.4$ deg$^2$ as early as $10$ hours before the merger. For these systems, ${\lesssim}\,10$ AGNs are in the LISA field of view. Finally, the LISA-Taiji network would guarantee the identification of an X-ray counterpart $10$ hours before merger for all binaries at $z\,{\lesssim}\,1$.

\end{abstract}


%%%%%%%%%%%%%%%%%%%%%%%%%%%%%%%%%%%%%%%%%%%%%%%%%%

%%%%%%%%%%%%%%%%% BODY OF PAPER %%%%%%%%%%%%%%%%%%
\section{Introduction}

Merging massive black hole binaries (MBHBs) with total masses of $10^4\,{-}\, 10^7 \msun$ are expected to emit gravitational waves (GWs) in the frequency interval between $0.1\,{-}\,100$ mHz, which will be probed by the upcoming space mission LISA, the  Laser Interferometer Space Antenna \citep{AmaroSeoane2017}. The gravitational signal associated to these systems carries information on the black hole masses and spins, and on the luminosity distance and sky-position of the source. Thanks to the current design of the mission, coalescing MBHBs can be detected out to cosmological distances ($z\,{\sim}\,20$) as the amplitude of GWs decays with the inverse of the luminosity distance. Hence, we will have access to the entire Universe, being possible to reconstruct for the first time the merger history of MBHBs from cosmic dawn to the present day and to infer, albeit indirectly, the origin and growth of the massive black holes (MBHs) lurking at the center of today galaxies  \citep{Colpi2019, Volonteri2021}. Besides, general relativity and gravity in the strong-field dynamical regime will be tested with unprecedented precision using the loudest and nearby GW sources. Therefore, the science related to LISA promises major advances in the domains of astrophysics, physics and cosmography \citep{LISA_Astrophysics_2022, LRR-physics2022, LRR-Cosmology2022}.\\

Coalescing MBHBs may not reside in vacuum, as they may form in the aftermath of gas-rich galaxy collisions \citep{Kocsis2006, Mayer2007, Colpi2014, Capelo2015, LISA_Astrophysics_2022}. In these circumstances, part of the gas can reach the proximity of the MBHB, triggering the birth of an Active Galactic Nucleus (AGN) and giving rise to an electromagnetic (EM) signature \citep{Roedig2014, Dascoli2018, Gutierrez2022A}. The properties of this signal are still unknown due to the lack of observations of transient broad-band emissions from AGNs compatible with the orbital motion related to sub-scale MBHBs \citep{Derosa2019}. Consequently, it is required to resort to theoretical models to understand how the lightcurves and spectra generated by MBHBs evolve during the inspiral and merger phases \citep{Bogdanovic2022}. During these stages, the gas reservoir is either under the form of a {\it circumbinary disc} fueling {\it mini-discs} that feed each individual black hole \citep[e.g.][]{Farris2012, Shi2016, Tang2018, Bowen2018, Paschalidis2021,Combi2022}, or under the form of low-angular momentum {\it gas clouds} \citep{Farris2012, Bode2012, Giacomazzo2012A, Cattorini2022}. Days to hours prior to coalescence, the precursor EM emission extends over different wavelengths and is expected to emerge from the circumbinary disc, the mini-discs, and from the hot gas and accretion streams that fill the cavity wall carved by the binary. For instance, the UV radiation produced by the disc components and the X-rays from coronal emission are expected to be the main contributions to the pre-merger EM signal  \citep{Tang2018, Dascoli2018,Gutierrez2022A}. These emissions show periodicities connected to the dynamical coupling of the binary with the circumbinary disc, providing a clear signature of an ongoing merger \citep{Bowen2018,Paschalidis2021,Combi2022}. Close to the merger time, the hard EM components of the spectrum may gradually disappear due to the evaporation of the mini-discs and resume in the post-merger phase \citep{Paschalidis2021}.  Accretion re-brightening may ensue a few months to years after the merger, depending on the mass and recoil velocity of the merger remnant \citep{2005ApJ...622L..93M,Rossi2010,Yuan2021}. Radio emission can also accompany the coalescence of MBHBs due to the interaction between the surrounding hot plasma and the magnetic field \citep{Cattorini2022}. Finally, we know that spinning black holes are powerful engines of jets \citep{BZ1977}. Therefore, it is possible that a dual jet during the inspiral phase \citep{Palenzuela2010} or a jet after the birth of the new MBH can launch a multi-wavelength beamed emission on timescales of weeks to years \citep{Yuan2021}. \\

The unequivocal detection of electromagnetic signatures coming from MBHBs can be an effective tool in cosmology. MBHBs are tagged as \textit{bright standard sirens} when an EM counterpart is detected during the rise and fall of the GW  signal. This joint observation informs us on the luminosity distance-redshift relation, without resorting to a cosmic distance ladder, providing an independent measure of the Hubble expansion parameter  \citep{Petiteau2011,Tamanini2016}. On one hand, the luminosity distance of the source is measured directly from the GW signal by combining the two polarization states of the wave and the variation of the GW frequency emitted during the inspiral \citep{Schutz1986,Holz2005}. On the other hand, the redshift is inferred after identifying the galaxy exhibiting the EM transient associated to the GW source. In order to detect an EM counterpart to a MBHB coalescence, the field of view (FOV) of the observatory dedicated to the search needs to be comparable to the LISA sky-localization uncertainty area. This requirement is not always guaranteed given that LISA sources can be localized within a sky-area that ranges from several hundreds of deg$^2$ to fractions of a deg$^2$, depending on the intrinsic properties of the binary and its luminosity distance \citep{Mangiagli2020}. Besides the sky-localization, the source of the GW signal must be bright enough to emit radiation above the detection limit of the EM observatory and be unambiguously identified among the other sources present in the joint LISA - observatory FOV. These last two requirements constitute a challenge as the MBHBs that can be well localized by LISA are the loudest GW emitters, which are however among the dimmest EM sources.\\

 %whenever a simultaneous detection of GWs and EM counterparts is possible. Therefore, the joint detection of GW and EM signals is required in order to build the redshift-luminosity distance relation without resorting to a cosmic distance ladder and to test the cosmological parameters describing our expanding Universe. The usage of bright standard sirens as cosmological tools depends on the capability of LISA to localize a source in the sky, either during the inspiral or at merger. Indeed, in order to detect an EM counterpart associated to a MBHB merger, the LISA sky-uncertainty of the event must be comparable to the field of view (FOV) of the observatory dedicated to the search of the EM signal. On top of this, the source of the GW signal must radiate with a flux above the detection limit of the EM observatory over a limited exposure time, and be unambiguously identified among the other sources present in the joint LISA - EM observatory FOV. Depending on the intrinsic properties of the binary and its luminosity distance, the LISA sources can be localized within a sky-error that ranges from several hundreds of deg$^2$ to fractions of a deg$^2$ \citep{Mangiagli2020}. This constitutes a tremendous challenge as only the loudest LISA GW sources, which are well localized and fully characterized in terms of masses and spins, are likely to be dim EM emitters.\\

Contemporary to the submission of this paper, Mangiagli et al. started to investigate the EM detectability of merging MBHBs extracted from  catalogs based on Press-Schechter galaxy formation and evolution models. Using a Markov Chain Monte Carlo (MCMC) Bayesian anlaysis, they inferred the sky-position uncertainty of the systems in their catalogs and explored their EM detectability in the optical, radio and soft X-ray bandwidths. In this way, Mangiagli et al. were able to draw predictions about the number of EM counterparts expected in 4 years of LISA observations. Their approach is complementary to the one followed in this paper, since our main focus is not  predicting the rate of EM counterparts, but rather exploring the galactic fields associated to LISA coalescing MBHBs under the assumption that they are X-ray emitters. More precisely, we aim at characterizing the number of galaxies lying inside the LISA error-box, along with the number of AGNs hosted in the same sky-region detectable by putative future X-ray missions based on the specifications of the proposed observatories Athena \cite[acronym of the Advanced Telescope for High-energy Astrophysics,][]{Nandra2013}, and Lynx \cite[][]{Lynx2018}. Answering this question will guide us in the identification of the true host galaxy in the LISA error-box. Besides, in this work we do not just confine our analysis to the time of the final coalescence but we also consider binaries in their late inspiral phase. Indeed, detecting  periodicities in the X-ray lightcurve commensurate to the periodicity of the GW chirp during the late inspiral would allow the unequivocal identification of the host galaxy of LISA coalescing MBHBs.\\

To this end, we used the \lgal{} semi-analytical model \citep[SAM,][]{IzquierdoVillalba2021} applied on top of the dark matter (DM) merger trees of the \texttt{Millennium-I} simulation \citep{Springel2005} to generate a synthetic lightcone of ${\sim}\,1000 \, \rm deg^2$ projected in the sky and redshift depth of ${\sim}\,3.5$, which is detailed enough to account for the cosmological evolution of galaxies and of their single and binary MBHs, either active or quiescent. To our knowledge, this is the first work aiming at studying the ability of future X-ray observatories - with designs close to that of Athena and Lynx - in the identification of the host galaxies of the LISA sources by creating a mock Universe where galaxies, MBHs and MBHBs are evolved self-consistently from redshift $z\,{\sim}\,3$ to the present. In this way, we have been able to explore the feasibility of the search for EM counterparts associated to binaries with different masses, mass ratios and inspiralling times and discuss the challenges of multi-messenger observations.\\

The paper is organized as follows: in Section \ref{sec:TheSimulatedUniverse} we describe the \lgal{} SAM and the lightcone construction. In particular, we shortly outline the physical processes implemented in the model to shape the population of galaxies, MBHs and MBHBs. In Section \ref{sec:LISA_Error_Boxes} we explain the strategy followed to associate LISA sky-uncertainties to MBHBs placed inside the lightcone. In Section \ref{sec:XrayObservatories} we describe our simulated X-ray observatories, whose properties mimic the capabilities of Athena and Lynx. In Section \ref{sec:Results} we present our results, focusing on the number of galaxies placed inside the sky-region delimited by LISA sky-uncertainties, and on the reduction of this number thanks to X-ray observations and the cooperation with future space-based GW detectors such as Taiji. A Lambda Cold Dark Matter $(\Lambda$CDM) cosmology with parameters $\Omega_{\rm m} \,{=}\,0.315$, $\Omega_{\rm \Lambda}\,{=}\,0.685$, $\Omega_{\rm b}\,{=}\,0.045$, $\sigma_{8}\,{=}\,0.9$ and $\rm H_0\,{=}\,67.3\, \rm km\,s^{-1}\,Mpc^{-1}$ is adopted throughout the paper \citep{PlanckCollaboration2014}.

\section{Tailoring catalogues for LISA and X-ray observatories}
\label{sec:TheSimulatedUniverse}

To perform a detailed study on the synergy between LISA and future X-ray observatories we simulated a {\it lightcone}, i.e. a Universe in which galaxies and single/binary MBHs are placed along the celestial coordinates: right ascension (RA), declination (DEC) and luminosity distance ($d_{\rm L}$ or redshift, $z$). For that, we made use of the \lgal{} SAM, which allows to construct lightcones with a reliable population of galaxies and MBHs/MBHBs thanks to the inclusion of detailed physical models that are able to track their assembly and evolution over a wide range of masses and scales. In the following, we summarize the main physics included in the SAM, but we refer the reader to \cite{Henriques2015} and \cite{IzquierdoVillalba2019,IzquierdoVillalba2020,IzquierdoVillalba2021} for further details about the version of the model used here.

\subsection{The backbone of \lgal{}: Dark matter merger trees}

To model the physics involved in galaxy formation and evolution \lgal{} uses, as a foundation, the merger trees extracted from the \texttt{Millennium} suite of DM N-body simulations. Specifically, in this paper we use the merger trees of the \texttt{Millennium-I} simulation (hereafter MS), which follows the evolution of $2160^3$ DM particles of mass ${\sim}\,8.6\,{\times}\,10^8 \rm M_{\odot}/\mathit{h}$ from redshift $z = 127$ to redshift $z=0$ inside a periodic box of side $500 \rm \, Mpc/\mathit{h}$ \citep{Springel2005}. By using a friend-of-friend group-finder and the \texttt{SUBFIND} and \texttt{L-HALOTREE} algorithms \citep{Springel2005}, DM halos are identified, stored and arranged as progenitors and descendants at $63$ different epochs or \textit{snapshots}. For this work, the original cosmology of the MS was re-scaled with the procedure of \cite{AnguloandWhite2010} to match the one provided by the Planck first-year data release \citep{PlanckCollaboration2014}.

\subsection{The galaxy formation model: From galaxies to massive black hole binaries}

Once the assembly history of DM halos has been provided, \lgal{} includes different analytical prescriptions to describe the formation and evolution of galaxies and MBHs. In the following, we give an overview of the main physics involved.

\subsubsection{The galaxy population}

\lgal{} follows the \cite{WhiteandRees1978} paradigm, assuming that DM halos act as the birthplaces of galaxies. To this end, the SAM assigns to each collapsed halo a fraction of baryons in the form of a diffuse hot gas atmosphere which gradually cools down. During this process, the gas settles into a disc-like structure promoting star formation episodes which give rise to the formation of a stellar disc. To regulate these star formation events, the model includes two feedback mechanisms: one associated to supernovae explosions and another triggered by the continuous and sub-Eddington hot gas accretion onto a central AGN. Besides discs, galaxies are able to develop galactic bulges through internal and external processes. In the former case, massive discs are prompt to instabilities that are able to redistribute the stellar matter, triggering in this way the formation of a pseudobulge structure. In the latter case, galaxy encounters are the triggers of the bulge formation. According to the baryonic merger ratio of the two interacting galaxies, these processes are divided in two different flavours: major and minor. During major encounters the discs of both galaxies are completely destroyed and the remnant system is an elliptical galaxy. On the other hand, minor mergers leave untouched the stellar disc of the central galaxy, while causing the formation of a dense pack of stars, also referred to as classical bulge, thanks to the incorporation of the stellar matter of the smaller progenitor.

\subsubsection{The massive black hole population} \label{sec:BHPopulation}

\begin{figure*}
\centering  
\includegraphics[width=2\columnwidth]{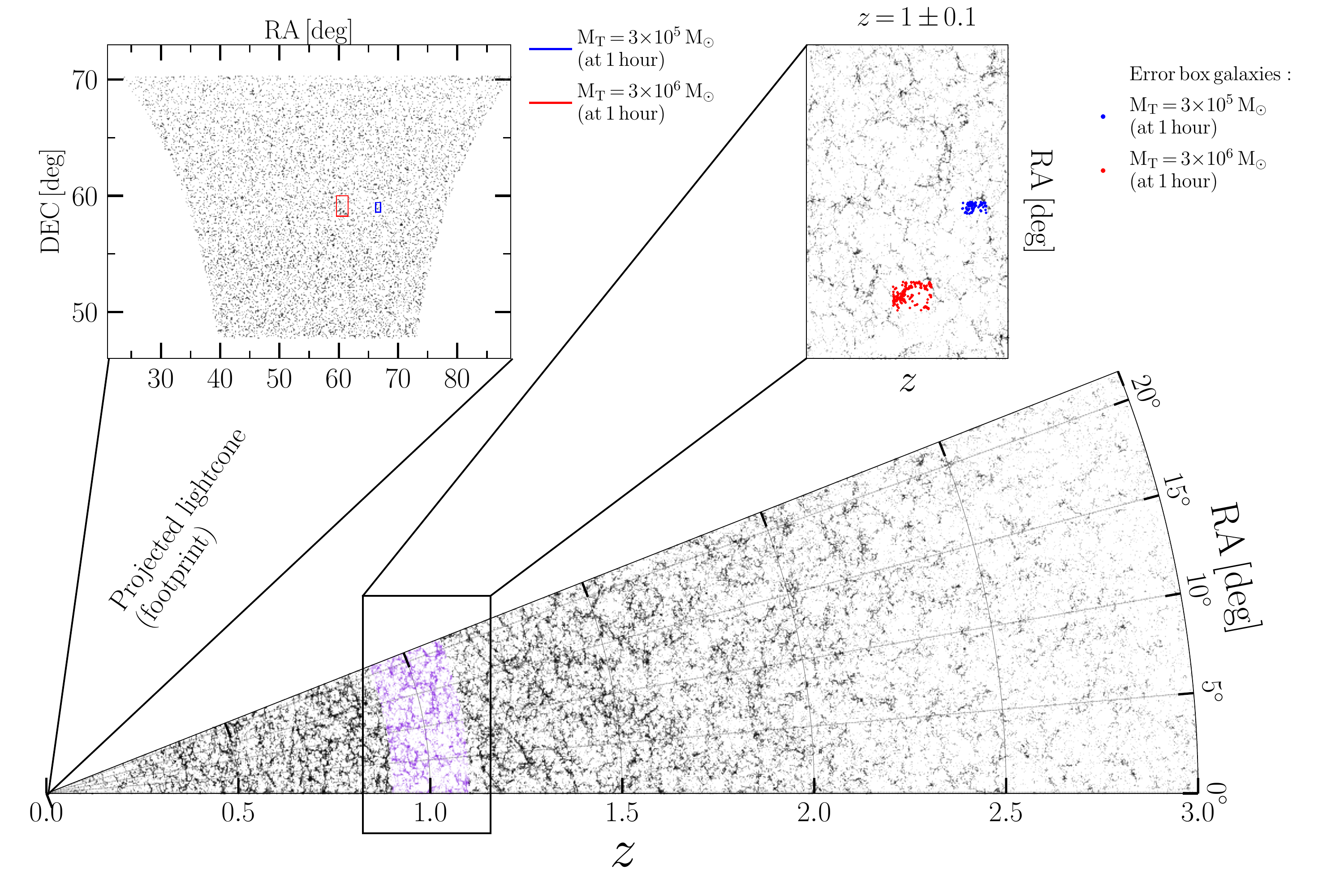}
\caption[]{\textbf{Lower panel}: Spatial distribution (right ascension $\rm RA$ and redshift $z$) of galaxies with $\rm M_{stellar} \,{>}\,5\,{\times}\,10^9\, \msun$ inside the lightcone generated for this work. To improve the clarity of the plot, we have only selected  galaxies inside the declination range $\rm DEC\,{=}\,58.9\,{\pm}\,0.15 \, \rm deg$. For convenience, the values of $\rm RA$ have been transformed in such a way that $\rm RA\,{=}\,0\, \rm deg$ corresponds to $\rm RA\,{=}\,56.3\, deg$. With purple color we have highlighted the galaxies laying inside the redshift shell $z\,{=}\,1\,{\pm}\,0.1$. \textbf{Upper left panel}: Projection of our 1027 deg$^2$ lightcone in the $\rm RA\,{-}\,DEC$ space. The red (blue) box corresponds to the angular extent of the LISA error-box at $1\, \rm hour$ before merger associated with a binary of $3\,{\times}\,10^6\, \msun$ ($3\,{\times}\,10^5\, \msun$) as extracted randomly from the \protect{\cite{Mangiagli2020}} catalogue. \textbf{Upper right panel}: Plane  $z\,{-}\,\rm RA$ for galaxies at $z\,{=}\,1\,{\pm}\,0.1$ with colored points corresponding to the galaxies associated with the LISA error-box of the upper left panel.}
\label{fig:LightCone}
\end{figure*}
%\monica{can we spend two words to explain why the AGN activity is triggered later on. Obscuration-- the read could not understand this.} 

%As an important part of the galaxy evolution theory, 

\lgal{} includes different physical models to describe the formation, evolution and assembly of MBHs along cosmic history. Specifically, in each newly resolved DM halo a {\it black hole seed} of $10^4\,\msun$ with random spin is implanted in its centre (see further improvements in \citealt{Spinoso2020}). The subsequent growth of the seed can follow different pathways: \textit{cold gas accretion}, \textit{hot gas accretion} and \textit{mergers} with other black holes. The first channel is triggered by both galaxy mergers and disc instabilities and accounts for the bulk of the black hole growth. Instead of assuming an instantaneous growth,  all the cold gas available for accretion is stored in a reservoir that is progressively consumed by the black hole through two different phases. During the first one, the black hole accretes gas at the Eddington limit until it consumes a given fraction of the total reservoir. Once this limit is reached, the black hole enters in a self-regulated phase, characterized by a sub-Eddington accretion rate. This phase is motivated by the hydro-dynamical simulation of \cite{Hopkins2005}, which showed a fading of the quasar light curve caused by the quasar feedback removing gas from the MBH surroundings. Based on these phases, \lgal{} computes at each time the accretion rate onto the MBH, $\dot{\rm M}_{\rm BH}$, and determines the associated bolometric luminosity as
\begin{equation}\label{eq:Lbol}
 \mathrm{L_{bol}}\,{=}\,\epsilon \dot{\rm M}_{\rm BH} c^2, 
\end{equation}
where $c$ is the speed of light and $\epsilon$ the radiative efficiency which depends on the black hole spin. We refer the reader to \cite{IzquierdoVillalba2020} for further details about the computation of $\dot{\rm M}_{\rm BH}$ and $\epsilon$. Based on Eq.~\ref{eq:Lbol}, in this work we estimate the contribution of the bolometric  luminosity in the soft ($0.5\,{-}\,2\,\rm keV$, $\rm L_{Sx}$) and hard ($2\,{-}\,10\, \rm keV$, $\rm L_{Hx}$) X-ray bands using the bolometric corrections of \cite{Shen2020}:
\begin{equation}\label{eq:BolometricCorrection_hard}
\rm log_{10}\left(L_{Hx}/L_{bol}\right) \,{=}\, -1.69 - 0.257\mathcal{L} - 0.0078\mathcal{L}^2 + 0.0018\mathcal{L}^3,\\
\end{equation}
\begin{equation}\label{eq:BolometricCorrection_soft}
\rm log_{10}\left(L_{Sx}/L_{bol}\right) \,{=}\, -1.84 - 0.260\mathcal{L} - 0.0071\mathcal{L}^2 + 0.0020\mathcal{L}^3,
\end{equation}
where $\rm \mathcal{L}\,{=}\,log_{10}(L_{bol}/L_{\sun})\,{-}\,12$. Since X-rays suffer photoelectric absorption by intervening cold gas, we assume that the X-ray spectrum of a source affected by this phenomenon is described as:
\begin{equation}\label{eq:ColumDensityAttenuation}
  f_{\rm E} \,{\propto}\, e^{-\rm N_{H} \, \sigma(E) }  \, \mathrm{E}^{-\alpha}  \,\,\,\,\,\,\,\,\,\,\,\,\,\,\,\,\rm [erg \,\, s^{-1}\,cm^{-2}\,keV^{-1}],
\end{equation}
where $\rm E$ is the photon energy, $\sigma(\rm E)\,{=}\,9\,{\times}\,10^{-23}\,(E/1\,\rm keV)^{-2.5} \, \rm cm^2$ is the photoelectric cross-section, $\alpha\,{=}\,0.7$ is the typical slope of an AGN spectrum, and $\rm N_{H}$ is the hydrogen column density, a parameter that we will use to define the level of flux absorption. Given that \lgal{} does not provide any information about $\rm N_{H}$ and the orientation of the AGN with respect to the observer, we assign to each AGN a random value according to the distribution presented in \cite{Masoura2021}. Specifically, the authors showed that the $\rm N_{H}$ distribution of a large sample of $z\,{<}\,3$ X-ray AGNs displays a maximum at $\rm N_{H}\,{\sim}\,10^{20.5}\,\rm cm^{-2}$ and a secondary peak centered at $\rm N_{H}\,{\sim}\, 10^{22.5}\,\rm cm^{-2}$. Throughout the paper we use this distribution of $\rm N_{H}$ as our fiducial approach to assign a level of absorption to all the X-ray AGNs generated by the SAM. An analysis that accounts for heavy absorption ($\rm N_{H}\,{\sim}\,10^{23}\,\rm cm^{-2}$, typical of {\it Compton tick} AGNs) is tackled in  Appendix~\ref{appendix:ComptonThinScenario}. Therefore, taking into account Eq.~\ref{eq:ColumDensityAttenuation}, a source affected by photoelectric absorption has a flux, $f^{\rm A}$, given by:
\begin{equation}
    f^{\rm A} \,{=}\, \rm \frac{\mathit{f}^{\rm U} (1-\alpha)}{E_2^{(1-\alpha)}-E_1^{(1-\alpha)}}  \int_{E_1}^{E_2} E^{-\alpha} e^{-N_{H}\,\sigma(E)} dE \;\; \rm [erg\,s^{-1}\,cm^{-2}],
\end{equation}
where $f^{\rm U}$ corresponds to the unabsorbed source flux, while $E_1$ and $E_2$ to the minimum and maximum energy of the X-ray band, respectively.\\

%The occurrence of heavy absorption processes know as {\it Compton thin} ($\rm N_{H}\,{\sim}\,10^{23}\,\rm cm^{-2}$) which refers to {\it Compton thin} AGN is the Appendix XXXXX. We recall that {\it Compton thick} AGN with  $\rm N_{H}\,{>}\,10^{24}\,\rm cm^{-2}$ would not be detectable by any of our observatories.\\

Finally, along with the mass growth, \lgal{} keeps track of the evolution of the spin modulus, which can evolve during gas accretion events and through black hole mergers. The SAM employs the approach of \cite{Dotti2013} and \cite{Sesana2014}, which correlates the  spin-up/spin-down episodes experienced by the black hole with the properties of the galactic bulge. After black hole coalescences, the spin of the remnant is computed according to the analytical expressions of  \cite{BarausseANDRezzolla2009}.

\subsubsection{The massive black hole binary population}

\lgal{} deals also with the dynamical formation and growth of MBHBs within the framework of \cite{Begelman1980}, for which the evolutionary pathway of these systems can be divided into three stages. Initially, during the so-called \textit{pairing phase}, the two MBHs are at a distance of the order of kpc and their dynamics inside the galaxy is dominated by the dynamical friction exerted by the gas and stellar components onto each individual MBH. This force slows down the MBHs, driving them towards the nucleus of the remnant galaxy, eventually leading to the formation of a gravitationally bound binary system. For simplicity, the model assumes that, after a galaxy merger, the MBH of the most massive galaxy is sited at the center of the remnant structure, whereas the MBH deposited by the satellite galaxy is the one undergoing the dynamical friction phase \citep{BinneyTremaine2008}. Once the pairing phase is over, the satellite MBH reaches the galaxy nucleus and binds ($\sim$ sub-pc separation) with the nuclear MBH, making it possible for the \textit{hardening phase} to take over. While the initial eccentricity of the newly formed binary is assumed to be random, the initial separation is set at the radius containing twice the mass of the lighter MBH in stars \citep{Colpi2014}. Once the hardening is started, the separation and eccentricity of the binary system are evolved depending on the environment in which it is embedded. If the gas reservoir around the binary system is larger than the total mass of the binary ($\rm M_{tot}$), the evolution of the system is guided by the interaction with the circumbinary gaseous disc (following \citealt{Dotti2015}). On the other hand, in gas-poor environments the system evolves due to the interaction with single stars (following \citealt{Quinlan1997,Quinlan1996} and \citealt{Sesana2015}). Once the hardening phase reduces the binary separation down to mpc scales, the emission of GWs becomes the principal shrinking mechanism until the final plunge \citep{Sesana2015}. Finally, besides tracing their dynamics, \lgal{} follows the mass growth of binaries. For that, the model assumes that the accretion rate of the lighter MBH is fixed to the Eddington limit, while the one of the most massive black hole is related to it through the mass ratio of the binary system \citep{Duffell2020}. For further details about the modelling of MBHBs in \lgal{} we refer the reader to \cite{IzquierdoVillalba2021}.

\subsection{Positioning galaxies and black holes within a lightcone} 

To explore the future synergy between LISA and upcoming X-ray observatories such as Athena and Lynx, we constructed a simulated {\it lightcone} where all the galaxies and MBHs generated by \lgal{} are placed along the celestial coordinates: right ascension (RA), declination (DEC) and luminosity distance ($d_{\rm L}$). The main limitation to generate such a lightcone with \lgal{} comes from the $500\, \rm Mpc/\mathit{h}$ side-length of the MS, which is insufficient to represent a lightcone with a redshift depth bigger than $0.1$. To overcome this issue and reach larger redshifts, we follow the procedure presented in \cite{IzquierdoVillalba2019LC}. In brief, by exploiting the periodic boundary conditions of the MS, the simulated box is replicated $14$ times along each Cartesian coordinate. This is enough to generate a lightcone with a depth of $z\,{\sim}\,3.5$. Once the box of the MS is repeated, the location of the observer, the orientation and the angular extent must be set. In our specific case, the observer is placed at the origin of the first replica with a line of sight of $\rm (RA , DEC) \,{=}\, (56.3,58.9) \, deg$, which was chosen following \cite{Kitzbichler2007} in order to minimise the structure repetition. Concerning the angular extent ($\delta$), it was set by imposing that no more than $2$ replicas are needed to represent the universe at $z\,{=}\,1$ and that the footprint of the lightcone in polar coordinates is a square. In this way, $\rm (\delta RA , \delta DEC) \,{=}\, (22.5/\cos(DEC), 22.5) \, deg$, corresponding to a total area of ${\sim}\,1027\, \rm deg^2$. Once all these steps were done, galaxies were placed inside the lightcone by determining the moment at which they (and their corresponding single and binary MBHs) cross the observer past lightcone. For that, the galaxy merger trees provided by \lgal{} were used since they accurately follow in time the evolution of individual galaxies between the DM snapshots with a time step resolution of $\rm {\sim} 5\,{-}\,20\, Myr$. In Fig.~\ref{fig:LightCone} we show a thin slice of the lightcone in which only galaxies above $\rm M_{stellar}\,{>}\,5\,{\times}\,10^{9}\, \msun$ are included.  As we can see, the simulated Universe reaches $z\,{\sim}\,3$ with no visible discreteness effects caused by the finite number of simulation snapshots. As expected, our procedure to construct the lightcone leaves untouched the large scale structures, with galaxies arranged in filaments and clusters. In the upper left panel of Fig.~\ref{fig:LightCone} we also show the $\rm RA\,{-}\,DEC$ projection. Clearly, the line of sight is centered at $(56.3,58.9) \, \rm deg$ and the footprint has a bell shape as a consequence of the fact that the lightcone in polar coordinates is characterized by a square shape.\\ 

\begin{figure}
\centering  
\includegraphics[width=1\columnwidth]{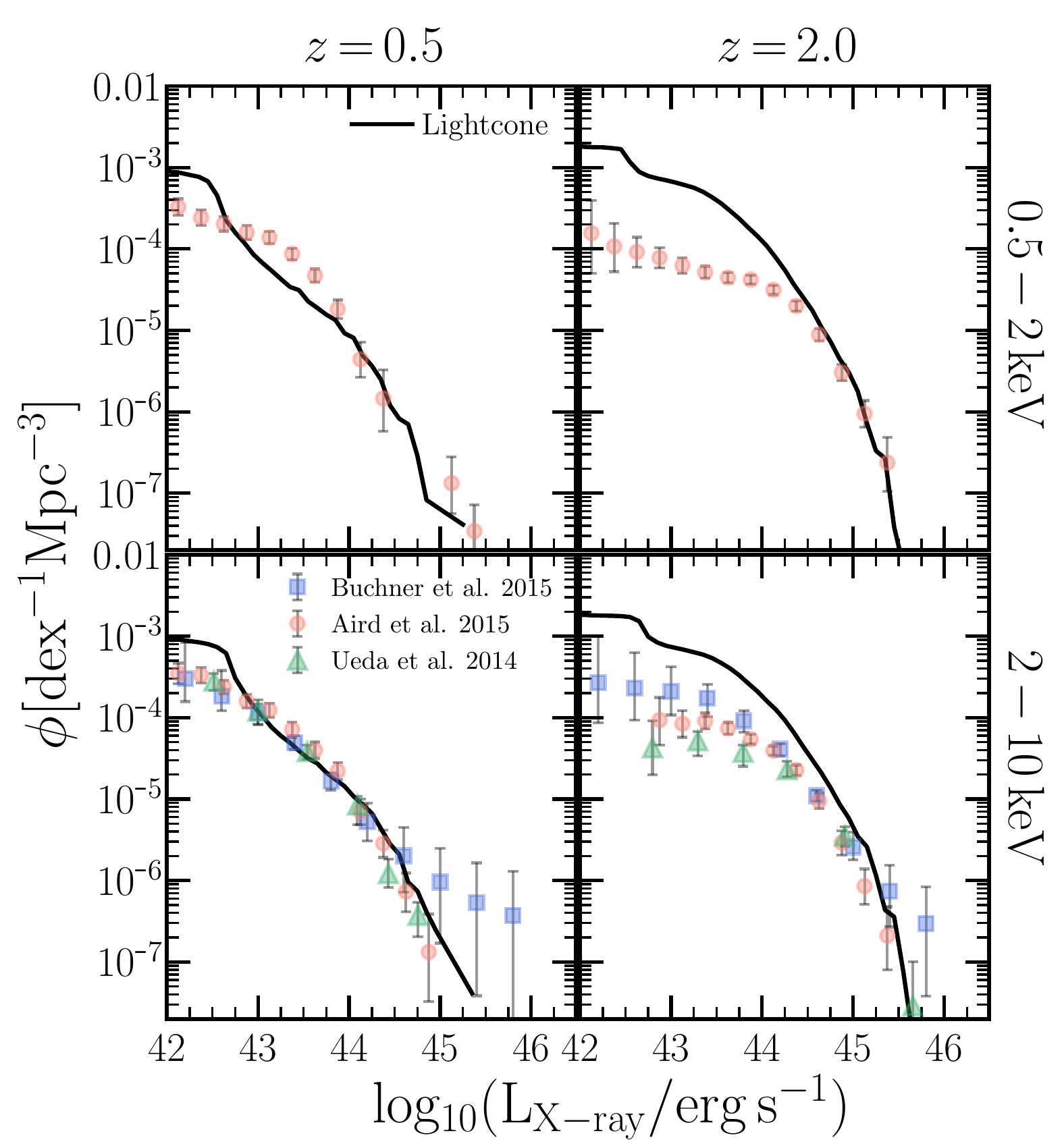}
\caption[]{Soft ($0.5\,{-}\,2\,\rm keV$, upper panels) and hard ($2\,{-}\,10\,\rm keV$, lower panels) X-ray luminosity functions at $z\,{=}\,0.5$ (left panels) and $z\,{=}\,2.0$ (right panels). While solid black lines display the predictions of the lightcone, different points represent observational constraints provided by \protect\cite{Buchner2015,Aird2015} and \protect\cite{Ueda2014} (squares, circles and triangles, respectively).}
\label{fig:Xray_LFs}
\end{figure}

Given that in this work we are particularly interested in the population of X-ray AGNs, it is important to verify that the lightcone is able to correctly predict the evolution of the X-ray AGN number density. Hence, in Fig.~\ref{fig:Xray_LFs} we compare the soft and hard X-ray luminosity functions (LFs) predicted by the model with the observational constraints of \cite{Buchner2015}, \cite{Aird2015} and \cite{Ueda2014}. Both observations and model corresponds intrinsic X-ray luminosity, i.e when the effect of absorption in the luminosity has been removed. We highlight that the LFs of the model are not fits to the data but they arise naturally from our model as a consequence of coupling the MBH growth model shown in Seciton~\ref{sec:BHPopulation} with the the hierarchical assembly of galaxies and secular processes taking place inside the galactic discs. As we can see, the predictions are generally in good agreement with the observations. However, large discrepancies are found at $z\,{=2}$ for low luminosities (${<}\,10^{44}\rm erg/s$). The nature of this inconsistency is still an open issue given the current limitations on both observational and theoretical models. From one side, observational studies lack a robust sampling of high-$z$ faint AGNs, resulting from current challenges of covering wide sky areas with large depth \citep{Siana2008,Masters2012,McGreer2013,Niida2016,Akiyama2018}. On the other hand, some theoretical works have made attempts to suppress the large excess of faint AGNs seen in most of the SAMs and hydro-dynamical simulations \citep[see e.g][]{Hirschmann2014,Griffin2018,Habouzit2022}. For instance, invoking empirical relations for obscuring accreting black holes or varying the efficiency of the seeding process could be plausible mechanisms \cite[see e.g][]{Degraf2010,Fanidakis2012,DeGraf2020,Spinoso2021}. Nevertheless, no clear answer has been proposed yet and further investigations are needed.

\section{LISA error-box: True host and host candidates} \label{sec:LISA_Error_Boxes}

The MBHBs detected by LISA will enter its sensitivity band from the low-frequency end (${\sim}\,0.1 \, \rm mHz$), moving towards higher frequencies as they reduce their separation and approach the final plunge. For long-lived GW signals (i.e. signals detected weeks before the merger), the information about the source position is encoded in the relative amplitudes and phases of the two polarization components, in the periodic Doppler shift imposed on the signal by the detector's motion around the Sun, and in the additional modulation of the signal caused by the detector's time-varying orientation \citep{Cutler1998}. On the other hand, for shorter-lived GW signals (i.e. signals detected days to hours before the merger), two additional effects that intervene close to and at merger can improve the final localization of the source. The first corresponds to the LISA pattern response, which becomes frequency-dependent in a way that informs us about the signal's position \citep{Rubbo2004, Marsat21}. The second is the contribution of the higher-order harmonics (beyond the quadrupole) present in the signal during the late-inspiral, merger and ringdown phases, which are crucial in breaking degeneracies in the source parameter estimation  \citep{Baibhav2020, Marsat21}. \\

In the last years, several works have been performed to explore LISA capabilities to localize a GW source in the sky. By means of a Bayesian analysis, \cite{Marsat21} have shown that the LISA sky-localization displays multi-modality patterns during the inspiral phase due to the high level of degeneracy between the luminosity distance and inclination angle, and the phase and polarization angles. Nevertheless, these degeneracies can be broken as the signal-to-noise ratio (SNR) increases with time thanks to the information stored in the higher-order harmonics of the signal. A few show-case studies demonstrated that, even eight days before merging, binaries with total masses of $3\,{\times}\,10^5\msun$ and mass ratio $q=1/3$ at $z\,{=}\,0.3$ do not show any multi-modality in the sky-localization pattern, owing to the long duration of the signal and the importance of the higher-order harmonics that emerge loud at merger. On the contrary, binaries heavier than a few $10^6\msun$ at $z\,{=}\,1$, which are shorter-lived, can be uniquely localized in a single area of the sky only at merger (Piro et al. 2022, in preparation). On the other hand, the Fisher information matrix (FIM) approach, which can not capture the  multi-modality in the sky-pattern of the sources, was recently explored by \cite{Mangiagli2020}. The authors analyzed LISA sky-localization capabilities \textit{on the fly} (i.e. as a function of the time to merger), considering a family of MBHBs with total masses of $3\,{\times}\,10^5 \, \msun$, $3\,{\times}\,10^6 \,\msun$ and $10^7\msun$, varying the mass ratio, the spins, the sky-position, and the polarization and inclination angles. The binaries were placed at three different redshifts, and the analysis was carried on using a waveform  which includes  higher-order modes in the inspiral-only phase. The results were extrapolated down to the time of merger using the PhenomC waveform. As expected, the authors found that the  median of the sky-localization error decreases with time for all sources, while the dispersion around it increases with decreasing time up to merger. At merger, the sky-localization  improves, but the dispersion remains still large (${\sim}\,1\rm dex$)\footnote{Ongoing studies concerning the comparison of Bayesian and Fisher Information  Matrix (FIM) analyses over the same sample of MBHBs are now being performed by Marsat and collaborators (privit communication). The comparison shows that when using waveforms including  full inspiral-merger-ringdown higher-order harmonics and frequency-dependent LISA responses, both methodologies predict comparable  values of SNR and related dispersion distribution as in \cite{Mangiagli2020}. However, the FIM analysis by \cite{Mangiagli2020} shows a faster decrease of the median of the sky-localization uncertainties with passing time and a wider spread at merger (Piro et al. 2022, in preparation, S. Marsat private communication). For the purposes of this work, these differences are not going to impact our results}.\\

In this work, we adopt the approach presented in \cite{Mangiagli2020} to study LISA sky-localization capabilities. Among all the coalescing MBHBs that can be potentially detected by LISA, here we focus on binaries with masses $3\,{\times}\,10^{5}\,\msun$, $3\,{\times}\,10^{6}\,\msun$ and $3\,{\times}\,10^{7}\,\msun$. The galactic fields associated to these systems and determined by LISA sky-uncertainties are studied at five different epochs: $z\,{=}\,0.3,0.5,1,2$ and $3$ (see Table~\ref{tab:MBHBs}). These redshift are chosen to explore how likely is the identification of GW source by means of X-ray observations. Indeed, at $z\,{\leq}\,3$, the EM signature associated to Eddington-limited MBHBs of total mass ${\leq}\,3\,{\times}\,10^7\, \rm M_{\odot}$ is bright enough to be easily detected by X-ray observatories that might be contemporary to the LISA mission such as Athena  \citep{Nandra2013}, and the mission concpect Lynx \citep{Lynx2018}. We stress that for the case of $3\,{\times}\,10^{5}\,\rm M_{\odot}$ we have limited the analysis at $z\,{\leq}\,1$ given that at higher-$z$ the AGN emission associated with such low-mass MBHB systems hampers dramatically its detection.

 \subsection{The LISA error-boxes} \label{subsec:ErrorBox}
 
     \begin{table}
    \begin{adjustbox}{width=\columnwidth,center}
    \begin{tabular}{|c|c|c|c|}
        \cline{1-4}
        \diagbox[width=\dimexpr \textwidth/18+2\tabcolsep\relax, height=0.5cm]{ $z$ }{ $\rm M_{tot}$} & \hspace{0.3cm} $3\,{\times}\, 10^5 \msun$ \hspace{0.3cm} &  \hspace{0.3cm} $3 \,{\times}\, 10^6 \msun$ \hspace{0.3cm} & \hspace{0.3cm} $ 3 \,{\times}\, 10^7 \msun$ \hspace{0.3cm} \\ \hline
        0.3       &  \cmark   &   \cmark  &  \cmark     \\
        0.5       &  \cmark   &   \cmark  &   \cmark  \\
        1         &  \cmark   &   \cmark   &   \cmark \\
        2         &  \xmark   &   \cmark   &   \cmark   \\
        3         &  \xmark   &   \cmark  &   \cmark \\ \hline
    \end{tabular}
    \end{adjustbox}
    \caption{Total mass ($\rm M_{tot}$) and redshift ($z$) of the MBHBs employed in this project.}
    \label{tab:MBHBs}
    \end{table}

In this section we summarize the procedure used to assign to the binaries placed inside the lightcone a LISA sky-uncertainty. Besides, we outline the methodology employed to assembly, during the inspiral phase and the time of the final coalescence, the so-called LISA \textit{error-box}, defined as the volume centered on the GW event whose extension is given by the error associated to the LISA sky-localization.

%In this section we summarize the procedure used to assembly, during the inspiral phase and the time of the final coalescence, the so-called LISA error-box defined as the volume centered on the GW event whose extension is given by the error associated to the LISA sky-localization. %After building these boxes, we outline the methodology by which we linked them to the MBHBs self-consistently simulated inside the lightcone.
 
\subsubsection{Synthetic binaries and their associated sky-localization errors: Monte Carlo realizations} \label{sec:MangiagliRealizations}

The LISA sky-localization errors used in this paper are determined according to the work of \cite{Mangiagli2020}. For each combination of binary mass and redshift listed in Table~\ref{tab:MBHBs}, $10^4$ different Monte Carlo simulations were performed by keeping fixed the total mass and redshift of the binary and randomizing the other parameters that describe a MBHB in a quasi-circular orbit: the mass ratio, the spin moduli and orientations, the sky-position, the polarization and inclination angles. Throughout the whole paper we will refer to all these Monte Carlo simulations as \textit{realizations}. For each of them, the so-called \textit{covariance matrix} (CM) was built at seven different times during the inspiralling phase: $1$ month, $1$ week, $3$ days, $1$ day, $10$ hours, $5$ hours and $1$ hour prior to merger. The resulting matrices had dimension $15\,{\times}\,15$ with the diagonal elements representing the variances ($\sigma^2$) related to the aforementioned binary parameters. Given that our interest concerns uniquely to the LISA sky-localization capabilities, we decided to work with the reduced $3\,{\times}\,3$ matrix whose diagonal values match the variances related to right ascension, declination and luminosity distance. We want to emphasise that the methodology of \cite{Mangiagli2020} retrieves variances corresponding to $\cos(\rm DEC)$ and $\ln(d_{\rm L})$, i.e. $\sigma^{2}_{\rm \cos(DEC)}$ and  $\sigma^{2}_{\ln(d_{\rm L})}$. Therefore, the errors associated with the sky-coordinates can be written as:
\begin{equation}\label{eq:Errors_RA_DEC}
  \begin{array}{ll}
    \Delta \mathrm{RA}  \,\,\,\,= \sigma_{\rm RA}, \\
    \Delta \mathrm{DEC} = \sigma_{\rm \cos(DEC)}/\sin(\rm DEC), \\
    \Delta d_{\rm L} \,\,\,\,\,\,= \sigma_{\rm \ln(d_{\rm L})} d_{\rm L},
  \end{array}
\end{equation}
 where we neglected any covariance between the parameters, i.e. the off-diagonal elements of the matrix are assumed to be small. From hereafter, we define $\Delta \Omega\,{=}\,(\Delta \mathrm{RA},\Delta \mathrm{DEC})$.\\
 %$\Delta \Omega\,{=}\,(3\,{\times}\,\Delta \mathrm{RA},3\,{\times}\,\Delta \mathrm{DEC})$ and $\Delta D_{\rm L} \,=\,3\,{\times}\,\Delta d_{\rm L}$.\\
 
  \begin{figure}
    \centering
    \includegraphics[width=1\columnwidth]{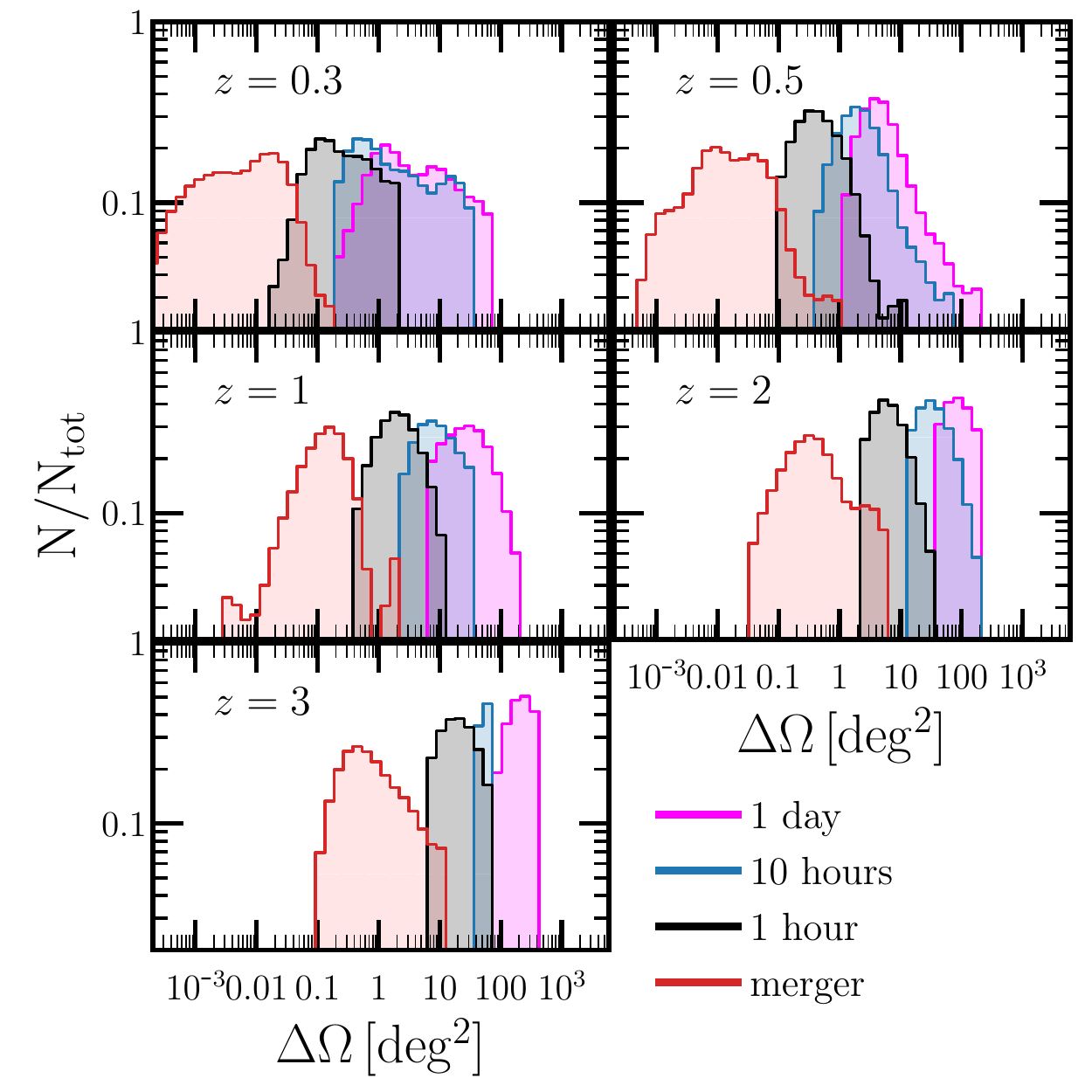}
    \caption[]{Distribution of the $\Delta \Omega$ associated to the true hosts with mass $\rm M_{\rm tot} \,{=}\, 3 \times 10^6 \msun$ at $z = 0.3, 0.5, 1, 2$ and  $3$. Different colors represent different times: $1$ day (magenta), $10$ hours (blue) and $1$ hour (black) before merger and at the coalescence time (red).}
    \label{fig:DeltaOmega_M3e6}
\end{figure}

In the original work of \cite{Mangiagli2020}, the CMs corresponding to the merger time were not computed. Here, we produced these matrices by making use of the information that the \cite{Mangiagli2020} methodology provides about the SNR of the binary during its inspiral and merger phases. Specifically, the CM at merger was determined by dividing the elements of the $1$ hour matrix by the squared ratio between the SNR at merger and that at $1$ hour \citep{Klein2016}.

\subsubsection{Associating LISA sky-localization uncertainties to the inspiralling binaries of the lightcone: True hosts} \label{sec:TrueHosts}

Once determined $\Delta \Omega$ and $\Delta d_{\rm L}$ of each realization of \cite{Mangiagli2020}, we assigned them to the binaries inside the lightcone. For that, we selected all the galaxies inside the lightcone hosting a MBHB in the inspiral (hardening) phase and sought a counterpart among all the realizations satisfying the following conditions:

    \begin{enumerate}
    \item[i)]    $z^{'}\,{-}\,0.2\,\,{<}\,\,z\,\,{<}\,z^{'}\,{+}\,0.2\,$,
    \item[ii)]   $0.5\,q^{'}\,\,\,\,{<}\,q\,\,{<}\,\,\,\,2\,q^{'}\,$,
    \item[iii)]  $\rm 0.5\, M_{\rm tot}^{'} \,\,\,{<}\,\,\,M_{\rm tot} \,\,\,{<}\,\,\,\,2 \,M_{\rm tot}^{'}\,$,
    \item[iv)]   $\rm |\,DEC^{'}\,{-}\,DEC\,|\,{<}\,2.25 \, \rm deg\,$,
    \end{enumerate}
where $z$ ($z^{'}$), $q$ ($q^{'}$), $\rm M_{\rm tot}$ ($\rm M_{tot}^{'}$) and $\rm DEC$ ($\rm DEC^{'}$) refer to the redshift, merger ratio, total mass and declination of the MBHB placed inside the lightcone (realization). Notice that we did not impose any constraint on the right ascension given that $\Delta \Omega$ is independent on its specific value (see Eq.~\ref{eq:Errors_RA_DEC}). On top of the previous conditions, an extra check was done to prevent that the $\Delta \Omega$ associated to the MBHBs overpasses the lightcone boundaries. For that, we imposed that the $\Delta \Omega$ at 10 hours prior to merger (the largest time explored in Section~\ref{sec:Results}) must be totally contained inside the lightcone footprint (see Fig.~\ref{fig:LightCone}). In case this condition was not fulfilled, we dropped the lightcone binary from our analysis. This procedure was performed for each combination of total mass and redshift listed in Table~\ref{tab:MBHBs}. From hereafter, we will refer to the lightcone MBHBs matching one of the \cite{Mangiagli2020} realizations as \textit{true hosts}. After applying these selection criteria, we found for the MBHBs of total masses $\rm M_{tot}\,{=}\,3\,{\times}\,10^{5}\,\msun$, $3\,{\times}\,10^{6}\,\msun$ and $3\,{\times}\,10^{7}\,\msun$ a median number of true hosts equal to ${\sim}\,3\,{\times}\,10^3$, ${\sim}\,5\,{\times}\,10^3$ and ${\sim}\,1\,{\times}\,10^4$. However, these values depend on the specific redshift, being smaller towards higher-$z$. We highlight that with the aforementioned procedure, we were not able to find any true host for the MBHB of $3\,{\times}\,10^{7}\,\msun$ at $z{=}3$. This fact is related to their associated value of $\Delta \Omega$ which are large enough (${>}\,1000\,\rm deg^2$) to be fully contained within the lightcone footprint \citep[see Figure 2 of][]{Mangiagli2020}. Finally, instead of using all the $\,{>}\,10^3$ true hosts found, we employed $200$ random cases at each redshift and binary mass. This number allowed us to reach a good balance between having a large enough sample to perform statistical studies and reducing the computational time of our analysis. Therefore, unless the contrary is stated, the total number of true hosts used in the analysis at each redshift and binary mass will be 200 cases.\\

\begin{figure}
    \centering
    \includegraphics[width=1\columnwidth]{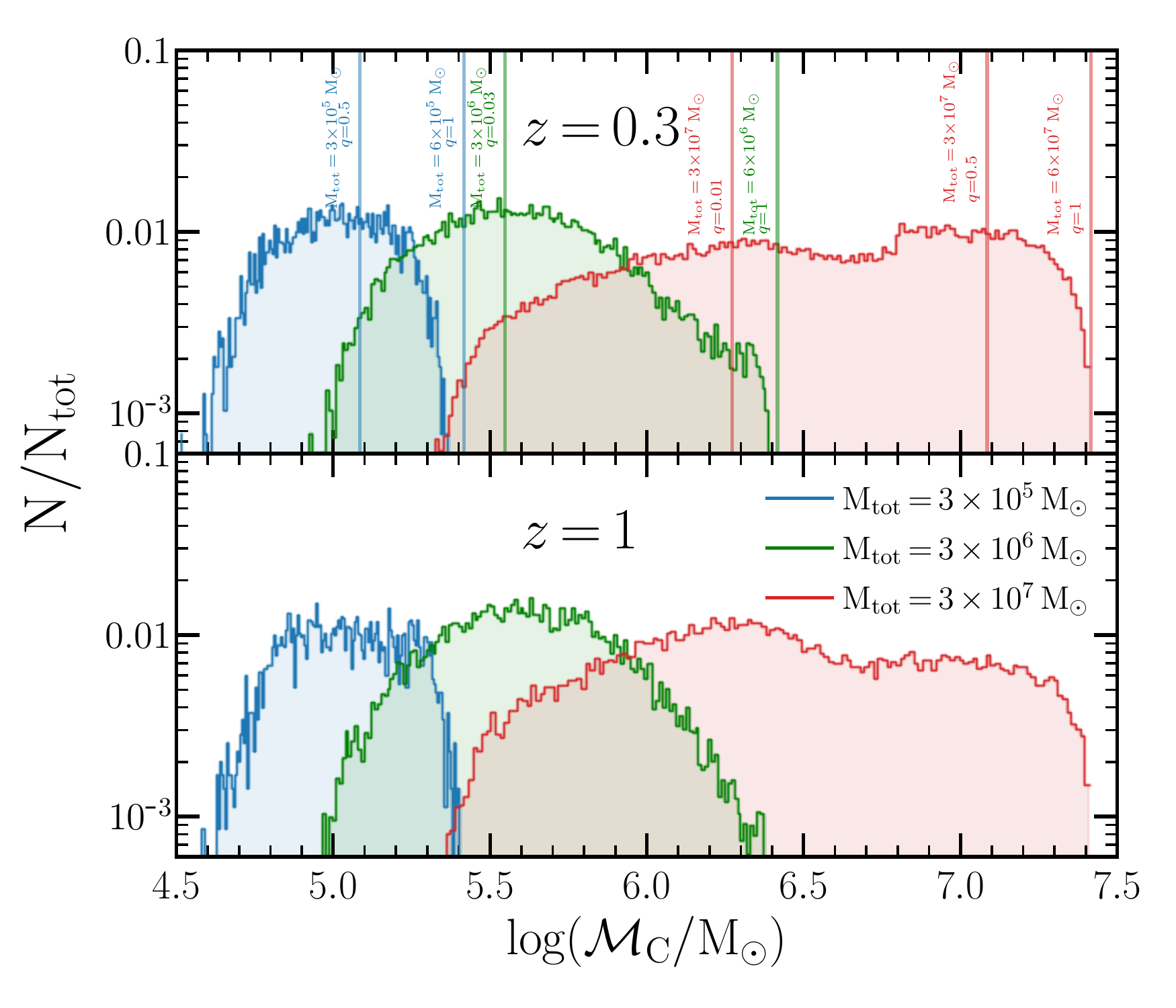}
    \caption[]{Distribution of the chirp mass, $\mathcal{M}_{\rm C}$, associated to the true hosts with mass $\rm M_{\rm tot} \,{=}\, 3 \,{\times}\, 10^5 \, \msun$ (blue), $\rm M_{\rm tot} \,{=}\, 3 \,{\times}\, 10^6 \, \msun$ (green) and $\rm M_{\rm tot} \,{=}\, 3 \,{\times}\, 10^7 \, \msun$ (red) at $z\,{=}\,0.3$ (upper panel) and $z\,{=}\,1$ (lower panel). The other redshifts studied in this work display very similar distributions.}
    \label{fig:ChirpMass_z1}
\end{figure}

\begin{figure*}
    \centering
    \includegraphics[width=2\columnwidth]{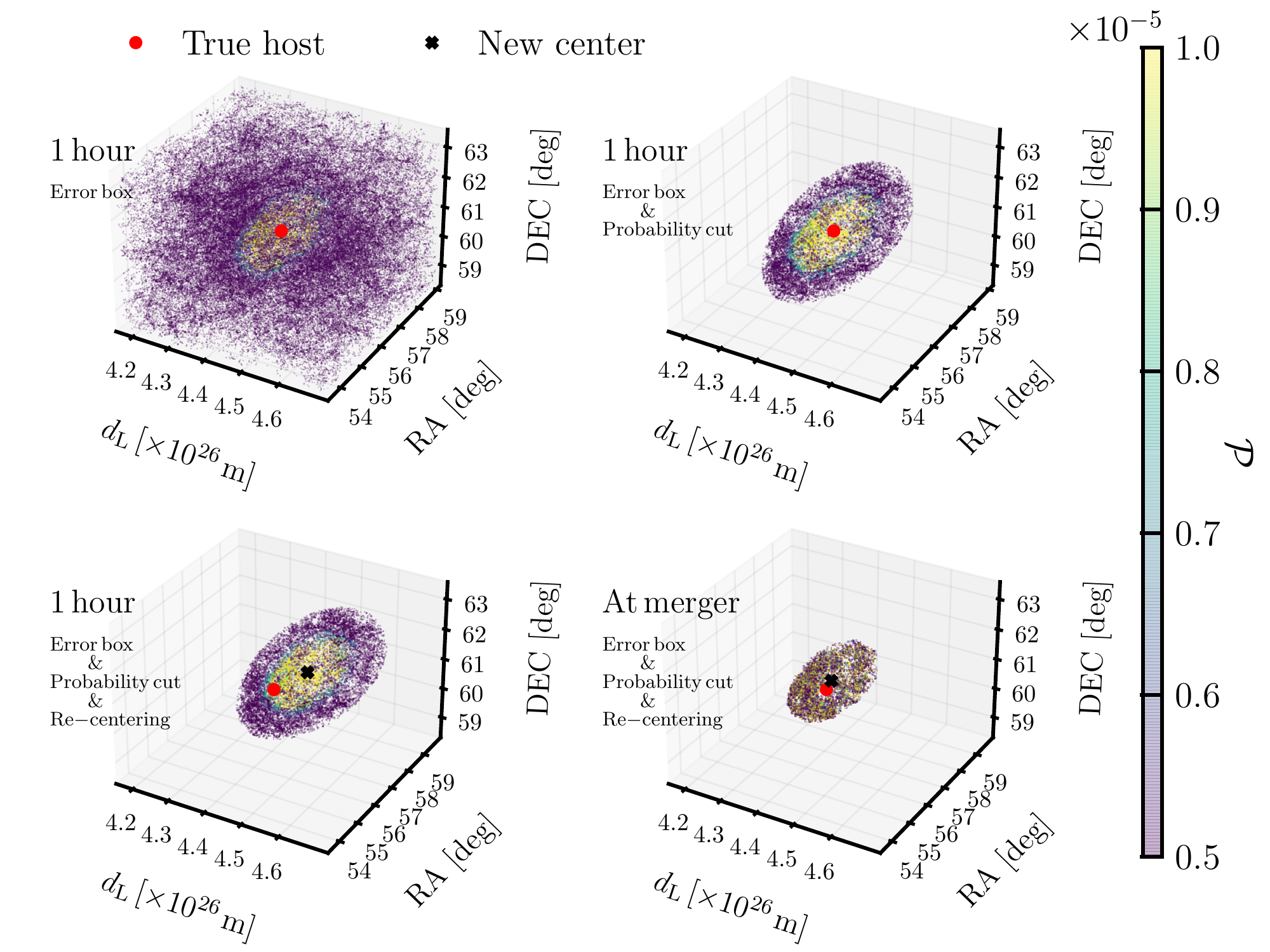}
    \caption[]{Schematic representation of the procedure followed in this project to construct the error-box associated to a LISA event. This example refers to a realization of the MBHB of $\rm M_{\rm tot} = 3 \times 10^6 \, M_{\odot}$ at $z\,{=}\,2$ at $1$ hour before merger. The error-box built at the time of the final coalescence is also shown in the lower right panel.}
    \label{fig:ErrorBoxConstruction}
\end{figure*}
    
The distribution of $\Delta \Omega$ for the true hosts at 1 day, 10 hours and 1 hour before merger and at the coalescence time is presented in Fig.~\ref{fig:DeltaOmega_M3e6}. For the sake of brevity, we only present the results for the $3\,{\times}\,10^6\, \msun$ MBHBs. Lighter and more massive binaries display similar trends (see Appendix~\ref{appendix:LISA_sky_localization}). As expected from the work of \cite{Mangiagli2020}, the value of $\Delta \Omega$ decreases towards the merger time. At times larger than 10 hours, the typical LISA sky-localization of the true hosts displays values ${>}\,10\, \rm deg^2$. On the contrary, at shorter times, $\Delta \Omega$ drops of almost $1\, \rm dex$ and rarely overpasses $1 \rm \, deg^2$. From Fig.~\ref{fig:DeltaOmega_M3e6} we can also observe that, at fixed time the distribution of $\Delta \Omega$ shifts towards larger values as the redshift increases, implying a more challenging sky-localization at high-$z$. Taking into account all of this, we can conclude that future observatories will struggle in the coverage of the sky-region delimited by LISA. For instance, Athena, Lynx or LSST with FOV of $0.4\,\rm \deg^2$, $0.1\,\rm \deg^2$ and $10\,\rm \deg^2$ will need more than $10$ pointings to cover the $\Delta \Omega$ associated to inspiralling MBHBs at 1 hour before the merger (see Section~\ref{sec:DetectionInspirallingBinaries}).\\

In addition to the sky-localization uncertainties, in Fig.~\ref{fig:ChirpMass_z1} we present the distribution of the source-frame chirp mass, $\mathcal{M}_c$, associated to the true hosts. For simplicity, we only present the results at $z\,{=}\,0.3$ and $z\,{=}\,1$. As shown, the distribution of $\mathcal{M}_c$ does not display an important redshift evolution, regardless of the binary mass explored. The true hosts associated to the $3\,{\times}\,10^5\, \msun$ realizations display a peak at $\mathcal{M}_c\,{\sim}\,10^{5.1}\,\msun$ corresponding to binary systems with mass ratio $q\,{\sim}\,0.5$. On the other hand, the binaries related to $3\,{\times}\,10^6\, \msun$ are mainly characterized  by $\mathcal{M}_c\,{\sim}\,10^{5.5}\,\msun$ and $q\,{=}\,0.03$. A long tail towards larger values is also displayed, pointing out that a considerable fraction of the true hosts have $q\,{>}\,0.1$. For the systems of $3\,{\times}\,10^7\, \msun$, a bimodal distribution is seen centered at $\mathcal{M}_c\,{\sim}\,10^{6.4}\,\msun$ and $\mathcal{M}_c\,{\sim}\,10^{7.1}\,\msun$. While the first corresponds to $q\,{\sim}\,0.01$, the latter is related to $q\,{\sim}\,0.5$. Finally, we highlight that in all the three mass bins there are values of $\mathcal{M}_c$ larger than the value expected for $3\,{\times}\,10^{5-6-7}\,\msun$ with $q\,{=}\,1$ (i.e. the maximum value expected). This is just a consequence of the fact that we allow a certain margin when matching lightcone binaries with the realizations produced by \cite{Mangiagli2020}. Indeed, the largest values of the $3\,{\times}\,10^{5-6-7}\,\msun$ distributions correspond to binaries with $q\,{=}\,1$ and mass $6\,{\times}\,10^{5-6-7}\,\msun$ (i.e. the largest masses allowed to link realizations to binaries inside the lightcone). Finally, and for the sake of brevity, in Appendix~\ref{appendix:Ncycles} we show the number of cycles covered by the binary systems at different times before the final coalescence. For instance, binaries of $3\,{\times}\,10^5 \, \msun$  at $z\,{\sim}\,0.5$ 
the number of orbital cycles is ${\sim}\,85$ (${\sim}\,20$)  10 hours (1 hour) before the final merger. This quantity is particularly important for observational studies, given that the orbital period of a binary can imprint distinctive variations in the luminosity emitted by active MBHBs \citep{Gutierrez2022A,Cattorini2022}.

\subsubsection{Building the LISA error-boxes for the inspiralling binaries: Host candidates ($N_{90}$)}

After selecting the true hosts, we assigned their associated LISA error-boxes as the volume centered on the true host  whose extension is given by ${\pm}\,3 \Delta \Omega$ and ${\pm}\,3 \Delta d_{\rm L}$. All the galaxies within this volume are considered as potential hosts of the GW source. To give an idea about the 3 dimensional extension of a LISA error-box and show how crowded of galaxies it is, in the upper left of Fig.~\ref{fig:ErrorBoxConstruction} we show the error-box associated to a random true host of mass $\rm M_{\rm tot} \,{=}\,3\,{\times}\,10^6 \, \msun$ located at $z\,{=}\,2$ and detected $1$ hour before the merger. As we can see, the true host lies at the center of the volume and it is surrounded by a large number of neighbor galaxies distributed in filaments and clusters, forming the well known \textit{cosmic web}. For all the error-boxes studied in this work, we reduced the number of potential hosts by removing the ones with low statistical relevance. To this end, we computed for each galaxy placed inside the error-box its probability, $\mathcal{P}$, of being the host of the GW signal:
\begin{equation}    \label{eq:HostProbability}
    \mathcal{P} \,{\sim}\,  e^{-\frac{1}{2}(\underline{\theta} - \underline{\hat{\theta}}) \Gamma (\underline{\theta} - \underline{\hat{\theta}})^T}
\end{equation}
where $\hat{\underline{\theta}}$  and $\underline{\theta}$ are the vectors containing the right ascension, declination and luminosity distance associated to the true host and each individual galaxy inside the volume, respectively. On the other hand, $\Gamma$ is the $3\,{\times}\,3$ FIM, defined as the inverse of the CM. Once the values of $\mathcal{P}$ have been quantified, we normalized the total probability inside the error-box and ranked the galaxies from the most to the less probable. Then, we computed the cumulative probability of the whole galaxy sample and selected only those objects contributing with the $90\%$. From hereafter, we define these galaxies as the \textit{host candidates} (or just $N_{90}$) of the GW event. All this procedure can be summarized in the the upper left and right panels of Fig. \ref{fig:ErrorBoxConstruction}. As we can see, the galaxies displaying a larger probability correspond to the ones closer to the true host. Moreover, the probability cut leads the error-box to shrink, reshaping it into an ellipsoidal. \\

Finally, to take into account a more realistic scenario in which the LISA error-box is not exactly centered on the true host of the source of the signal, we re-centered the volume on a point randomly chosen from a multivariate Gaussian distribution centered on the true host. As shown in the lower left panel of Fig.~\ref{fig:ErrorBoxConstruction} the new center of the error-box does not coincide anymore with the true host position. Nevertheless, the true host remains inside the region delimited by LISA sky-uncertainties. For completeness, in the lower right panel of Fig.~\ref{fig:ErrorBoxConstruction} we present the final LISA error-box for the same binary but at merger. As we can see, we have fewer number of galaxies thanks to the improvement of LISA sky-localization capabilities as the MBHB approaches to the final plunge.

\section{The simulated X-ray observatories}\label{sec:XrayObservatories}

One of the main goals of this paper is studying the possibility to localize the host of a LISA GW event by associating it with the presence of an X-ray AGN. To explore the detection of such objects by means of X-ray emission, in this section we build three different simulated X-ray observatories by specifying their sensitivity curves. As we will see, their specific characteristics mimic the capabilities of the future Athena and Lynx space detectors.\\

The sensitivity curve of an X-ray observatory can be defined as the minimum integration time (hereafter exposure time, $t_{\rm exp}$) required to reach a specific flux in a given band (hereafter flux limit, $f_{\rm lim}^{\rm band}$). For the purpose of the project, we focused on the soft (0.5-2 keV, Sx) and hard (2-10 keV, Hx) X-ray bands. Based on \cite{McGee2020}, the sensitivity curves in the aforementioned energetic windows scale with the exposure time as:
    \begin{equation}
        f^{\rm Sx}_{\rm lim}(t_{\rm exp}) \,{=}\, A_{\rm Sx} \left(\frac{T_{\rm norm}}{t_{\rm exp}}\right)^{1/2},
    \label{eq:FSx_lim}
    \end{equation}
    
    \begin{equation}
        f^{\rm Hx}_{\rm lim} (t_{\rm exp}) \,{=}\, A_{\rm Hx} \left(\frac{T_{\rm norm}}{t_{\rm exp}}\right)^{1/2},
    \label{eq:FHx_lim}
    \end{equation}

\noindent where $A_{\rm Sx}$, $A_{\rm Hx}$ and $T_{\rm norm}$ are free parameters that fully determine the capabilities of the X-ray observatories. Besides sensitivity curves, another key property is the so-called \textit{confusion limit}, $f_{ \rm c}$, which defines the flux level below which a detector can no longer distinguish a given source from the astronomical background. To determine the confusion limits of the simulated observatories, we followed the methodology presented in \cite{Griffin2020}, which is based on the \textit{source density criterion} of \cite{Condon1974}. In brief, the procedure is a two step process. The first one computes the cumulative number count per solid angle at the confusion limit, $N({>}\,f_{ \rm c})$. The second step determines the value of $f_{ \rm c}$ associated with these number counts by using the empirical model of \cite{Lehmer2012}. Following \cite{Griffin2020}, the value of $N({>}\,f_{ \rm c})$ is computed as:
\begin{equation}
    N({>}\,f_{ \rm c}) \,{=}\, \dfrac{1}{\mathcal{N}_{\rm beam} \, \Omega_{\rm beam}},
\label{eq:CumulativeNumberCount}
\end{equation}
where $\mathcal{N}_{\rm beam} \,{=}\, 30$ corresponds to the number of beams per source \citep[taken from][]{Hogg2001,Vaisanen2001} and $\Omega_{\rm beam}$ is the beam solid angle. By assuming a Gaussian beam pattern, $\Omega_{\rm beam}$ can be expressed as:
\begin{equation}
    \Omega_{\rm beam} \,{=}\, \dfrac{\pi}{4\,\ln(2)} \dfrac{\theta_{\rm FWHM}^2}{(\gamma - 1)},
\end{equation}
where $\theta_{\rm FWHM}$ corresponds to the angular resolution of the X-ray observatory and is defined as the point where the power received from a point source is half its peak value. On the other hand, $\gamma$ is the slope of the power law describing the relation between differential number count and flux, taken from \cite{Lehmer2012}.\\

\begin{table} 
\begin{adjustbox}{width=\columnwidth,center}
\begin{tabular}{cccc|}
\cline{2-4}
\multicolumn{1}{c|}{} & \multicolumn{1}{c|}{ \hspace{0.25cm} $O_1$ \hspace{0.25cm}  } & \multicolumn{1}{c|}{\hspace{0.25cm} $O_2$ \hspace{0.25cm} } & \multicolumn{1}{c|}{\hspace{0.25cm} $O_3$ \hspace{0.25cm} } \\ \hline
\multicolumn{1}{|l|}{$\theta_{\rm FWHM}\,[\rm arcsec]$ }          &               5                &        10                       &                   0.5          \\ 
\multicolumn{1}{|l|}{$T_{\rm norm}\, \rm [ks]$ } &           $100$                    &           $100$                    &  $1$                              \\
\multicolumn{1}{|l|}{$\rm FOV\,[\rm deg^2]$ }          &               0.4                &        0.4                      &                   0.1         \\ \hline
\multicolumn{4}{|c|}{\cellcolor[HTML]{C0C0C0}  \hspace{1.8cm} $\mathbf{0.5\,{-}\,2}$ \textbf{keV}}                                                                \\ 
\multicolumn{1}{|l|}{$A_{\rm Sx}\, \rm [erg\,\, cm^{-2} \, s^{-1}]$}              &        $1.00\,{\times}\,10^{-16}$                       &                  $1.00\,{\times}\,10^{-16}$             &                $1.10\,{\times}\,10^{-16}$              \\ 
\multicolumn{1}{|l|}{$\gamma$}          &            1.5                   &             1.5                  &              2.22                 \\ 
\multicolumn{1}{|l|}{$f_{ \rm c} \, \rm [erg \,\, cm^{-2} \, s^{-1}]$ }    &         $1.69\,{\times}\,10^{-17}$                      &               $2.24\,{\times}\,10^{-16}$                &     $7.80\,{\times}\,10^{-20}$                          \\ \hline 
\multicolumn{4}{|c|}{\cellcolor[HTML]{C0C0C0} \hspace{1.8cm} $\mathbf{2\,{-}\,10}$ \textbf{keV}}                                                                  \\ 
\multicolumn{1}{|l|}{$A_{\rm Hx}\, \rm [erg\,\, cm^{-2} \, s^{-1}]$}              &          $1.50\,{\times}\,10^{-15}$                     &                 $1.50\,{\times}\,10^{-15}$              &           $3.98\,{\times}\,10^{-16}$                    \\ 
\multicolumn{1}{|l|}{$\gamma$}          &          1.32                     &               1.32                &          2.29                     \\ 
\multicolumn{1}{|l|}{$f_{ \rm c} \, \rm [erg \,\, cm^{-2} \, s^{-1}]$}     &         $1.73\,{\times}\,10^{-16}$                      &                    $1.91\,{\times}\,10^{-15}$           &                $1.00\,{\times}\,10^{-19}$               \\ \cline{1-4}
\end{tabular}
\end{adjustbox}
\caption{Values of the parameters employed to derive, in the soft and hard X-ray bands, the sensitivity curves and the confusion fluxes of the three simulated X-ray observatories: $O_1$, $O_2$ and $O_3$.}
\label{tab:X-Observatories_Parameters}
\end{table}

Among all the possible X-ray observatories that can be build based on Eq.~\ref{eq:FSx_lim}, Eq.~\ref{eq:FHx_lim} and Eq.~\ref{eq:CumulativeNumberCount}, in this work we explore three different configurations whose values of $A_{\rm Sx}$, $A_{\rm Hx}$, $T_{\rm norm}$, $\theta_{\rm FWHM}$, $\gamma$ and FOV are listed in Table~\ref{tab:X-Observatories_Parameters}:
\begin{itemize}
    \item  \textit{Observatory 1} ($O_1$): Its properties are those of the Athena space observatory equipped with a Wide Field Imager (WFI) of $0.4\, \rm deg^2$ FOV and a spatial angular resolution of 5 arseconds \citep{Nandra2013}.\\

    \item  \textit{Observatory 2} ($O_2$): This observatory corresponds to the Athena design but with  a poorer angular resolution (10 arcseconds) leading to a larger value of the confusion limit (see Table~\ref{tab:X-Observatories_Parameters}). As we will see, this condition imposes shallower observations than $O_1$. The FOV of this observatory is assumed to be $0.4\, \rm deg^2$.\\
    
    \item \textit{Observatory 3} ($O_3$): Its properties are those of the concept-mission Lynx, with a FOV of 0.1 deg$^2$ and a spatial resolution of 0.5 arseconds \citep{Lynx2018}.

\end{itemize}

As we can see in Table~\ref{tab:X-Observatories_Parameters}, $O_1$ and $O_2$ have similar characteristics with the unique difference concerning the value of $\theta_{\rm FWHM}$. The larger value assumed for $O_2$ is motivated by current challenges in reaching small angular resolutions. Finally, $O_3$ is the configuration with the largest sensitivity and the smallest $f_{\rm c}$.

%\davcoment{We want to stress that no conditions on the field of view (FOV) of $O_1$, $O_2$ and $O_3$ have been imposed. For instance, according to \cite{Nandra2013} an observatory like Athena (i.e. similar to $O_1$) would have a maximum FOV of $0.4\, \rm deg^2$. This value, when compared to the $\Delta \Omega$ expected for LISA (see Fig.~\ref{fig:DeltaOmega_M3e6}) points out that more than one pointing would be required to scan the full LISA error-box. This translates in serious restrictions on the search for the EM counterpart of a GW source. For simplicity, throughout this work, we assume that $O_1$, $O_2$ and $O_3$ would have a large enough FOV to cover the full LISA error-box with one single pointing. However, the reader must keep in mind that this might be too optimistic in many cases and would be only accessible for low-$z$ sources detected at few hours before the merger or at the merger time.} \monica{I am not sure about this statment. This could be critized by the referee. We know that Athena has 0.4 and Lynx 0.1 so we can say that for $O_2$ we assume the same of Athena.... but when we compute the number of galaxies in the field of view we need to take of the value of the field of view of the instrument...and understand when tailing is possible}

\begin{figure}
\centering
\includegraphics[width=1\columnwidth]{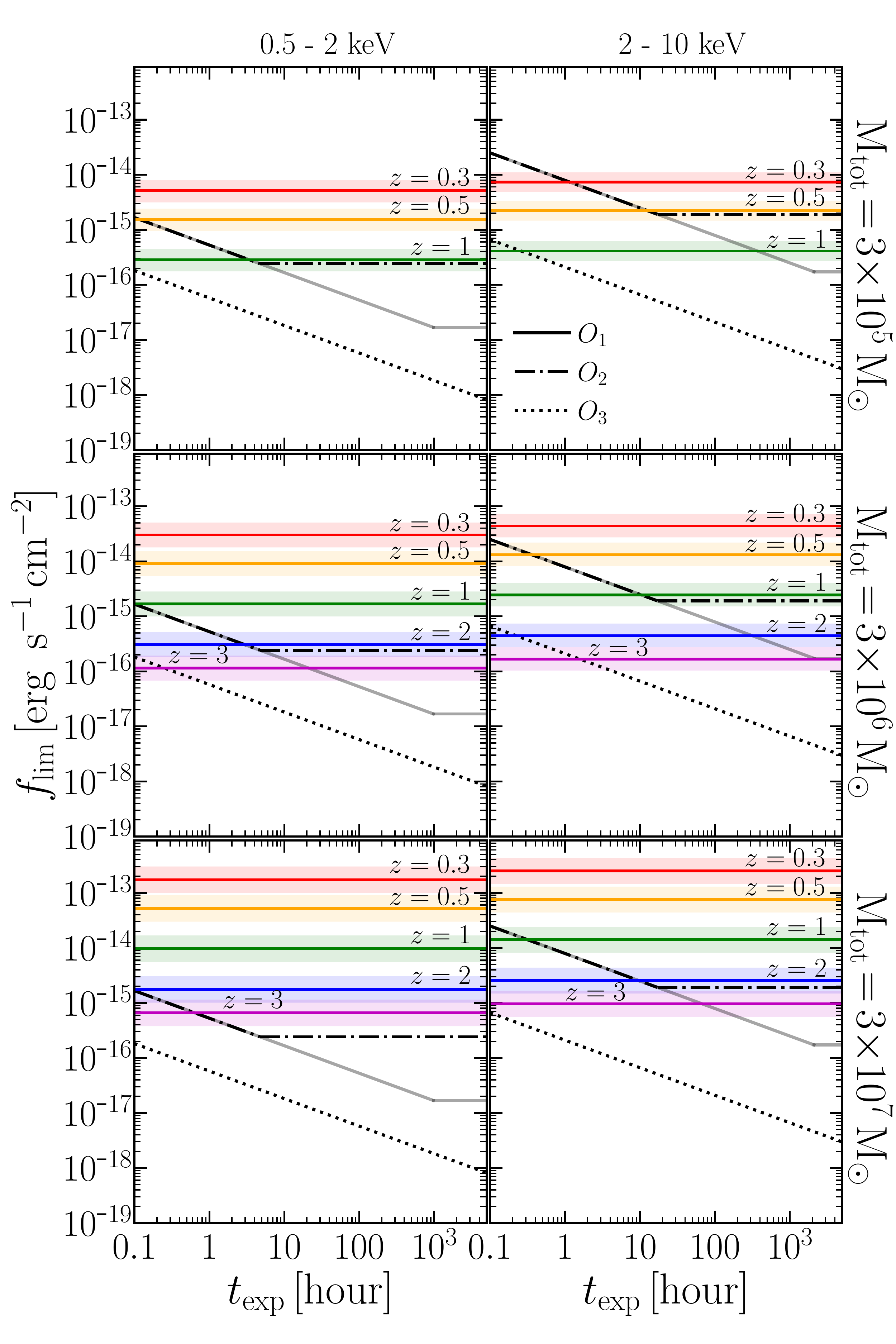}
\caption[]{Sensitivity curves of the three simulated X-ray observatories employed in this project to study a potential cooperation between LISA and an X-ray detector: $O_1$ (solid line), $O_2$ (dash-dotted line) and $O_3$ (dotted line). These functions have been compared with the fluxes at the Eddington limit (horizontal solid lines) associated to the MBHBs, which have been computed by using the bolometric corrections of \cite{Shen2020} and an hydrogen column density $\rm N_{\rm H} = 10^{21.5} \rm cm^{-2}$. Shaded areas are delimited by the maximum and minimum fluxes corresponding to the maximum and minimum values of the bolometric corrections.}
\label{fig:SensitivityCurves}
\end{figure}

\subsection{The X-ray detection of inspiralling and merging LISA binaries} \label{sec:DetectionInspirallingBinaries}

In this section we explore the capabilities of the three X-ray observatories to detect the AGNs associated to the MBHBs during their inspiral and merger phases. To this end, we assume the \textit{most optimistic scenario} possible in which all the binaries of Table~\ref{tab:MBHBs} are related with AGNs radiating at the Eddington limit. Their corresponding X-ray fluxes and luminosities, shown in Table~\ref{tab:Lum_and_flux_Limit_AllBinaries}, have been computed using  Eqs.~\ref{eq:Lbol}-\ref{eq:BolometricCorrection_hard}-\ref{eq:BolometricCorrection_soft} to account for the bolometric corrections in the hard and soft X-ray bands and the galactic absorption related to $\rm N_{\rm H} = 10^{21.5} \rm cm^{-2}$ (i.e. the median value of the hydrogen column density distribution presented in \cite{Masoura2021}). We refer the reader to Appendix~\ref{appendix:ComptonThinScenario} for the same analysis carried out for sources whose X-ray emission is highly absorbed ($\rm N_{H}\,{=}\,10^{23}\,\rm cm^{-2}$).\\
% Then, these fluxes have been attenuated by obscuration assuming $\rm N_{\rm H} = 10^{23.2} \, \rm cm^{-2}$, which is the median value of the hydrogen column density distribution presented in \cite{Masoura2021} (see Section~\ref{sec:BHPopulation} for further details).\\

\begin{table}
\begin{adjustbox}{width=\columnwidth,center}
\begin{tabular}{lcccc|} \cline{2-5}
                       & \multicolumn{1}{|c|}{\diagbox[width=\dimexpr \textwidth/18+2\tabcolsep\relax, height=0.5cm]{ $z$ }{ $\rm Mass$}} & $3\,{\times}\,10^5\,[\msun]$   &  $3\,{\times}\,10^6\,[\msun]$  &  $3\,{\times}\,10^7\,[\msun]$   \\ \hline
\multicolumn{5}{|c|}{\cellcolor[HTML]{C0C0C0} \hspace{1.8cm}  $\mathbf{0.5\,{-}\,2}$ \textbf{keV}} \\
\multicolumn{1}{|l|}{$\rm L_{\rm X-ray}\,[erg\,s^{-1}]$}   &    -    &   $1.59\,{\times}\,10^{42}$    &   $9.35\,{\times}\,10^{42}$   &   $5.35\,{\times}\,10^{43}$  \\ \cline{1-5}
 \multicolumn{1}{|l|}{}                       & 0.3    &  $5.13\,{\times}\,10^{-15}$     &  $3.02\,{\times}\,10^{-14}$    &   $1.73\,{\times}\,10^{-13}$   \\
 \multicolumn{1}{|l|}{}                      & 0.5    &  $1.55\,{\times}\,10^{-15}$     &   $9.09\,{\times}\,10^{-15}$    &  $5.20\,{\times}\,10^{-14}$    \\
 \multicolumn{1}{|l|}{}                      & 1     &   $2.87\,{\times}\,10^{-16}$    &   $1.68\,{\times}\,10^{-15}$    &   $9.63\,{\times}\,10^{-15}$   \\
 \multicolumn{1}{|l|}{}                      & 2     & -     &  $3.07\,{\times}\,10^{-16}$   &  $1.76\,{\times}\,10^{-15}$   \\
\multicolumn{1}{|l|}{\multirow{-5}{*}{$f \, \rm [erg \,\, cm^{-2} \, s^{-1}]$}}    & 3      & -     &  $1.15\,{\times}\,10^{-16}$    &   $6.58 \,{\times}\,10^{-16}$  \\ \cline{1-5}
\multicolumn{5}{|c|}{\cellcolor[HTML]{C0C0C0} \hspace{1.8cm} $\mathbf{2\,{-}\,10}$ \textbf{keV}}  \\
\multicolumn{1}{|l|}{$\rm L_{\rm X-ray}\,[erg\,s^{-1}]$}  &    -    &   $2.28\,{\times}\,10^{42}$    &   $1.36\,{\times}\,10^{43}$   &   $7.77\,{\times}\,10^{43}$   \\ \cline{1-5}
 \multicolumn{1}{|l|}{}                       & 0.3    &   $7.35\,{\times}\,10^{-15}$    &  $4.39\,{\times}\,10^{-14}$    &    $2.51\,{\times}\,10^{-13}$  \\
 \multicolumn{1}{|l|}{}                      & 0.5    &    $2.22\,{\times}\,10^{-15}$   &   $1.32\,{\times}\,10^{-14}$   &    $7.56\,{\times}\,10^{-14}$  \\
 \multicolumn{1}{|l|}{}                      & 1      &    $4.11\,{\times}\,10^{-16}$   &   $2.45\,{\times}\,10^{-15}$   &   $1.40\,{\times}\,10^{-14}$   \\
 \multicolumn{1}{|l|}{}                      & 2      &   -   &     $4.47\,{\times}\,10^{-16}$  &  $2.55\,{\times}\,10^{-15}$    \\
\multicolumn{1}{|l|}{\multirow{-5}{*}{$f \, \rm [erg \,\, cm^{-2} \, s^{-1}]$}}    & 3      &  -     &  $1.68\,{\times}\,10^{-16}$     &  $9.56\,{\times}\,10^{-16}$ \\  \hline
\end{tabular}
\end{adjustbox}
\caption{Luminosities and fluxes in the $0.5\,{-}\,2\,\rm keV$ and $2\,{-}\,10\,\rm keV$ bands emitted by MBHBs accreting at the Eddington limit, assuming an hydrogen column density $\rm N_{\rm H} = 10^{21.5} \rm cm^{-2}$. We refer the reader to Table~\ref{tab:Lum_and_flux_Limit_AllBinaries_ComptonThin} for the case of sources with $\rm N_{\rm H}\,{=}\,10^{23} \, \rm cm^{-2}$.}
\label{tab:Lum_and_flux_Limit_AllBinaries}
\end{table}

In Fig.~\ref{fig:SensitivityCurves} we present for the $0.5\,{-}\,2\,\rm keV$ and $2\,{-}\,10\,\rm keV$ bands a comparison between the sensitivity curves of the observatories and the fluxes associated to all the binary systems. As we can see, $O_3$ is able to detect with exposure times $t_{\rm exp}\,{<}\,0.1$ hours all the MBHBs studied in this work, regardless of redshift. The observatory $O_1$ can also detect all the binaries, but the values of $t_{\rm exp}$ display important differences. For instance, in the $0.5\,{-}\,2$ keV band, $O_1$ needs ${\sim}\,3$ (${\sim}\,20$) hours to reach the minimum flux required to detect an Eddington-limited binary of $3\,{\times}\,10^5 \, \msun$ ($3\,{\times}\,10^6 \, \msun$) at $z\,{=}\,1$ ($z\,{=}\,3$). For the $2\,{-}\,10$ keV band, the same binaries require longer $t_{\rm exp}$. Observatory $O_2$ behaves similarly to $O_1$, but its larger $f_c$ prevents the detection of some high-$z$ binaries. This is the case of MBHBs with $3\,{\times}\,10^5 \, \msun$ at $z\,{>}\,0.5$ in the $2\,{-}\,10\,\rm keV$ band or MBHBs of $3\,{\times}\,10^6 \, \msun$ at $z\,{>}\,2$ in the $0.5\,{-}\,2\,\rm keV$ band.\\

From Fig.~\ref{fig:SensitivityCurves} it is clear that the three X-ray observatories display a large potential to detect all the MBHBs studied in this work. However, our goal does not only concern their detection, but also the minimum time required to do so. When LISA will detect and constrain the sky-position of a GW signal coming from an inspiralling MBHB, the X-ray observatories will have a limited time before the merger event to scan the sky and detect the associated AGN. After that time, the possibility of investigating the pre-merger phase will be lost and only merger and post-merger studies will be accessible. Therefore, the feasibility of studying MBHBs during their last inspiral phase through multi-messenger astronomy depends on the minimum time required by the observatories to detect these systems. Taking into account the importance of this, in Table~\ref{tab:Exposure_times} we characterize the minimum exposure time needed by the X-ray observatories to detect the target binaries.  
%\as{Following on my comment about obscuration, can't we show also a table where we assume average levels of obscuration from the Masoura et al distribution? I don't expect those will affect the hard-X part of the spectrum at all, but they might have a significant impact on the soft-X exposure time requested.}
As we can see, the large sensitivity reached by $O_3$ allows the detection of all the AGNs associated with binaries of $3{\times}\,10^7\,\msun$ in less than ${<}\,0.1$ hours, both in the soft and hard X-ray bands. This trend also applies to most of the other combinations of $\rm M_{\rm tot}$ and $z$, with only few AGNs needing exposure times lasting from tens of minutes to an hour (see e.g $3\,{\times}\,10^6\,\msun$ at $z\,{=}\, 3$ in the hard X-ray band). Thus, $O_3$ will be able to monitor the whole inspiralling phase of binaries with masses in between $10^5\,{-}\,10^7\, \msun$. Similar capabilities are reached by $O_1$ and $O_2$ when targeting systems with $3\,{\times}\,10^7\, \msun$. This is caused by the fact that these binaries, radiating at the Eddington limit, are bright X-ray sources (${>}\,5\,{\times}\,10^{43}\rm \, erg/s$) easily accessible by any observatory. For $3\,{\times}\,10^5\, \msun$ and $3\,{\times}\,10^6\, \msun$ binaries, the smaller flux sensitivity of $O_1$ and $O_2$ (especially in the hard X-ray band) imposes larger exposure times needed to reach a detection. For example, observations lasting hundreds of hours (tens of hours) will be required by $O_1$ in order to detect MBHBs of $3{\times}\,10^6\,\msun$ at $z=2$ ($z=1$) in the hard X-ray band. These times are reduced in the soft band where less than three hours are needed. Similar trends characterize MBHBs of $3{\times}\,10^5\,\msun$ at $z\,{=}\,1$. Taking into account this, $O_1$ and $O_2$ will be blind to the hard (soft) X-ray signatures of $3{\times}\,10^{5-6}\,\msun$ MBHBs at $z\,{>}\,1$ in the last 20 to 3 hours (${<}\,2$ hours) prior to merger. The most promising cases of $O_1$ and $O_2$ concern the low-$z$ systems ($z\,{\leq}\,0.5$) in which detections require less than $0.1$ hours for both $3\,{\times}\,10^5\, \msun$ and $3\,{\times}\,10^6\, \msun$ binaries in the soft X-ray band.\\

\begin{table}
\begin{adjustbox}{width=\columnwidth,center}
\begin{tabular}{cccccccccc|}\cline{2-10}
                          & \multicolumn{3}{|c|}{$3\,{\times}\,10^5\,[\msun]$} & \multicolumn{3}{c|}{$3\,{\times}\,10^6\,[\msun]$} & \multicolumn{3}{c|}{$3\,{\times}\,10^7\,[\msun]$} \\ \cline{2-10}
\multicolumn{1}{c|}{}       & $O_1$     & $O_2$    & \multicolumn{1}{c|}{$O_3$}    & $O_1$     & $O_2$    & \multicolumn{1}{c|}{$O_3$}    & $O_1$     & $O_2$    & $O_3$    \\ \cline{1-10}
\rowcolor[HTML]{C0C0C0} 
\multicolumn{1}{|l|}{\diagbox[width=\dimexpr \textwidth/12+2\tabcolsep\relax, height=0.85cm]{ $z$ }{ $t_{\rm exp} \rm [h]$}} & \multicolumn{9}{c|}{\cellcolor[HTML]{C0C0C0}$\mathbf{0.5\,{-}\,2}$ \textbf{keV}}                    \\
\multicolumn{1}{|l|}{0.3}                       &  ${<}\,0.1$    &  ${<}\,0.1$     &  \multicolumn{1}{l|}{${<}\,0.1$}     &    ${<}\,0.1$    &   ${<}\,0.1$    &   \multicolumn{1}{l|}{${<}\,0.1$}     &   ${<}\,0.1$     &    ${<}\,0.1$   &     ${<}\,0.1$  \\
\multicolumn{1}{|l|}{0.5}                       &    0.116    &   0.116    &   \multicolumn{1}{l|}{${<}\,0.1$}    &     ${<}\,0.1$   &     ${<}\,0.1$  &   \multicolumn{1}{l|}{${<}\,0.1$}     &   ${<}\,0.1$     &   ${<}\,0.1$    &   ${<}\,0.1$    \\
\multicolumn{1}{|l|}{1}                       &    3.37     &    3.37    &   \multicolumn{1}{l|}{${<}\,0.1$}    &    ${<}\,0.1$    &   ${<}\,0.1$    &  \multicolumn{1}{l|}{${<}\,0.1$}     &   ${<}\,0.1$     &  ${<}\,0.1$     &   ${<}\,0.1$    \\
\multicolumn{1}{|l|}{2}                       &    -        &   -        &   \multicolumn{1}{l|}{-}            &    2.95    &    2.95   &   \multicolumn{1}{l|}{${<}\,0.1$}    &   ${<}\,0.1$     &   ${<}\,0.1$     &   ${<}\,0.1$    \\
\multicolumn{1}{|l|}{3}                       &    -        &    -       &    \multicolumn{1}{l|}{-}           &    21.0    &  \xmark     &  \multicolumn{1}{l|}{0.25}     &  0.64      &    0.64   &    ${<}\,0.1$   \\ \cline{1-10}
 
\multicolumn{1}{c|}{}                          & $O_1$     & $O_2$    & \multicolumn{1}{c|}{$O_3$}    & $O_1$     & $O_2$    & \multicolumn{1}{c|}{$O_3$}    & $O_1$     & $O_2$    & $O_3$    \\ \cline{1-10}
\rowcolor[HTML]{C0C0C0} 
\multicolumn{1}{|l|}{\diagbox[width=\dimexpr \textwidth/12+2\tabcolsep\relax, height=0.85cm]{ $z$ }{ $t_{\rm exp} \rm [h]$}}& \multicolumn{9}{c|}{\cellcolor[HTML]{C0C0C0}$\mathbf{2\,{-}\,10}$ \textbf{keV}}    \\
\multicolumn{1}{|l|}{0.3}                       &  1.16    &  1.16     &  \multicolumn{1}{l|}{${<}\,0.1$}     &    ${<}\,0.1$    &   ${<}\,0.1$    &   \multicolumn{1}{l|}{${<}\,0.1$}     &   ${<}\,0.1$     &    ${<}\,0.1$   &     ${<}\,0.1$  \\
\multicolumn{1}{|l|}{0.5}                       &    12.68       &   12.68         &   \multicolumn{1}{l|}{${<}\,0.1$}    &      0.36            &    0.36   &    \multicolumn{1}{l|}{${<}\,0.1$}    &   ${<}\,0.1$     &  ${<}\,0.1$     &   ${<}\,0.1$   \\
\multicolumn{1}{|l|}{1}                       &     370.0      &   \xmark        &   \multicolumn{1}{l|}{0.261}         &      10.41            &   10.41    &   \multicolumn{1}{l|}{${<}\,0.1$}     &  0.32      &   0.32    &   ${<}\,0.1$    \\
\multicolumn{1}{|l|}{2}                       &     -          &    -            &   \multicolumn{1}{l|}{-}             &       312.8           &    \xmark   &     \multicolumn{1}{l|}{0.22}   &  9.61      &   9.61    &   ${<}\,0.1$    \\
\multicolumn{1}{|l|}{3}                       &    -           &     -           &    \multicolumn{1}{l|}{-}            &       \xmark           &     \xmark  &   \multicolumn{1}{l|}{1.56}     &   68.38     &   \xmark    &   ${<}\,0.1$   \\ \hline
\end{tabular}
\end{adjustbox}
\caption{Average minimum exposure times, $t_{\rm exp}$ (in hours) required to detect the binaries emitting at the Eddington limit with an hydrogen column density of $\rm N_{\rm H} = 10^{21.5} \rm cm^{-2}$. The values of $t_{\rm exp}$ are computed for the three different X-rays observatories ($O_1$, $O_2$ and $O_3$) at five different redshifts ($z \,{=}\,0.3$, $0.5$, $1$, $2$ and $3$). The values have been presented for two different bands: $\rm 0.5\,{-}\,2\, keV$ (soft X-rays) and $\rm 2\,{-}\,10\, keV$ (hard X-rays). Crosses are drawn when the detection of a MBHB is not possible. These values correspond to the intersection between the horizontal solid lines in Fig~\ref{fig:SensitivityCurves} and the sensitivity curves of the three observatories. We refer the reader to Table~\ref{tab:Exposure_times_ComptonThin} for the case of sources in which a hydrogen column density of $\rm N_{\rm H}\,{=}\,10^{23} \rm \, cm^{-2}$ is assumed.}
\label{tab:Exposure_times}
\end{table}

\begin{table}
\begin{adjustbox}{width=\columnwidth,center}
\begin{tabular}{ccccccc|}\cline{2-7}
                          & \multicolumn{2}{|c|}{$3\,{\times}\,10^5\,[\msun]$} & \multicolumn{2}{c|}{$3\,{\times}\,10^6\,[\msun]$} & \multicolumn{2}{c|}{$3\,{\times}\,10^7\,[\msun]$} \\ \cline{2-7}
\multicolumn{1}{c|}{}       & $O_1$/$O_2$  & \multicolumn{1}{c|}{$O_3$}    & $O_1$/$O_2$    & \multicolumn{1}{c|}{$O_3$}    & $O_1$/$O_2$    & $O_3$    \\ \cline{1-7}
\rowcolor[HTML]{C0C0C0} 
\multicolumn{1}{|l|}{\diagbox[width=\dimexpr \textwidth/12+2\tabcolsep\relax, height=0.85cm]{ $z$ }{ $N_{\rm p}$}} & \multicolumn{6}{c|}{\cellcolor[HTML]{C0C0C0} \textbf{10 hours prior to merger} }                    \\
\multicolumn{1}{|l|}{0.3}   &  \textcolor{mygreen}{$2.056$}      &  \multicolumn{1}{l|}{\textcolor{mygreen}{$8.226$}}     &    \textcolor{mygreen}{$2.888$}    &   \multicolumn{1}{l|}{\textcolor{mygreen}{$11.553$}} &  \textcolor{mygreen}{$5.516$}  &   \textcolor{mygreen}{$22.065$}  \\
\multicolumn{1}{|l|}{0.5}           &    \textcolor{red}{9.980}    &   \multicolumn{1}{l|}{\textcolor{mygreen}{$39.921$}}    &     \textcolor{mygreen}{$5.159$}  &   \multicolumn{1}{l|}{\textcolor{mygreen}{$20.635$}}     &   \textcolor{mygreen}{$13.668$}    &   \textcolor{mygreen}{$54.671$}    \\
\multicolumn{1}{|l|}{1}                       &    \textcolor{red}{24.992}      &   \multicolumn{1}{l|}{\textcolor{red}{$99.966$}}    &    \textcolor{red}{$20.406$}      &  \multicolumn{1}{l|}{\textcolor{mygreen}{$81.625$}}     &   \textcolor{red}{$111.942$}       &   \textcolor{red}{$447.768$}    \\
\multicolumn{1}{|l|}{2}                       &    -            &   \multicolumn{1}{l|}{-}            &    \textcolor{red}{71.400}     &   \multicolumn{1}{l|}{\textcolor{red}{$285.601$}}    &   \textcolor{red}{$928.436$}      &   \textcolor{red}{$3713.743$}    \\
\multicolumn{1}{|l|}{3}                       &    -          &    \multicolumn{1}{l|}{-}           &    \textcolor{red}{171.229}     &  \multicolumn{1}{l|}{\textcolor{red}{684.916}}     &  -    &  -   \\ \cline{1-7}
 
\multicolumn{1}{c|}{}                          & $O_1$/$O_2$    & \multicolumn{1}{c|}{$O_3$}    & $O_1$/$O_2$    & \multicolumn{1}{c|}{$O_3$}    & $O_1$/$O_2$    & $O_3$    \\ \cline{1-7}
\rowcolor[HTML]{C0C0C0} 
\multicolumn{1}{|l|}{\diagbox[width=\dimexpr \textwidth/12+2\tabcolsep\relax, height=0.85cm]{ $z$ }{ $N_{\rm p}$}} & \multicolumn{6}{c|}{\cellcolor[HTML]{C0C0C0} \textbf{1 hour prior to merger} }                    \\
\multicolumn{1}{|l|}{0.3}                       &  \textcolor{mygreen}{$0.118$}       &  \multicolumn{1}{l|}{\textcolor{mygreen}{$0.472$}}      &  \textcolor{mygreen}{$0.479$}  & \multicolumn{1}{l|}{\textcolor{mygreen}{$1.918$}}     &   \textcolor{mygreen}{$5.131$}     & \textcolor{mygreen}{$20.524$}  \\
\multicolumn{1}{|l|}{0.5}                       &    \textcolor{red}{0.843}      &   \multicolumn{1}{l|}{\textcolor{mygreen}{$3.374$}}    &     \textcolor{mygreen}{$1.454$}    &   \multicolumn{1}{l|}{\textcolor{mygreen}{$5.818$}}     &   \textcolor{mygreen}{$13.435$}     &   \textcolor{mygreen}{$53.741$}    \\
\multicolumn{1}{|l|}{1}                       &    \textcolor{red}{1.577}      &   \multicolumn{1}{l|}{\textcolor{mygreen}{$6.309$}}   &    \textcolor{red}{$3.556$}    &  \multicolumn{1}{l|}{\textcolor{mygreen}{$14.223$}}    &   \textcolor{red}{$103.722$}   &    \textcolor{red}{$414.888$}    \\
\multicolumn{1}{|l|}{2}                       &    -          &   \multicolumn{1}{l|}{-}            &    \textcolor{red}{15.544}    &   \multicolumn{1}{l|}{\textcolor{red}{$62.176$}}    &   \textcolor{red}{$928.436$}      &   \textcolor{red}{$3713.743$}    \\
\multicolumn{1}{|l|}{3}                       &    -          &    \multicolumn{1}{l|}{-}           &    \textcolor{red}{39.150}   &  \multicolumn{1}{l|}{\textcolor{red}{156.601}}     &  -     &    -   \\ \cline{1-7}
 
\multicolumn{1}{c|}{}                          & $O_1$/$O_2$    & \multicolumn{1}{c|}{$O_3$}    & $O_1$/$O_2$    & \multicolumn{1}{c|}{$O_3$}    & $O_1$/$O_2$    & $O_3$    \\ \cline{1-7}

\rowcolor[HTML]{C0C0C0} 
\multicolumn{1}{|l|}{\diagbox[width=\dimexpr \textwidth/12+2\tabcolsep\relax, height=0.85cm]{ $z$ }{ $N_{\rm p}$}}& \multicolumn{6}{c|}{\cellcolor[HTML]{C0C0C0} \textbf{At merger}}    \\
\multicolumn{1}{|l|}{0.3}                       &  \textcolor{mygreen}{0.027}         &  \multicolumn{1}{l|}{\textcolor{mygreen}{$0.109$}}     &    \textcolor{mygreen}{$0.015$}     &   \multicolumn{1}{l|}{\textcolor{mygreen}{$0.059$}}     &   \textcolor{mygreen}{$0.046$}     &     \textcolor{mygreen}{$0.185$}  \\
\multicolumn{1}{|l|}{0.5}                       &    \textcolor{mygreen}{0.173}       &   \multicolumn{1}{l|}{\textcolor{mygreen}{$0.691$}}    &      \textcolor{mygreen}{0.032}             &    \multicolumn{1}{l|}{\textcolor{mygreen}{$0.131$}}    &   \textcolor{mygreen}{$0.061$}        &   \textcolor{mygreen}{$0.245$}   \\
\multicolumn{1}{|l|}{1}                       &     \textcolor{mygreen}{0.200}        &   \multicolumn{1}{l|}{\textcolor{mygreen}{0.800}}         &       \textcolor{mygreen}{0.308}            &   \multicolumn{1}{l|}{\textcolor{mygreen}{$1.231$}}     &   \textcolor{mygreen}{0.340}      &   \textcolor{mygreen}{$1.362$}    \\
\multicolumn{1}{|l|}{2}                       &     -            &   \multicolumn{1}{l|}{-}             &       \textcolor{mygreen}{0.778}           &     \multicolumn{1}{l|}{\textcolor{mygreen}{3.112}}   &  \textcolor{mygreen}{0.879}       &   \textcolor{mygreen}{$3.517$}    \\
\multicolumn{1}{|l|}{3}                       &    -             &    \multicolumn{1}{l|}{-}            &       \textcolor{mygreen}{1.850}          &   \multicolumn{1}{l|}{\textcolor{mygreen}{7.402}}     &   -  &  -  \\ \hline
\end{tabular}
\end{adjustbox}
\caption{Average number of pointings, $N_{\rm p}$, needed by the X-ray observatories $O_1$, $O_2$ and $O_3$ to cover the $\Delta \Omega$ associated to the true hosts of $\rm M_{tot}\,{=}\, 3\,{\times}\,10^{5-6-7}$ at $z \,{=}\,0.3$, $0.5$, $1$, $2$ and $3$. Three different times were considered: $10$ and $1$ hours before merger and the time of the final coalescence. The results for observatories $O_1$ and $O_2$ are gathered together since they share the same FOV. The numbers in \textcolor{mygreen}{green} (\textcolor{red}{red}) color correspond to the cases where the number of pointings multiplied by the exposure time of Table~\ref{tab:Exposure_times} required to detect the binary in the hard X-ray band is smaller (larger) than the corresponding time to merger. We refer the reader to Table~\ref{tab:Exposure_times_ComptonThin} for the case of sources in which a hydrogen column density of $\rm N_{\rm H}\,{=}\,10^{23} \rm \, cm^{-2}$ is assumed.}
\label{tab:Pointings}
\end{table}

%Another important property that determines the efficiency of the simulated X-ray observatories in detecting inspiralling and merging MBHB event corresponds to the number of pointings, $N_{\rm p}$ required to cover the LISA error-box. Here we define $N_{\rm p}$ as the ratio between the $\Delta \Omega$ associated to the binaries with the FOV of the X-ray observatory. needed to scan the full LISA projected error-box. These numbers are shown in Table~\ref{tab:Pointings} at $10$ and $1$ hours before the merger and at the time of the final plunge. They have been defined as the ratio between the median value of the $\Delta \Omega$ determined by LISA sky-localization capabilities for a given combination of $\rm M_{\rm tot}$ and $z$ and the FOV of $O_1$, $O_2$ and $O_3$. 

Besides exposure times, the feasibility of the X-ray observatories in detecting inspiralling MBHBs is also determined by their efficiency in covering the full angular extension of the LISA error-box. As we discussed in Section~\ref{sec:LISA_Error_Boxes}, at high redshifts the values of $\Delta \Omega$ associated with the MBHBs detected by LISA can be very large (up to $10^3\, \rm deg^2$), forcing the X-ray observatories to do multiple pointings to cover the full projection of the LISA error-box in the DEC-RA plane. As a consequence, the values of $t_{\rm exp}$ summarized in Table~\ref{tab:Exposure_times} will increase proportionally to the number of pointings,  hampering even more the possibility of detecting inspiralling MBHBs. Here we define $N_{\rm p}$ as the ratio between the $\Delta \Omega$ associated to the binaries (see Figure~\ref{fig:DeltaOmega_M3e6}) and the FOV of the X-ray observatory.  In Table~\ref{tab:Pointings} we report the values of $N_{\rm p}$ required by each observatory to cover the projected LISA error-box. Besides, in Table~\ref{tab:Pointings} we have colored in green (red) the cases in which the number of pointings permit (prevent) the full coverage of the LISA projected error-box within the inspiral time of the binary. To determine these cases, we have multiplied $N_{\rm p}$ by the corresponding exposure times presented in Table~\ref{tab:Exposure_times}. For simplicity we have used the values of $t_{\rm exp}$ associated to the $2\,{-}\,10\, \rm keV$ band, but the same approach can be applied for the $0.5 \,{-}\,2\, \rm keV$ energy window. We stress that, regardless of the value of $N_{\rm p}$, at the merger time all the X-ray detectors are able to cover the full LISA projected error-box since they are not constrained by any inspiral time. As shown in Table~\ref{tab:Exposure_times}, as the time to merger increases, the LISA sky-localization get worse and the values of $N_{\rm p}$ raise. For instance, $O_1$ ($O_3$) needs $N_{\rm p}\,{\sim}\,2$ (${\sim}\,8$) to cover the full projected LISA error-box associated to a $z\,{=}\,0.3$ MBHB of $\rm 3 \,{\times}\,10^5\, \msun$ at $10$ hours before merger. For the same system, but at the time of the final coalescence, $O_1$ requires less than 5\% of its FOV to scan the sky-area delimited by LISA uncertainties. These value increase by a factor of $1.4$ and $2.6$ for a $3 \,{\times}\,10^6\,\msun$ and a $3 \,{\times}\,10^7\,\msun$ binary, respectively. This increase is caused by the fact that the LISA error-boxes associated to higher-mass systems are generally wider than those related to less massive binaries. Larger number of pointings are required not just for high-mass MBHBs, but also towards higher redshifts, as a consequence of the poorer LISA sky-localization. Indeed, in the pre-merger phase, the X-ray observatories are not able to cover the LISA sky-area associated to $z\,{>}\,1$ systems, requiring an amount of time exceeding the inspiral time of the detected binary. Thus, a complete covering of the FOV of LISA within the inspiralling time will be \textit{only} feasible for binary system of ${\leq}\,3\,{\times}\,10^7 \, \msun$ at $z\,{\lesssim}\,1$.\\

All the previous results highlight that characterizing LISA \textit{inspiralling} MBHBs through X-ray observations will be challenging. The necessity of deep sensitivities and wide FOV to detect in a short time-scale faint AGNs imposes strong limitations. These ones are particularly important at high-$z$, where most of the X-ray observatories are blind to inspiralling LISA sources, especially in the hard X-rays. Despite this, optimistic scenarios have been shown. The best candidates to perform multi-messenger astronomy in the inspiral phase are low-$z$ ($z\,{\lesssim}\,1$) binaries with $\rm M_{\rm tot}\,{\leq}\,3\,{\times}\,10^7\msun$. As we will see in the next sections, the worsening of LISA sky-localization capabilities towards higher redshifts will prevent, during the inspiral phase, feasible searches of EM counterparts associated to heavier systems. These systems can be bright X-ray sources and thus easily detected. However their associated error-boxes will exceed by several orders of magnitude the FOV of future X-ray observatories.

\section{Results}
\label{sec:Results}
In this section, for the binaries of Table~\ref{tab:MBHBs}, we compute  the number of galaxies, and X-ray AGNs observable either in the soft or in the hard band inside the corresponding LISA error-box. This will help us quantify whether an X-ray counterpart would be a unique signature of a  MBHB merger and, if not, how many potential \textit{false positive} will fall in the LISA error-box. Furthermore, under the assumption that the true host is associated to an AGN radiating at the Eddington limit (i.e the most optimistic scenario) we explore if the environment in which the source is embedded could display particular features not shared with other AGNs inside the LISA error-box. Among all the inspiral times accessible by the realizations of Section~\ref{sec:LISA_Error_Boxes}, we have focused on three reference times: 10 hours and 1 hour before the merger and at merger. As discussed in the same section, the error-boxes associated with inspiralling times larger than 10 hours are so wide to hamper the search for the host of the GW source.
    
    \begin{figure}
        \includegraphics[width=1\columnwidth]{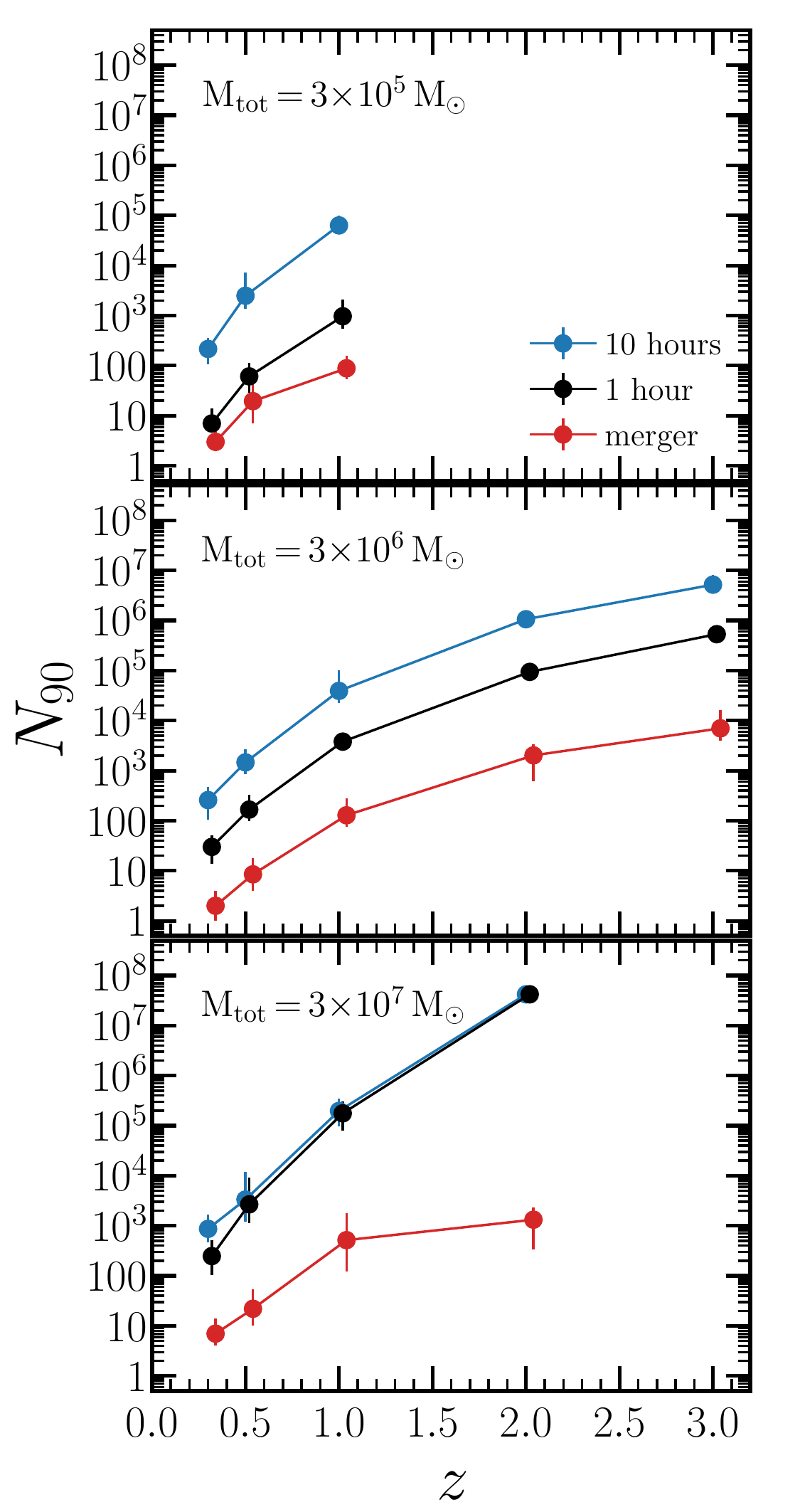}
        \caption{Median values of the total number of host candidates ($N_{90}$) laying inside the error-boxes associated to the true hosts and delimited by $\Delta \Omega$ and $\Delta d_{\rm L}$ values. The error-bars are taken from the $32^{\rm th}\,{-}\,68^{\rm th}$ percentiles. From top to bottom we show the results for $3\,{\times}\,10^5\msun$, $3\,{\times}\,10^6\msun$ and $3\,{\times}\,10^7\msun$. Different colors represent different times during the evolution of the systems: $10$ hours before merger (blue), $1$ hour before merger (green), and merger (red).}
        \label{fig:N90}
    \end{figure}
    
\subsection{Galaxies inside the LISA error-box}
\label{subsec:N90_LISA}

When considering multi-messenger astronomy, one of the key quantities that must be taken into account is the number of galaxies inside the LISA error-box that can be considered as potential hosts of the detected GW signal. To shed light on this, in this section we compute the total number of host candidates, $N_{\rm 90}$, associated to each of the true hosts. To this end, we employ the methodology of Section~\ref{subsec:ErrorBox}. Given that the extension of the LISA error-box varies with time, here we investigate $N_{90}$ at three different times: $10$ hours and $1$ hour before merger and at merger. We highlight the number of $N_{90}$ would vary depending on the specific wavelength in which the observations will be taken (see e.g. Section~\ref{sec:NAGNs}). Some galaxies would be bright in certain bands of the EM spectrum while too dim in others. In this section we aim to give the absolute number of $N_{90}$, regardless of the wavelength. A comprehensive study of how $N_{90}$ varies as a function of the observed frequency is deferred to a future paper.\\% Finally, the values of $N_{90}$ presented in this section can be considered as lower limits. Resolution effects caused by the underlying dark matter simulation might cause an underestimation on the exact vale of $N_{90}$.\\

The redshift evolution of $N_{90}$ for the three binary systems considered in this work is presented in Fig.~\ref{fig:N90}. As shown, the median values of $N_{90}$ display a strong time evolution, with the number of host candidates significantly decreasing as the binary approaches to the final merger. This behaviour is shared by all the MBHBs and is independent on redshift. For instance, the error-box associated to a GW event of $3\,{\times}\,10^5\, \msun$ at $z\,{=}\,1$ ($z\,{=}\,0.3$) displays on average $N_{90}\,{\sim}\,10^5$ (${\sim}\,10^2$) at 10 hours before the merger while this value drops down to ${\sim}10^2$ (${\sim}\,5$) at the coalescence time. Consequently, our results show that $N_{90}$ can exhibit up to ${\sim}\, 2 \, \rm dex$ variations between the last hours of the inspiral phase and the merger. This is an expected trend given the narrowing of the LISA error-box due to the improvement on the sky-localization capabilities as the binary system comes closer to the final coalescence (see Fig.~\ref{fig:DeltaOmega_M3e6}). We further notice that the redshift at which the GW source is placed has an important role in determing the values of $N_{90}$: all the binaries show an increase of the number of galaxies around the true hosts towards high-$z$, regardless the time. In more quantitative terms, at a fixed time of 10 hours prior to merger, an event related to a GW signal coming from a $3\,{\times}\,10^6\, \msun$ ($3\,{\times}\,10^7\, \msun$) MBHB displays $N_{90}$ typically spanning from ${\sim}\,200$ (${\sim}\,10^3$) to ${\sim}\,10^6$ (${\sim}\,10^8$) at $z\,{=}\,0.3$ and $z\,{=}\,2$, respectively. A similar trend can be seen at merger where the values can vary from ${\sim}\,2$ (${\sim}\,10$) to ${\sim}\,10^3$ (${\sim}\,10^3$) in the same redshift range as before. As discussed in Section~\ref{sec:TrueHosts}, this trend seen in the redshift is caused by the fact that LISA sky-position and luminosity distance uncertainties increase towards higher-$z$, corresponding to wider sky-areas and volumes and thus, larger number of galaxies within them.\\

%\monica{This feature leads back again to the fact that LISA sky-position an luminosity distance uncertainties increase for higher-$z$ GW sources,  corresponding to far larger sky-areas and volumes, even if selected in given redshift bins.  We need to discuss this???? correct?} \davcoment{We discussed that in Section 3 but we can remember this to the reader here again. Something like:} \davcoment{As discussed in Section~\ref{sec:TrueHosts}, this trend seen in the redshift is caused by the fact that LISA sky-position and luminosity distance uncertainties increase towards higher-$z$, corresponding to wider sky-areas and volumes and thus, larger number of galaxies within them.}\\

%\as{I know this goes beyond the scope of this paper, but here completeness effects will be important. Also, perhaps $N_{90}$ can be reduced for massive binaries. In the light cone we have all galaxies with $M_*$ down to $5\times 10^9$, but a $3\times10^7$ binary will likely be hosted in a rather massive galaxy (say MW type). Maybe we can just drop a comment on this...Monica?} \davcoment{Can you comment something on that monica?}\\

\begin{figure*}
    \hspace{-0.3cm}
    \includegraphics[width=2.\columnwidth]{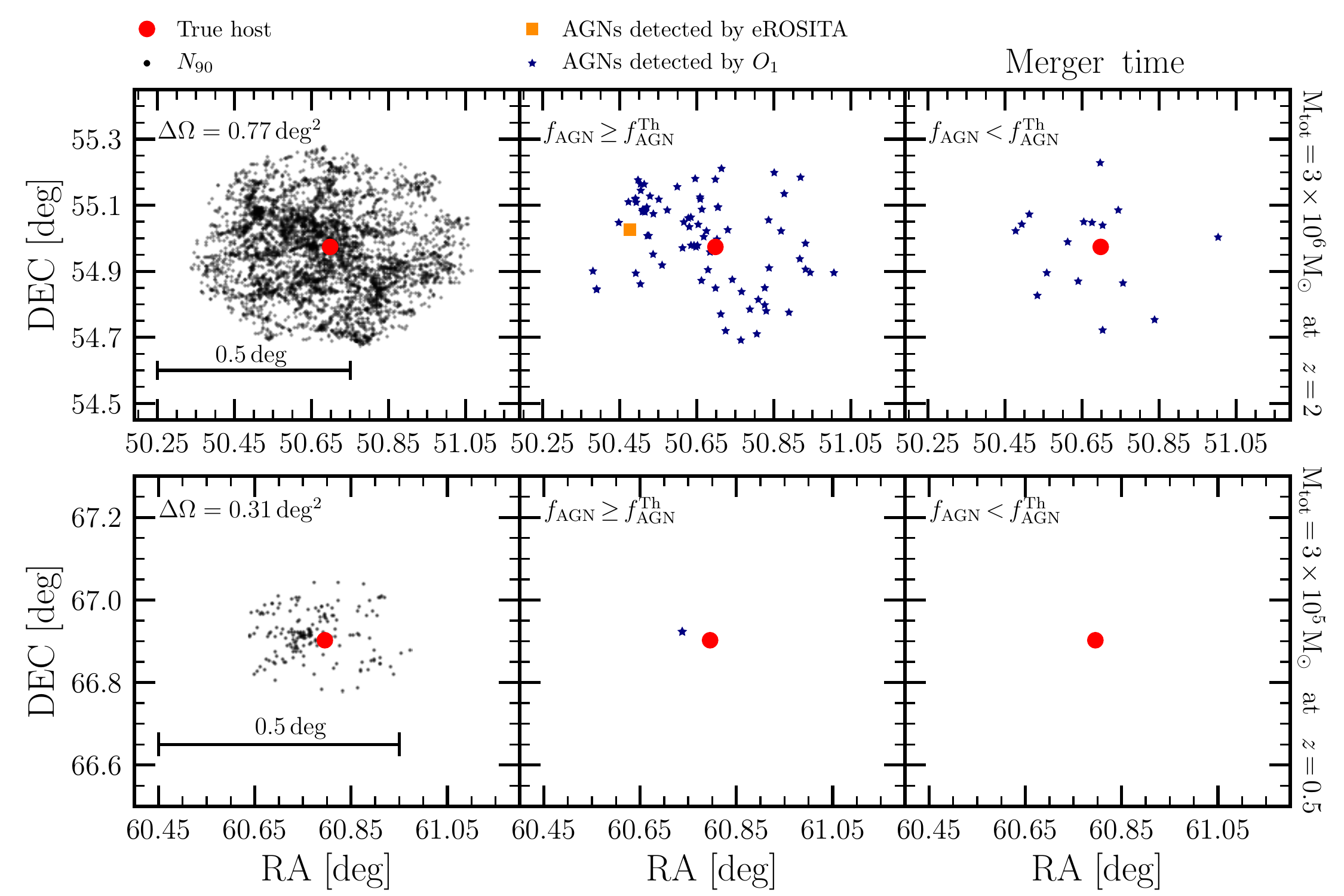}
    \caption{\textbf{Upper panel:} Projection of the LISA error-box in the RA-DEC plane at the time of merger for one random true host at $z\,{=}\,2$ with total mass $3\,{\times}\,10^6\, \msun$. If the true host is assumed to be active at the Eddington limit and $\rm N_{H}\,{=}\,10^{21.5}\, \rm cm^{-2}$, the corresponding soft X-ray flux is $f_{\rm AGN}^{\rm Th}\,{=}\,2.7\,{\times}\, 10^{-17} \rm erg \, cm^{-2} \, s^{-1}$ (central red point). In the left panel we present all the potential hosts ($N_{90}$, black points). In the central panel we show the $N_{90}$ galaxies associated with AGNs detected by $O_1$ (blue points) and eRosita (orange squares) whose flux ($f_{\rm AGN}^{\rm th}$) are larger than the one corresponding to the true host ($f_{\rm AGN}^{\rm th}$, central black dot). In the right panel we represent the same, but the AGNs have $f_{\rm AGN}\,{<}\,f_{\rm AGN}^{\rm th}$. In all the panels, the flux corresponds to the soft X-ray band ($0.5\,{-}\,2\, \rm keV$). \textbf{Lower panel:} The same as in the upper panels but for a random true host at $z\,{=}\,0.5$ with total mass $3\,{\times}\,10^5\, \msun$. If the true host is assumed to be active at the Eddington limit, the corresponding soft X-ray flux is $f_{\rm AGN}^{\rm Th}\,{=}\,1.6 \,{\times}\,10^{-15} \, \rm erg \, cm^{-2} \, s^{-1}$ (assuming $\rm N_{H}\,{=}\,10^{21.5}\, \rm cm^{-2}$).} %\monica{here is with no selection in redshift space gtso? we need to write this down and how fainter are the dots in the last panel compared to our true source?} }
    \label{fig:2D_ErrorBox_3e6_z2}
\end{figure*}

In addition to the previous dependencies, $N_{90}$ correlates with the total mass of the MBHB triggering the GW signal. For instance, at $z\,{=}\,1$ the error-boxes related to $3\,{\times}\,10^5\, \msun$ MBHBs contain on average ${\sim}\,10$ galaxies at merger time. Conversely, at the same redshift and time, $N_{90}$ rises up to  ${\sim}\,10^3$ for the $3\,{\times}\,10^7\, \msun$ MBHBs. This mass dependence is the result of low-mass MBHBs staying inside the LISA sensitivity band for longer times than higher-mass systems, allowing a better sky-localization. It is important to highlight that for systems with $3\,{\times}\,10^7\, \msun$ we do not find any difference on $N_{90}$ at 10 hours and 1 hour before merger at $z\,{\geq}\,0.5$. This peculiar feature is the result of the Fisher parameter estimation giving rise to uncertainties that remain constant from $10$ to $1$ hours prior to the merge. This inevitably implies that the LISA error-boxes in such time interval remain untouched, implying the same $N_{90}$ selection.\\

The results presented in Fig.~\ref{fig:N90} deliver two main messages. The first one concerns the fact that the LISA error-boxes are going to be crowded by galaxies, ranging between hundreds to thousands potential host candidates. This brings to light the challenges that multi-messenger studies will have to face when searching for the host of a LISA GW source. The second message refers to the most promising GW sources whose host can be potentially identified among background galaxies. Unsurprisingly, the lowest numbers of potential candidates correspond to GW sources at $z\,{<}\,0.5$ at merger. Nevertheless and for the more massive systems, the number of host candidates is of several tens, underlining that smart strategies are required to efficiently distinguish between fake and true hosts. In the next section we tackle this by using X-ray observations of AGNs. Finally, we underline that the values of $N_{90}$ reported in Fig.~\ref{fig:N90} correspond to the idealized case where a precise information about the redshift of all the galaxies inside the RA-DEC plane delimited by LISA is available. If this assumption is relaxed and no constraints on the redshift would be possible, the total number of galaxies within the projected LISA error-box would be several orders of magnitude larger than $N_{90}$ since all the galaxies within the line of sight should be considered (see for instance the right panel of Fig.~\ref{fig:Neighbors_z05}). Thus, the lack of a precise redshift information should be considered a serious challenge in the search for the galaxy hosting the GW source.% A specific analysis of how redshift inaccuracies will affect the detection of the GW host goes beyond the scope of this paper.

\subsection{X-ray AGNs inside the LISA error-box}
\label{sec:NAGNs}

As discussed in the previous section, even in the best case scenario the sky-localization capabilities of LISA imply that tens to thousands of host candidates can be present inside its FOV. In this section, under the assumption that the GW sources are associated to Eddington-limited AGNs (see Table~\ref{tab:Lum_and_flux_Limit_AllBinaries} for their associated fluxes and luminosity), we explore the possibility of reducing the number of potential hosts based on AGN selection. To this end, we use the X-ray observatories presented in Section~\ref{sec:XrayObservatories} whose characteristics are close to the design of future surveys of Athena \citep{Nandra2013} and the mission-concept Lynx \citep{Lynx2018}. We further include the predictions for the X-ray observatory eRosita \citep{Merloni2012}, which is expected to provide a full-sky map in X-rays\footnote{The flux limits associated to eRosita that are going to be used in this work correspond to $1.5\,{\times}\,10^{-14} \, \rm erg \, s^{-1}\, cm^{-2}$ and $1.8\,{\times}\,10^{-13}\, \rm erg \, s^{-1}\, cm^{-2}$ in the $0.5\,{-}\,2\, \rm keV$ (soft) and $2\,{-}\,10\,\rm keV$ (hard) X-ray bands, respectively \citep[see][]{Singh2016}.} by the time LISA will be launched. Finally, we compute the \textit{total} number of detected X-ray AGNs within the whole LISA error-box. However, the small FOV of the observatories will imply several pointings to cover the  X-rays the area associated to the LISA sky-position uncertainties. As already discussed in Section~\ref{sec:XrayObservatories}, scanning with X-rays the LISA projected error-box on timescales  smaller than the inspiral time of the binary will be feasible only for binaries at $z\,{<}\,1$. On contrary, at the end of coalescence, this task can be carried out at any redshift.\\

%A small number of pointings (${<}\,3$) will be feasible only for binaries at $z\,{<}\,1$ detected at times ${<}\,1$ hour before the merger. During the rest of the inspiral phase, the associated $\Delta \Omega$ values are very large and covering the LISA sky-area will require a mosaic of ${>}\,30$ pointings (see full discussion in Section~\ref{sec:XrayObservatories}). We will highlight and discuss these cases throughout this section.\\

To determine if the $N_{90}$ galaxies are detected by $O_1$, $O_2$ and $O_3$ as AGNs, we proceed as follows. Using Eq.~\ref{eq:BolometricCorrection_hard}, Eq.~\ref{eq:BolometricCorrection_soft} and Eq.~\ref{eq:ColumDensityAttenuation} we transform  the bolometric luminosity\footnote{We highlight that the bolometric luminosity of each MBH in our lightcone is computed self-consistently by \lgal{}, taking into account the galaxy evolution. We refer the reader to Section~\ref{sec:BHPopulation} for further details.} into X-ray luminosity ($\rm L_{Hx,Sx}$) for all the MBHs/MBHBs hosted by the $N_{90}$ galaxies\footnote{We refer the reader to Appendix~\ref{appendix:ComptonThinScenario} for an analysis in which the X-ray AGNs display values of $\rm N_{H}$ larger than the ones used in the fiducial approach presented in Section~\ref{sec:BHPopulation}.}. Then, we check their observability by imposing that the X-ray flux (i.e. $f_{\rm Hx/Sx}\,{=}\, \rm L_{Hx,Sx}/4 \pi \mathit{d}_{\rm L}^2$) must be above the flux limit of a given observatory. Since we are exploring LISA events at 10 hours, 1 hour and at merger, these observations are limited by different exposure times (see Fig.~\ref{fig:SensitivityCurves}). As a consequence, different flux limits will be reached for each of these times\footnote{We stress that at merger time, we selected a flux limit corresponding to the confusion limit of the X-ray observatory.}, causing that the number of $N_{90}$ detected as X-ray AGNs will vary. An illustrative example of this methodology is presented in the upper panels of Fig.~\ref{fig:2D_ErrorBox_3e6_z2} where we show the projected LISA error-box in the RA-DEC plane at the time of merger for one random true host at $z\,{=}\,2$ and total mass $3\,{\times}\,10^6\, \msun$. As we can see, the LISA error-box is crowded by galaxies. However, if we assume that only the galaxies seen as AGNs by $O_1$ and eRosita can be candidates for hosting the GW event, we reduce considerably the number of potential candidates. Besides this, the figure shows that the large majority of the AGNs associated with the LISA error-box displays fluxes in the soft band larger than the one expected for the GW source. This is not a surprising result, given that an active binary of $3\,{\times}\,10^6\, \msun$ at $z\,{=}\,2$ is a relatively dim source ($f_{\rm Sx}\,{=}\,2.7\,{\times}\, 10^{-17} \rm erg \, cm^{-2} \, s^{-1}$). %Concerning eRosita, a very small fraction of X-ray AGNs are detected inside the LISA error-box, displaying all of them fluxes above the one associated to the true host. 
In the lower panels of  Fig.~\ref{fig:2D_ErrorBox_3e6_z2} we present the same as before but for a random true host at $z\,{=}\,0.5$ and total mass $3\,{\times}\,10^5\, \msun$. As we can see, in this scenario the search of the true hosts is more feasible. The small area delimited by the LISA sky-localization uncertainties (${\sim}\,0.3\, \rm deg^2$) implies a small number of host candidates, with only one of them undergoing an active phase. \\

    \begin{figure}
        %\hspace{-0.7cm}
        \centering
        \includegraphics[width=1\columnwidth]{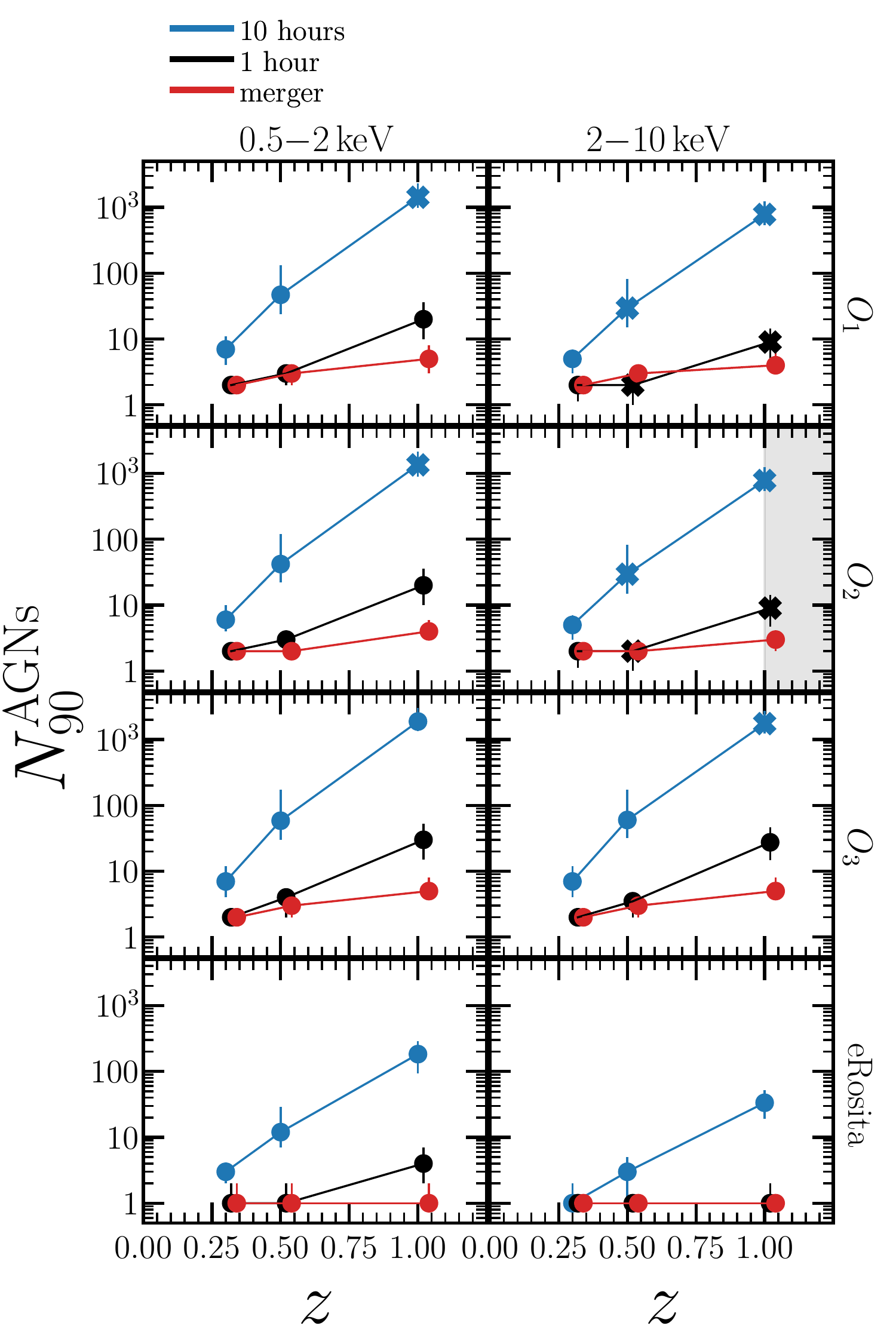}
        \caption{\textbf{Left panel:} Number of X-ray AGNs in soft (left panel) and hard bands (right panel) detected inside the LISA error-box associated to the true hosts of $\rm M_{\rm tot}\,{=}\,3\,{\times}\,10^5 \msun$ at $z\,{=}\,0.3,0.5$ and $1$. The error bars correspond to the $32^{\rm th}\,{-}\,68^{\rm th}$ percentiles. Crosses are drawn when the number of pointings needed by the X-ray observatories to cover the full LISA error-box require exposure times larger than the inspiral time of the binary. Different colors represent different times during the evolution of the systems: $10$ hours before merger (blue), $1$ hour before merger (black) and merger (red). Finally, shaded gray areas highlight the redshifts at which the observatories are not able to detect any AGN associated to the MBHBs.}
        \label{fig:NAGN_M3e5}
    \end{figure}

    \begin{figure}
        %\hspace{-0.7cm}
        \centering
        \includegraphics[width=1\columnwidth]{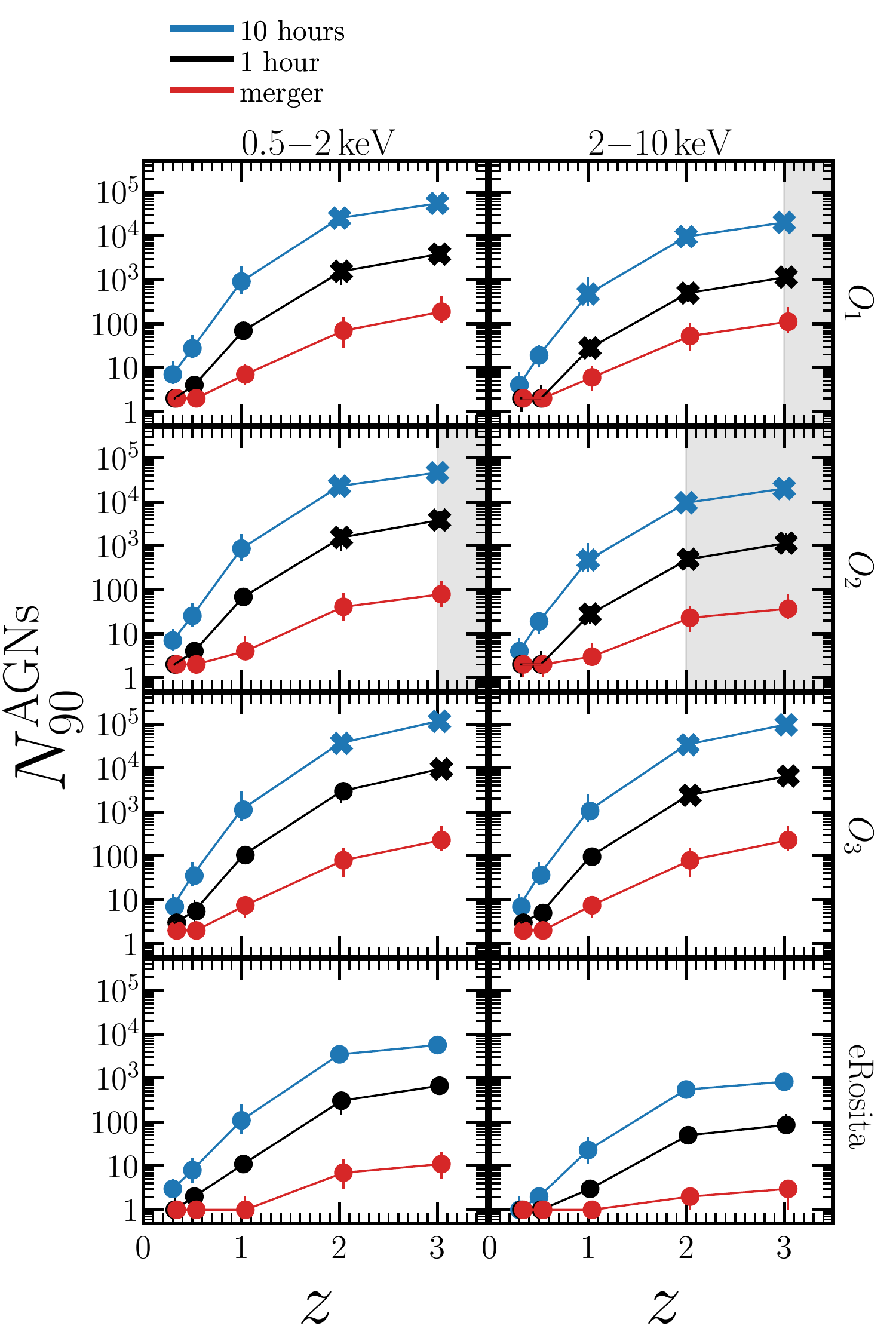}
        \caption{Number of X-ray AGNs in soft (left panel) and hard bands (right panel) detected inside the LISA error-box associated to the true hosts of $\rm M_{\rm tot}\,{=}\, 3\,{\times}\,10^6 \msun$ at $z\,{=}\,0.3,0.5,1,2$ and $3$. The error bars correspond to the $32^{\rm th}\,{-}\,68^{\rm th}$ percentiles. Crosses are drawn when the number of pointings needed by the X-ray observatories to cover the full LISA error-box require exposure times larger than the inspiral time of the binary. Different colors represent different times during the evolution of the systems: $10$ hours before merger (blue), $1$ hour before merger (black) and merger (red). Finally, shaded gray areas highlight the redshifts at which the observatories are not able to detect any AGN associated to the MBHBs.}
        \label{fig:NAGN_M3e6}
    \end{figure}

\begin{figure}
        %\hspace{-0.7cm}
        \centering
        \includegraphics[width=1\columnwidth]{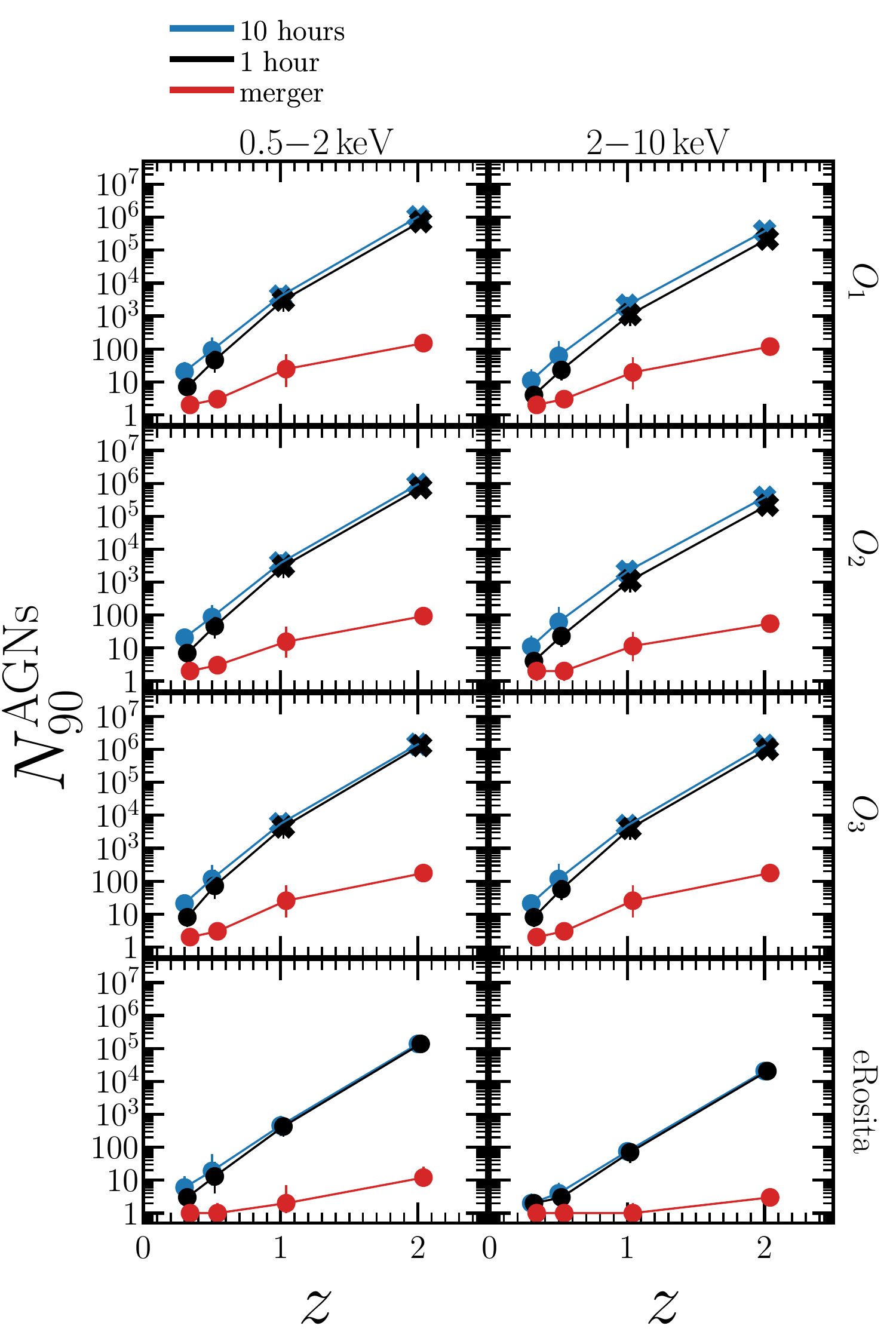}
        \caption{Number of X-ray AGNs in soft (left panel) and hard bands (right panel) detected inside the LISA error-box associated to the true hosts of $\rm M_{\rm tot}\,{=}\, 3\,{\times}\,10^7 \msun$ at $z\,{=}\,0.3,0.5,1$ and $2$. The error bars correspond to the $32^{\rm th}\,{-}\,68^{\rm th}$ percentiles. Crosses are drawn when the number of pointings needed by the X-ray observatories to cover the full LISA error-box require exposure times larger than the inspiral time of the binary. Different colors represent different times during the evolution of the systems: $10$ hours before merger (blue), $1$ hour before merger (black) and merger (red). Finally, shaded gray areas highlight the redshifts at which the observatories are not able to detect any AGN associated to the MBHBs.}
        \label{fig:NAGN_M3e7}
    \end{figure}

To quantify the previous results, in Fig.~\ref{fig:NAGN_M3e5}, Fig.~\ref{fig:NAGN_M3e6} and Fig.~\ref{fig:NAGN_M3e7} we depict the number of $N_{90}$ associated to X-ray AGNs (hereafter, $N_{90}^{\rm AGNs}$). In the same plots, we highlighted with crosses the cases for which the number of pointings needed by the X-ray observatories to cover the full LISA error-box require exposure times larger than the inspiral time of the binary (see Table~\ref{tab:Pointings}). For these combinations of total mass, redshift and time we assume that the search for the true host among the sample of host candidates hosting an AGN will not be possible. As shown, $N_{90}^{\rm AGNs}$ shares the same trends displayed by $N_{90}$, regardless of the performance of the X-ray observatory. On one hand, it increases towards high-$z$, independently of the mass and time. On the other hand, the values are larger as the mass of the binary increases. As explained before, these two behaviors are just a consequence of the improvement of the LISA sky-localization as the mass and redshift of the binary decrease. When comparing the values of $N_{90}^{\rm AGNs}$  with the ones of $N_{90}$, the former can be up to $2\rm \, dex$ smaller, highlighting that AGN selection can reduce considerably the number of potential host candidates. For instance, the LISA error-box related to a system of $3\,{\times}\,10^5\, \msun$ ($3\,{\times}\,10^7\, \msun$) at 10 hours before the merger has in the soft X-ray band values of $N_{90}^{\rm AGNs}\,{\sim}\,10$ ($20$) and $10^3$ ($10^4$) at $z\,{=}\,0.3$ and $z\,{=}\,1$, respectively. At $z\,{<}\,1$ these values are similar among $O_1$, $O_2$ and $O_3$ given that all of them have enough flux sensitivity to detect all the X-ray AGNs placed inside the LISA error-box. At higher redshifts, the different sensitivities achieved by each observatory leave an imprint in the values of $N_{90}^{\rm AGNs}$. For instance, at $z\,{=}\,2$ the observatory $O_3$ is able to detect up to $3\,{-}\,4$ times more AGNs than $O_1$ and $O_2$, regardless of the binary mass and X-ray band explored.\\
%\monica{I would have though a significantly larger number for O3.. are we sure? or is it compensated by the samller area???}\\

Another behavior shown by $N_{90}^{\rm AGNs}$ concerns its decrease as the inspiral time of the MBHB gets shorter, reaching the lowest values at merger. This trend is shared among all the observatories and X-ray bands. To give an example, at $z\,{=}\,1$ the error-box associated to a system of $3\,{\times}\,10^5\,\msun$ ($3\,{\times}\,10^6\,\msun$, $3\,{\times}\,10^7\,\msun$) has $N_{90}^{\rm AGNs}\,{\sim}\,10^3$ ($10^3$, $10^4$) at 10 hours before merger. For the same binary system, this value drops down to $8$ ($10$, $20$) at the time of merger \cite[slightly above the values reported by][]{Kocsis2006}. This significant drop has fundamental consequences in multi-messenger astronomy since follow-up studies at different epochs and wavelengths are more feasible with a low number of targets. Indeed, our results show that at $z\,{<}\,1$ all the error-boxes associated with the binary systems have $N_{90}^{\rm AGNs}\,{<}\,10$ at the time of merger, regardless of the observatory used. As a result, X-ray observations of low-$z$ GW events can be very promising for pinpointing the LISA GW hosts. At $z\,{>}\,1$, this conclusion does not apply anymore given that $N_{90}^{\rm AGNs}$ is systematically larger than 200, regardless of the binary mass.  Such behavior is the consequence of the strong positive correlation displayed between $N_{90}^{\rm AGNs}$ and redshift. \\

%\gaia{This last conclusion is particularly true for MBHBs with total masses $\leq 3\,{\times}\,10^6\,\msun$ at $z\leq1$. Indeed, as already stated in Section \ref{sec:DetectionInspirallingBinaries}, the LISA sky-area associated to these systems can be fully covered by the X-ray observatories within few pointings ($\sim 3$), starting from $1$ hour prior to merger. For all the other combinations of $\rm M_{\rm tot}$ and $z$, this is possible only at the time of merger. Therefore, we can conclude that the best candidates for multi-messenger observations are low-mass binaries at $z\leq1$ for which a full coverage of the projected error-box is guaranteed prior to merger.} \\

Besides a correlation with mass, redshift and inspiral time, $N_{90}^{\rm AGNs}$ shows a dependence on the X-ray band used to detect AGNs. As shown, observations taken in the soft band display larger values of $N_{90}^{\rm AGNs}$ than the ones performed in the hard band. As discussed in Section~\ref{sec:XrayObservatories}, the larger flux sensitivity reached by the X-ray observatories in the soft X-rays makes the detection of AGNs easier. Despite Fig.~\ref{fig:NAGN_M3e5}, Fig.~\ref{fig:NAGN_M3e6} and Fig.~\ref{fig:NAGN_M3e7} suggest that hard X-ray band observations might be convenient to reduce the number of potential hosts, it is important to keep in mind that the $O_1$ and $O_2$ observatories require too long exposure times (sometimes incompatibles with the MBHB inspiral time) to detect the AGNs associated to binaries of $3\,{\times}\,10^5\,\msun$ and $3\,{\times}\,10^6\,\msun$ (see Table~\ref{tab:Exposure_times}).\\

Finally, we have explored the possibility of reducing the number of host candidates by using a reference catalogue of AGNs. To this end, we have focused on the eRosita mission, which will provide a full-sky survey in different X-ray bands by the time LISA will be in operation. Since the detection threshold of eRosita is above the one associated with the Eddington-limited MBHBs, it will be possible to discard from the host candidate pool X-ray sources previously detected by eRosita. Consequently, the cross-match between eRosita AGNs and the ones detected by $O_1$, $O_2$, and $O_3$ has the potential of reducing the number of $N_{90}^{\rm AGNs}$. With this in mind, in the lower panels of  Fig.~\ref{fig:NAGN_M3e5}, Fig.~\ref{fig:NAGN_M3e6} and Fig.~\ref{fig:NAGN_M3e7} we present the number of eRosita detections at 10 hours, 1 hour and at merger. As we can see, the specific numbers depend on binary mass, time and redshift. Despite that, there is a general trend of detecting ${\sim}\,1\, \rm dex$ less number of AGNs than $O_1$, $O_2$ and $O_3$. In this way, the reference catalogue provided by eRosita could reduce up to ${\sim}\,10\%$ the number AGNs inside the LISA error-box that can be potential hosts of the GW event.\\

Based on the results presented above, we can draw the conclusion that reducing the number of LISA host candidates based on AGN selection is a promising approach. Indeed, less than 0.1\% of the galaxies falling inside the detector FOV are expected to be observed as X-ray AGNs by very sensitive observatories. We stress that the approach explored here is only valid under the optimistic assumption that the GW sources detected by LISA are associated to Eddington-limited sources. Sub-Eddington or inactive binaries will challenge the host identification.

\subsection{The environments of AGNs triggered by single and binary MBH systems}
\label{sec:Neighbors}
    
We have shown that AGN selection is a favorable approach to reduce the number of host candidates inside the LISA error-box. Despite promising, this methodology does not allow us to unequivocally pinpoint the true host, since the number of candidates with X-ray AGN signatures is still too large. Thus, extra information will be needed to distinguish between the AGN triggered by the LISA source and the ones \textit{placed inside the LISA error-box} and potentially detectable by the simulated X-ray observatories. With this aim, we explore the number of neighbors around these two populations of AGNs. This analysis can give us information about the environments in which the GW sources are embedded, allowing us to determine if they are different than those hosting normal AGNs. Among all the redshifts and inspiralling times, in this section we focus on the LISA error-boxes at the time of merger for $z\,{=}\,0.5$. This choice is motivated by the low number of AGNs detected at this specific redshift and time (${\lesssim}\,20$), which offers the best scenario possible to perform feasible follow-up studies for all the candidate galaxies.\\

\begin{figure}
    \centering
    \includegraphics[width=1\columnwidth]{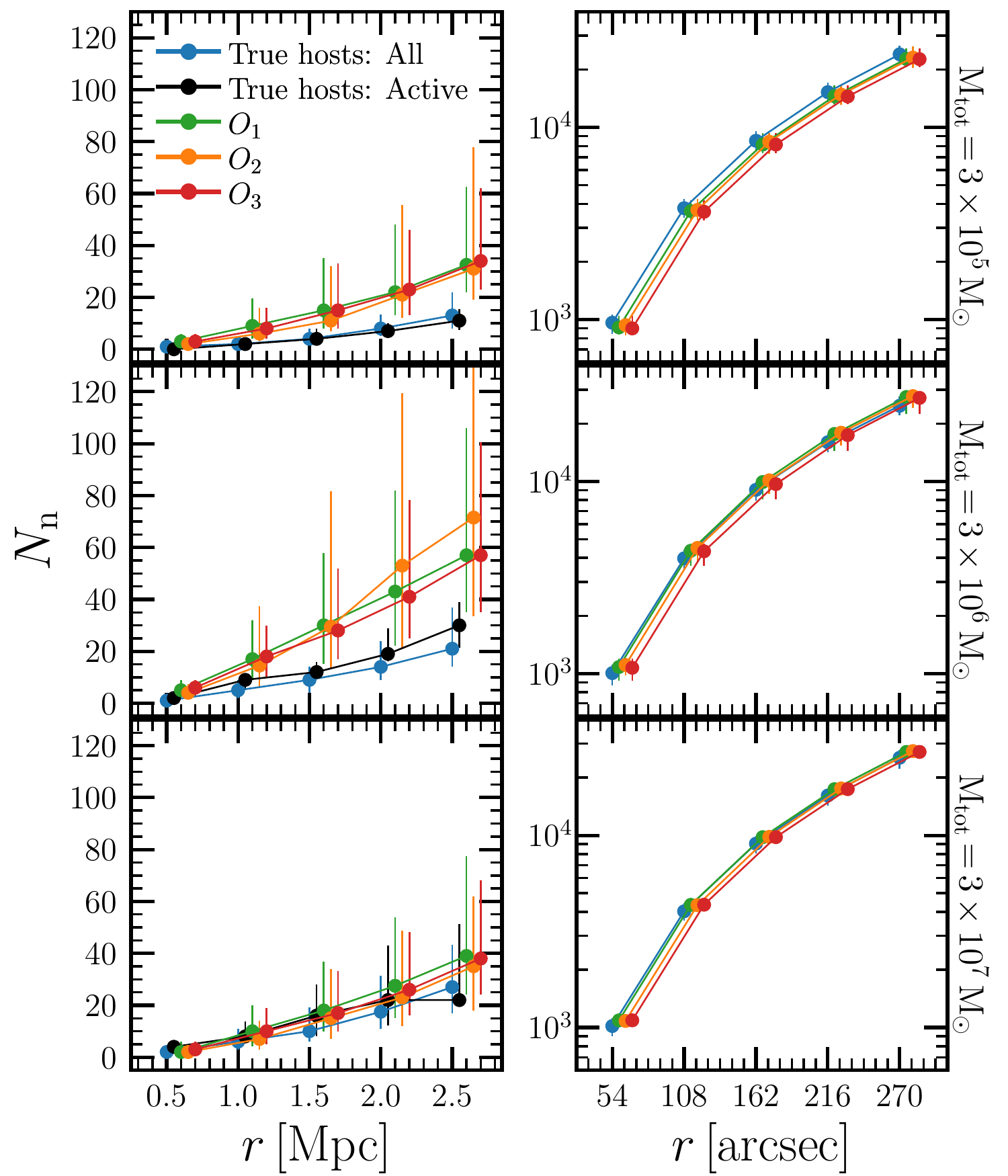}
    \caption{Number of neighboring galaxies, $N_{\rm n}$, within a given distance, $r$, around the true hosts (blue dots) and around AGNs placed inside the LISA error-box and detected by $O_1$. Different colors correspond to the number of neighbors around observable AGNs placed inside the LISA error-box centered on the true hosts at $z = 0.5$ at the time of merger. The observability of these AGNs is determined based on $O_1$ (green), $O_2$ (orange) and $O_3$ (red). The left panels present the results when considering comoving distances, the right ones correspond to the projected distances (i.e considering RA and DEC coordinates). In all the panels, dots correspond to the median values, while error bars display the $\rm 32^{th}\,{-}\,68^{th}$ percentiles. To avoid overlapping, all the results concerning AGNs have been shifted in by $0.05\rm \, Mpc$ ($5\, \rm arcsec$).}
    \label{fig:Neighbors_z05}
\end{figure}

In Fig.~\ref{fig:Neighbors_z05} we present the results for the three different binaries at $z\,{=}\,0.5$. The AGNs have been selected so that their fluxes are above the sensitivity limit\footnote{We highlight that, since we are considering the error-boxes at merger time, the flux limit of the X-ray observatories corresponds to the confusion limit.} of the X-ray observatories in the soft X-rays, but the same results hold true in the hard X-ray band. In the left panels we present the evolution of $N_{\rm n}$, defined as the number of galaxies within a sphere of comoving radius $r$  centered either on a true host or an observable AGN within the LISA error-box. %\monica{I do not understand the label of Figure 12: blue true active host, black nearby field AGN... woudl it be better???} \gaia{Actually, blue dots refer to the neighbors of all the true host, regardless of the fact that they (the true hosts) are active or inactive. Black dots, on the contrary, refer to the neighbors of the active true hosts only.} 
As we can see, $N_{\rm n}$ increases as a function of $r$, independent of the binary mass, and irrespective to the choice of the observatory. For instance, AGNs placed inside the error-boxes associated to a $3\,{\times}\,10^5\,\msun$ ($3\, {\times}\,10^6\,\msun$) binary display ${\sim}\,2$ (${\sim}\,4$) neighbors at $r\,{<}\,1\, \rm Mpc$ at the time of merger. At $r\,{>}\,2\, \rm Mpc$ the number increases by a factor of 10. These values barely vary among the AGN samples detected by the three X-ray observatories. This is caused by the fact that at $z\,{=}\,0.5$, the flux sensitivity of $O_1$, $O_2$ and $O_3$ are good enough to detect most of the AGNs inside the LISA error-box. Interestingly, differences are seen when comparing the values of $N_{\rm n}$ related to AGNs with the ones related to the true hosts with $3\,{\times}\,10^5\,\msun$ and $3\,{\times}\,10^6\,\msun$. While at $r\,{<}\,1\, \rm Mpc$ true hosts and AGNs seem to be surrounded by roughly the same amount of neighbors, at larger scales the two samples diverge and the values of $N_{\rm n}$ associated to the true hosts are systematically smaller than those related to the AGN populations. Since it is known that active MBHs reside in more crowded regions than inactive ones \citep[see e.g][]{Bonoli2009}, we have checked if these differences still hold when considering only the true hosts that are active according to \lgal{}. For that, we have performed the same analysis but only for the true hosts whose flux computed from the SAM (taking into account the galaxy merger history) is above the detection limit of the $O_1$ observatory. As shown, the differences remain.\\

\begin{figure}
    \centering
    \includegraphics[width=1\columnwidth]{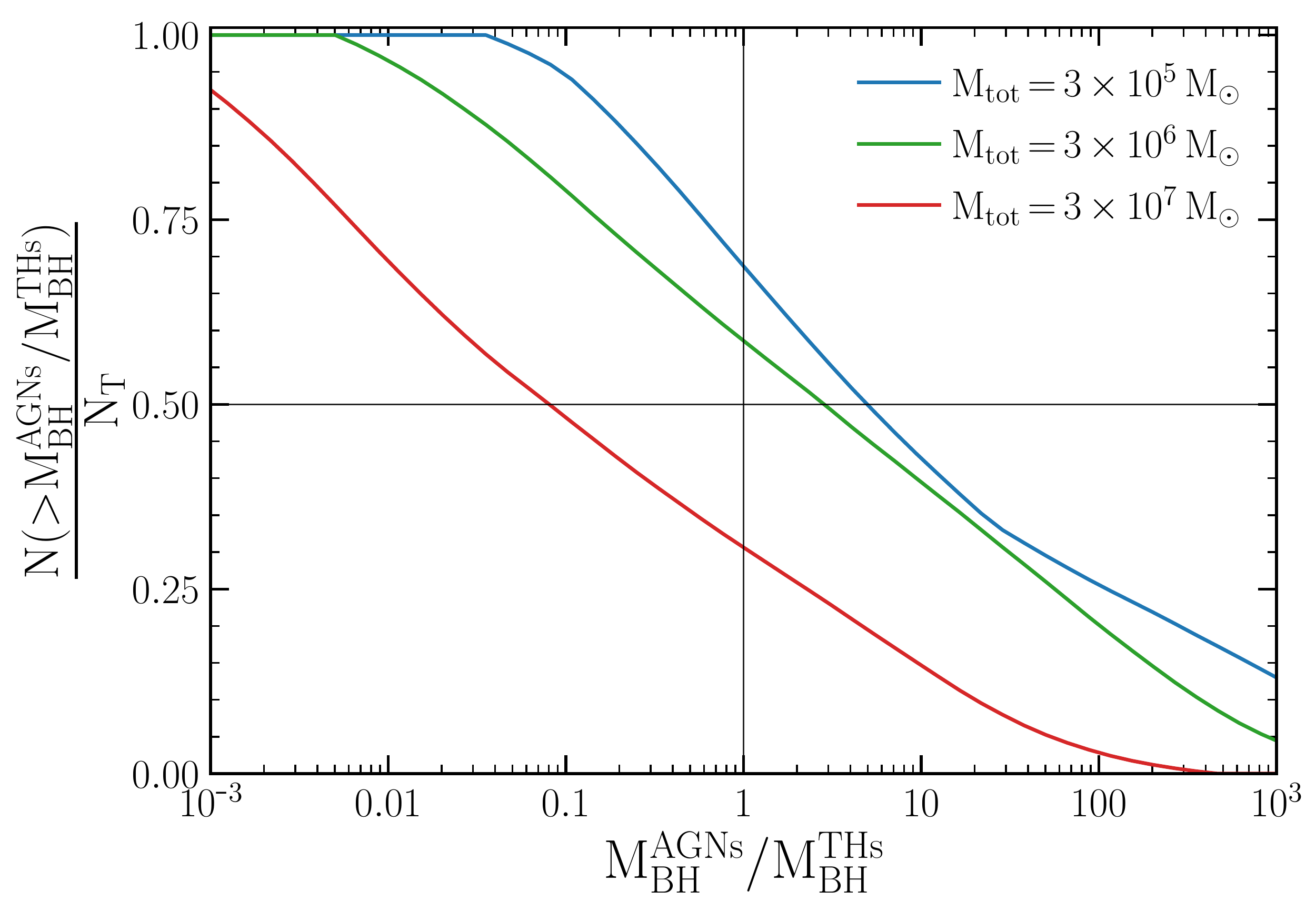}
    \caption{Cumulative distribution of the ratio between the black hole mass triggering the AGNs placed inside the LISA error-box ($\rm M_{BH}^{AGNs}$) and the total mass of the binary systems producing the GW signal ($\rm M_{BH}^{THs}$). The AGNs considered are those detected by $O_1$ and within the LISA error-box associated to $z\,{=}\,0.5$ at the merger time. Blue, green and red colors correspond to the case of $3\,{\times}\,10^5\,\msun$, $3\,{\times}\,10^6\,\msun$, and $3\,{\times}\,10^7\,\msun$, respectively. While the vertical black line highlights the position in which $\rm M_{BH}^{AGNs}\,{=}\,M_{BH}^{THs}$, the horizontal one corresponds to 0.5.}
    \label{fig:Mass_ratio_betwee_AGNS_TrueHosts}
\end{figure}

The differences in the values of $N_{\rm n}$ can be better understood by looking at Fig.~\ref{fig:Mass_ratio_betwee_AGNS_TrueHosts}, where we compare at $z\,{=}\,0.5$ the masses of the binary systems associated to the true hosts with the masses of the active MBHs within the LISA error-box at the merger time. For the sake of brevity, we only considered AGNs detected by $O_1$ given that similar results are seen for $O_2$ and $O_3$. As shown, for the case of $3\,{\times}\,10^5\,\msun$ and $3\,{\times}\,10^6\,\msun$, more than $50\%$ of the AGNs inside their LISA error-boxes are triggered by MBHs with masses $ {>}\,1\, \rm dex$ larger than the binaries associated to the true hosts. Given that the mass of the black hole correlates with the environment \citep[see e.g. Figure 15 of][]{Bonoli2009}, it is expected that the AGN population detected inside the LISA error-box will be placed, on average, in more crowded regions than the LISA sources. For the case of  $3\,{\times}\,10^7\,\msun$, Fig.~\ref{fig:Mass_ratio_betwee_AGNS_TrueHosts} shows that the AGNs are typically triggered by MBHs with masses similar to those of the MBHBs sourcing the GW signal. Thus, no large differences should be seen in $N_{\rm n}$, as shown by Fig.~\ref{fig:Neighbors_z05}. In view of these results, we can draw the conclusion that the number of neighbors at $r\,{>}\,2\, \rm Mpc$ could be a good indicator for pinpointing the host of ${<}\,3{\times}\,10^6\, \msun$ MBHBs.\\

Finally, in the right panels of Fig.~\ref{fig:Neighbors_z05} we have performed the same study discussed before but considering projected distances in the sky, i.e in the $\rm RA\,{-}\,DEC$ plane. This case would correspond to the observational scenario in which no information about redshift is available. As we can see, the projected value of $N_{\rm n}$ ranges between $10^3$ within $54\, \rm arcsec$ (corresponding to $0.5\,\rm Mpc$ at $z\,{=}\,0.5$) up to $3\,{\times}\, 10^4$ at $r\, \leq \,270\, \rm arcsec$ (corresponding to $2.5\,\rm Mpc$ at $z\,{=}\,0.5$). As happened before, similarities are seen in the AGN samples. However, the difference previously found in the behavior of the true hosts is washed out. Therefore, loosing the redshift information of the galaxies inside the LISA error-box can considerably make more difficult the selection of the true host based on the number of neighbors. 
%\as{But here you are considering the neighbors within degrees! On those scales all the AGNs in your errobox will share the same neighbors because the size of the error-box is of the same order of magnitude. If you want to show this, you should consider much smaller angular scales, corresponding to few Mpc in physical size at the redshift of the detected source. Perhaps the effect is going to be washed out anyway, but that would be a meaningful test.} \gaia{We have corrected Fig. 12 by considering projected distances corresponding to the comoving distances presented in the left panels.}

\subsection{The search of host candidates with dual detector networks: the LISA-Taiji case}

Another topic which is currently studied in the literature is the possibility of improving the sky-localization of GW sources by making use of dual networks of space-based detectors \citep[][]{Crowder2005,Ruan2021,Wang2021b,Shuman2022}. In particular, in this work we explore how the cooperation between  LISA and Taiji \citep[][]{Ruan2018} would change the expected number of potential host candidates, $N_{90}$ and $N_{90}^{\rm AGNs}$. More precisely, our analysis is based on the work of \cite{Shuman2022}, in which the effects of the LISA-Taiji cooperation on the sky-localization of GW sources are presented. Among the seven LISA-Taiji configurations explored by the authors, we chose to work with the LLF40 one, in which Taiji is assumed to be $40^{\circ}$ ahead and $40^{\circ}$ tilted with respect to the LISA orbit. The reason behind this choice refers to the large improvement achieved on the the sky-localization, with a decrease of three and one orders of magnitude for $\Delta \Omega$ and $\Delta d_{\rm L}$, respectively (see their Table 2). Based on these results, we built the LISA-Taiji error-boxes by following the procedure outlined in Section~\ref{sec:LISA_Error_Boxes}, but re-scaling LISA sky-localization errors ($\Delta \Omega$, $\Delta d_{\rm L}$) according to the Table 2 of \cite{Shuman2022}. We stress that the minimum value associated the new $\Delta d_{\rm L}$ is limited to the weak lensing error, taken from \cite{Petiteau2011}. This choice was done in view of the fact that the extended distribution of DM halos between us and the source causes fluctuations in the amplitude of the GW signal, raising for sources at redshift higher than $z\,{\sim}\,0.25$ an uncertainty in the luminosity distance that can be larger than the one achieved by the LISA-Taiji network.\\ %This choice was done in view of the fact that weak lensing is known to magnify signals, imposing a limit on the precision with which it is possible to measure distances \monica{are you shore that it magnify...this is strong lensing. I think if has both magnification and demagnification. I would just delete the sentence..."this choice..."}\footnote{The extended distribution of DM halos between us and the source causes fluctuations in the amplitude of the GW signal, raising an uncertainty in the luminosity distance, for sources at redshift higher than $z\,{\sim}\,0.25$.}. \\ %Therefore, with this shrewdness we simply wanted to guarantee that the accuracy of the LISA+Taiji estimation of the luminosity distance of a GW source was not smaller than that defined by the weak lensing limit\footnote{This check was not done when building the LISA error-boxes due to the fact that the $\Delta d_{\rm L}$ in that case were of the same order of magnitude of the weak lensing limit or above it.}.
%\as{What is the typical $\Delta{d_l}$? Of the sources? If it shrinks by a factor of 10, it will be likely below the weak lensing limit. Did you consider that? For example you can have a look here https://arxiv.org/pdf/1102.0769.pdf. You can take for example the yellow curve as a proxy for the weak lensing error. I think the error on $d_l$ you have from Mangiagli's code should be of the same order of this weak lensing error, so we can perhaps ignore the WL in that case, but if we divide it by 10, we'll be limited by weak lensing, I presume. This will increase the number of sources is figure 13, but still there will be less than 100 galaxies at any redshift at merger and likely no contaminating AGN among them.} \gaia{We've re-done Figure 14 by taking into account this fact.} \\

\begin{figure}
    \includegraphics[width=1.\columnwidth]{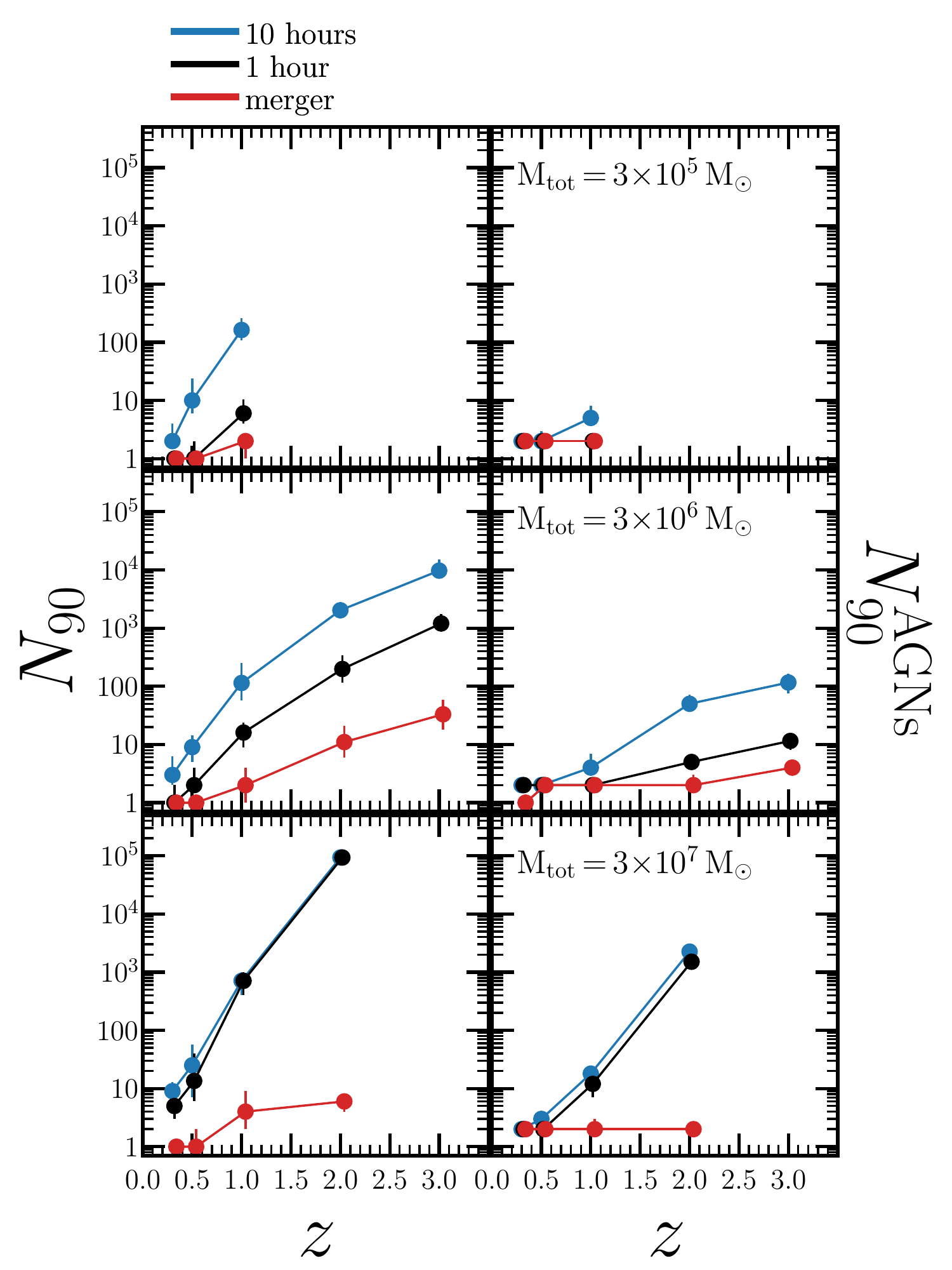}
    \caption{Median values of the total number of host candidates ($N_{90}$, left panels) and AGNs detectable by the observatory $O_1$ in the soft X-ray band ($N_{90}^{\rm AGNs}$, right panels) laying inside the error-boxes associated to the true hosts, after re-scaling the errors on the parameters in order to simulate a potential LISA-Taiji cooperation \citep[][]{Shuman2022} and taking into account the weak lensing limit \citep[][]{Petiteau2011}. The error-bars are taken from the $32^{\rm th}\,{-}\,68^{\rm th}$ percentiles. From top to bottom we show the results for $3\,{\times}\,10^5\msun$, $3\,{\times}\,10^6\msun$ and $3\,{\times}\,10^7\msun$. Different colors represent different times during the evolution of the systems: $10$ hours before merger (blue), $1$ hour before merger (green) and merger (red). We highlight that at difference of Fig.~\ref{fig:NAGN_M3e5}, \ref{fig:NAGN_M3e6} and \ref{fig:NAGN_M3e7} no crosses are shown. This is because the the large decrease of $\Delta \Omega$ implies that the three X-ray observatories studied in this work will cover the sky area delimited by the LISA-Taiji cooperation in a time smaller than the corresponding time to the binary merger.}
    \label{fig:N90_LISA_Taiji}
\end{figure}
    
In the left panels of Fig.~\ref{fig:N90_LISA_Taiji} we show the redshift evolution of $N_{90}$ inside the LISA-Taiji error-boxes associated to the true hosts. As we did in Fig.~\ref{fig:N90}, three different times were investigated: $10$ and $1$ hours prior to the merger and the merger time. As expected, the synergy between the two GW detectors does not modify the dependencies of $N_{\rm 90}$ on the time to the merger, redshift and total mass of the binary system, being the same ones already discussed in Section~\ref{subsec:N90_LISA}. On the contrary, the specific values of $N_{\rm 90}$ display a decrease of ${\sim}\,3\,\rm dex$, regardless of $\rm M_{\rm tot}$ and $z$. These results have an important impact on the search of the GW hosts, especially at the time of merger, where ${\lesssim}\,20$ galaxies lie inside the error-box, regardless of redshift and MBHB mass. Concerning the times prior to the merger, they are still limited by the presence of several potential hosts but the considerable reduction caused by the cooperation with Taiji, makes more feasible the detection of the true GW hosts at these times. Therefore, we can conclude that the use of dual networks of space-based detectors, such as the LISA-Taiji configuration, would allow to reduce the number of host candidates at the time of merger to few galaxies.\\ %Moreover, if the MBHB sourcing the GW signal is active, then it would be the only AGN inside the error-box. In conclusion, the LISA-Taiji combination would likely guarantee an unequivocal identification of the true host. \\%inside the LISA error-box without having to rely on the AGN selection discussed in Subsection \ref{sec:NAGNs}. 
%\as{Here the conclusion should be stronger. the Network allows to down select few galaxies at merger. If the MBHB is active, that would be the only AGN in the errorbox. So the combination of LISA and Taiji is likely to allow univocal determination of the true host.} 

Finally, in the right panels of Fig.~\ref{fig:N90_LISA_Taiji}, we show the redshift evolution of $N_{\rm 90}^{\rm AGNs}$ associated to the LISA-Taiji error-boxes. The results are only presented for $O_1$ in the soft X-rays given that no significant differences are seen between X-ray bands and observatories. As expected, the behavior of $N_{\rm 90}^{\rm AGNs}$ resembles the one of $N_{90}$. However, the values of $N_{\rm 90}^{\rm AGNs}$ can be up to ${\sim}\,2\, \rm dex$ smaller, diminishing down to $10$ AGNs at $z\,{<}\,1$ for any binary mass and inspiralling time. Besides, we have checked that the large decrease on $\Delta \Omega$ implies that the three X-ray observatories studied in this work will cover the sky area delimited by the LISA-Taiji cooperation in a time smaller than the corresponding time to the binary merger. In view of this results, we can conclude that the LISA-Taiji dual network will make it possible to extend up to $z\,{\sim}\,1$ the search of an EM counterpart of a GW source detected at ${\leq}\,10$ hours before the merger.

\section{Summary and Conclusions}
In this paper, we studied from a theoretical point of view the galactic fields of LISA merging MBHBs and how these fields are seen by three X-ray observatories whose designs are  close to the Athena and Lynx observatories. To this end, we generated a lightcone using the \lgal{} SAM applied on top of the \texttt{Millennium-I} merger trees. The version of the SAM used in this work is the one presented in \cite{IzquierdoVillalba2020,IzquierdoVillalba2021} in which different physical models were included to tackle the growth of MBHs and the dynamical evolution of MBHBs. The resulting lightcone featured an area of ${\sim}\,1027\, \rm deg^2$ with a line of sight $\rm (RA,DEC) \,{=}\, (56.3,58.9) \, deg$ and a redshift-depth large enough to encompass all the galaxies, MBHs and MBHBs up to $z\,{\sim}\,3.5$. Among all the binary systems placed inside the lightcone that can be potentially detected by LISA, we focused on the ones in the hardening phase at $z\,{<}\,3$ with total masses of ${\sim}\,3\,{\times}\,10^{5}\,\msun$, ${\sim}\,3\,{\times}\,10^{6}\,\msun$ and ${\sim}\,3\,{\times}\,10^{7}\,\msun$ and mass ratio between [0.01,1]. Their associated LISA sky-localization uncertainties $\Delta \Omega$ and $\Delta d_{\rm L}$ were estimated with the methodology of  \cite{Mangiagli2020} at two different times prior to merger (10 hours and 1 hour) and at merger. According to these uncertainties, we built the LISA fields (or just error-boxes) defined as the volumes centered on the galaxy hosting the GW source whose extension is given by ${\pm}\,3 \Delta \Omega$ and ${\pm}\,3 \Delta d_{\rm L}$. Once the error-boxes for each MBHB were assembled, we studied the number of galaxies placed within their boundaries, being considered as potential \textit{host candidates} of the GW event. %We remind that the search of a counterpart in real galaxy catalogues requires the knowledge of photometric/spectroscopic redshift for a match with the numbers of host candidates present in our simulated Universe. 
The main results can be summarized as follows: 

\begin{itemize}

    \item The number of host candidates can be very large during the \textit{inspiral phase}, reaching values of $10^5$ for a binary of $3\,{\times}\,10^5\, \msun$, $10^7$ for a binary of $3\,{\times}\,10^6\, \msun$ and $10^8$ for a binary of $3\,{\times}\,10^7\, \msun$. These numbers depend on redshift, being the largest at $z\,{\sim}\,3$. Such behaviour is caused by the worsening of the LISA sky-localization for sources at high-$z$.\\
    
    \item Concerning the \textit{merger} time, the number of host candidates drops to a handful (${\sim}\,10$) for sources at $z\,{<}\,0.5$, irrespective of the binary mass. On the contrary, at larger redshifts the number of host candidates is very large with values ${\sim}\,10^3$.

\end{itemize}

In light of these results and considering that the LISA MBHBs might display an EM counterpart, we considered the possibility of reducing the number of potential hosts limiting observations to the X-ray sky in search of an AGN counterpart. To this end, we only considered as host candidates the ones placed inside the LISA error-box whose nuclear MBH is active above a limiting X-ray flux. Instead of assuming arbitrary limits, we considered three different observatories whose flux sensitivity curves are close to those of Athena and of the concept-mission Lynx. While Lynx can detect almost instantaneously the dim sources associated to the LISA MBHBs used in this work thanks to its large sensitivity, Athena with its wider FOV requires less pointings to cover the LISA sky-area. The main results can be summarized as follows:

\begin{itemize}

  \item During the \textit{inspiral phase}, the number of host candidates can be reduced up to $2\rm \, dex$ when an X-ray AGN selection is performed based on Athena/Lynx-like capabilities. However, the number of potential hosts is prohibitive (${>}\,100$), especially at high-$z$. \\
  
  \item At the \textit{merger} time, the number of X-ray AGNs placed inside the LISA error-box is ${<}\,10$ at $z\,{<}\,1$, regardless of the mass of the system. At $z\,{>}\,1$ these X-ray candidates can be up to $100$, especially inside the FOV of a Lynx-like observatory thanks to its large sensitivity flux.\\
 
  \item The small FOV of the X-ray observatories and the large LISA error-box lead to the need of a large numbers of X-ray pointings (${\gtrsim}\,20$) hampering the detection of $z\,{>}\,0.5$ MBHB during the {\it the inspiral phase}.\\
   
  \item If obscuration is accounted, at the level of an hydrogen column density of $10^{23} \, \rm cm^{-2}$, the detection of AGNs associated to MBHB in soft X-rays requires exposure times longer than the one associated to the inspiral phase, irrespective of the binary mass. This large obscuration has a smaller impact in the hard X-ray band, where inspiralling MBHBs can be detected at $z\,{<}\,0.5$ as early as 10 hours before the merger.\\

  \item The AGN population detected inside the LISA error-box by Athena/Lynx-like observatory is typically placed in slightly more crowded regions than the ones of GW sources. These differences can be found primarily in binaries with total mass ${<}\,3\,{\times}\,10^{6}\, \msun$, vanishing at $3\,{\times}\,10^{7}\, \msun$. Thus, galaxies hosting LISA sources are more likely to be found in low density galaxy regions.
  
  %\item The number of host candidates can be reduced up to $2\rm \, dex$ when an X-ray AGN selection is performed based on Athena/Lynx-like capabilities. However, the number of potential hosts is prohibitive (${>}\,100$), especially at high-$z$.  This number can be reduced up to 10\% when combining future X-ray observations with data extracted from legacy surveys such as eRosita.\\
  
  %\item The number of host candidates with a detectable X-ray AGN increases with the binary mass, displaying a $2\,{\rm dex}$ variation between $3\,{\times}\,10^5\, \msun$ and $3\,{\times}\,10^7\, \msun$ binaries. Besides, this number decreases during the binary evolution, reaching its minimum at the time of the merger, when ${<}\,10$ AGNs are found for all the binary masses at $z\,{<}\,1$. \\
  
  %\item At $z\,{<}\,1$ all the X-ray observatories display a similar value \monica{allow for a comparable number of }of detectable AGNs within the LISA error-box (${<}\,10^3$). At higher-$z$, their different flux \monica{limiting} sensitivities leave an imprint in the number of detections, being a Lynx-like observatory more successful than an Athena-like (up to $3\,{-}\,4$ times). Despite this, the large uncertainty in the LISA sky-localization at $z\,{>}\,1$ and the small FOV of the observatories lead to the need of ${\gtrsim}\,20$ pointings to cover the full LISA error-box, hampering the detection of inspiralling MBHBs before the final merger.\\
  
\end{itemize}

Given all the results summarized above, we can conclude that the LISA error-boxes are going to be crowded with a number of potential host candidates ranging from hundreds to thousands. Thus, multi-messenger studies will have to face important challenges when searching for the host of a LISA GW source. Future X-ray observatories could help with this task, reducing the number of potential hosts by several orders of magnitude. However, this improvement is based on the optimistic assumption that the GW sources detected by LISA are radiating at the Eddington limit. Sub-Eddington or inactive binaries will make the host identification more challenging. Despite this, our results show a bright side in the unequivocal detection of GW hosts at $z\,{<}\,0.5$. In the very last hours prior to the merger and at the merger time, the error-box associated with these low-$z$ MBHBs contains a number of detectable X-ray sources small enough (${\lesssim}\,10$) to perform feasible follow-up studies at different epochs and wavelengths. For higher-$z$ sources, the employment of dual networks of space-based detectors (such as the LISA-Taiji configuration) will reduce by several orders of magnitude the number of potential hosts of the GW event, making the identification of the host easier, especially for sources at $z\,{<}\,1$ detected at ${\leq}\,10$ hours before the final merger.

\section*{Acknowledgements}
M.C. and A.S. thank Sylvain Marsat for enlightning discussions on LISA sky-position uncertainties and sky multi-modality patterns.  G.L and M.C acknowledges funding from MIUR under the Grant No. PRIN 2017-MB8AEZ. D.I.V. and A.S. acknowledges financial support provided under the European Union’s H2020 ERC Consolidator Grant ``Binary Massive Black Hole Astrophysics'' (B Massive, Grant Agreement: 818691). D.I.V. acknowledges also financial support from INFN H45J18000450006. S.B. acknowledges partial support from the project PGC2018-097585-B-C22, MINECO/FEDER, UE of the Spanish Ministerio de Economia, Industria y Competitividad. A.M. acknowledges support from the postdoctoral fellowship of IN2P3 (CNRS). Numerical computations were partly performed on DANTE platform, APC, France.


\section*{DATA AVAILABILITY}

The simulated data underlying this article will be shared on reasonable request to the corresponding author. This work used the 2015 public version of the Munich model of galaxy formation and evolution: \lgal{}. The source code and a full description of the model are available at http://galformod.mpa-garching.mpg.de/public/LGalaxies/.

\bibliographystyle{mnras}
\bibliography{references} % if your bibtex file is called example.bib

%\begin{thebibliography}{99}
%\end{thebibliography}

%%%%%%%%%%%%%%%%%%%%%%%%%%%%%%%%%%%%%%%%%%%%%%%%%%

%%%%%%%%%%%%%%%%% APPENDICES %%%%%%%%%%%%%%%%%%%%%

\appendix

\section{LISA sky-localization} \label{appendix:LISA_sky_localization}
\begin{figure}
    \centering
    \includegraphics[width=0.8\columnwidth]{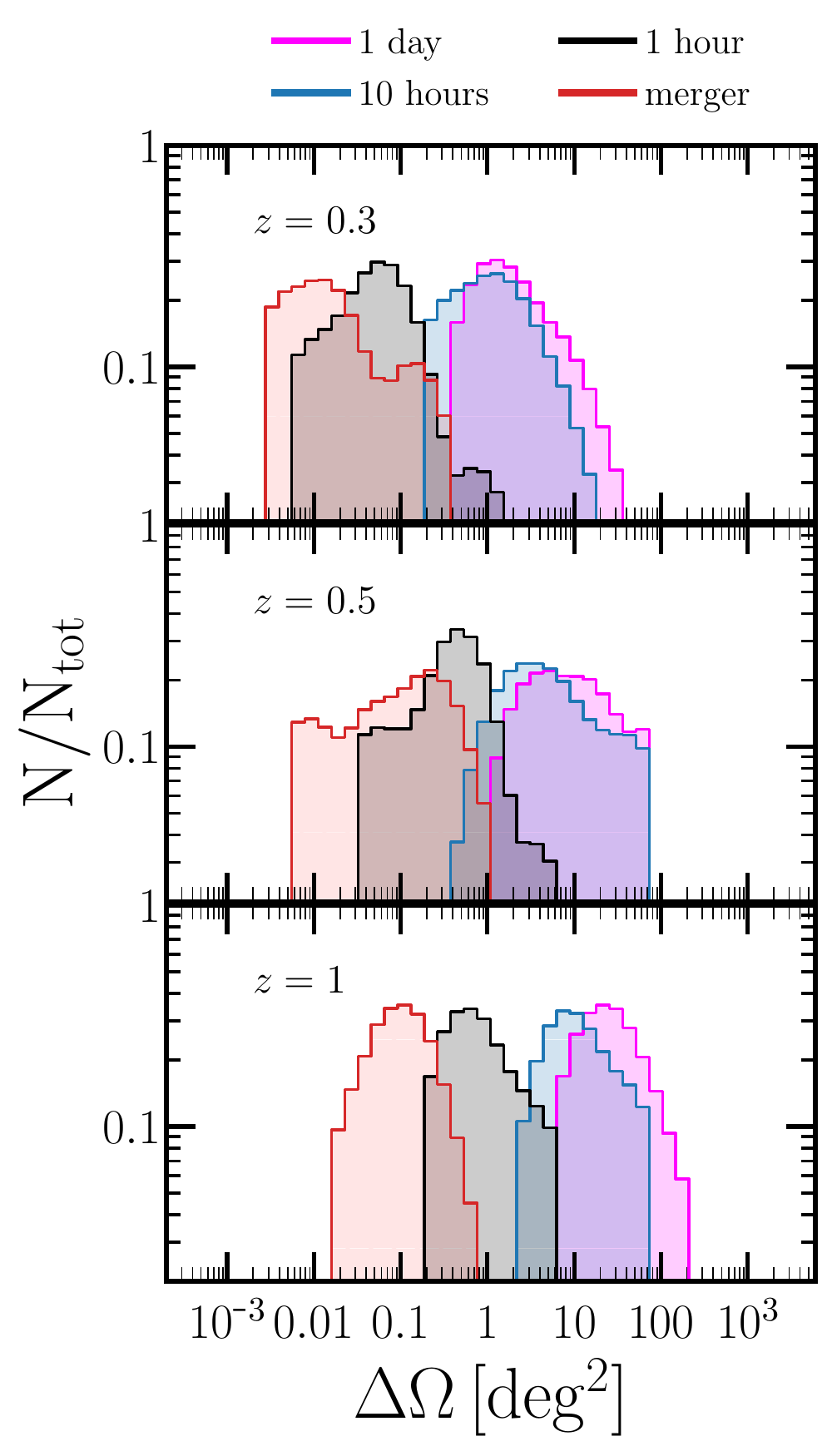}
    \caption[]{$\Delta \Omega$ associated to the true hosts with mass $\rm M_{\rm tot}\,{=}\, 3\,{\times}\,10^5\, \msun$ at $z\,{=}\,0.3, 0.5, 1$. Different colors represent different times: $1$ day (magenta), $10$ hours (blue) and $1$ hour (black) before merger and the time of the final coalescence (red).}
    \label{fig:DeltaOmega_M3e5}
    \end{figure}

\begin{figure}
    \centering
    \includegraphics[width=1\columnwidth]{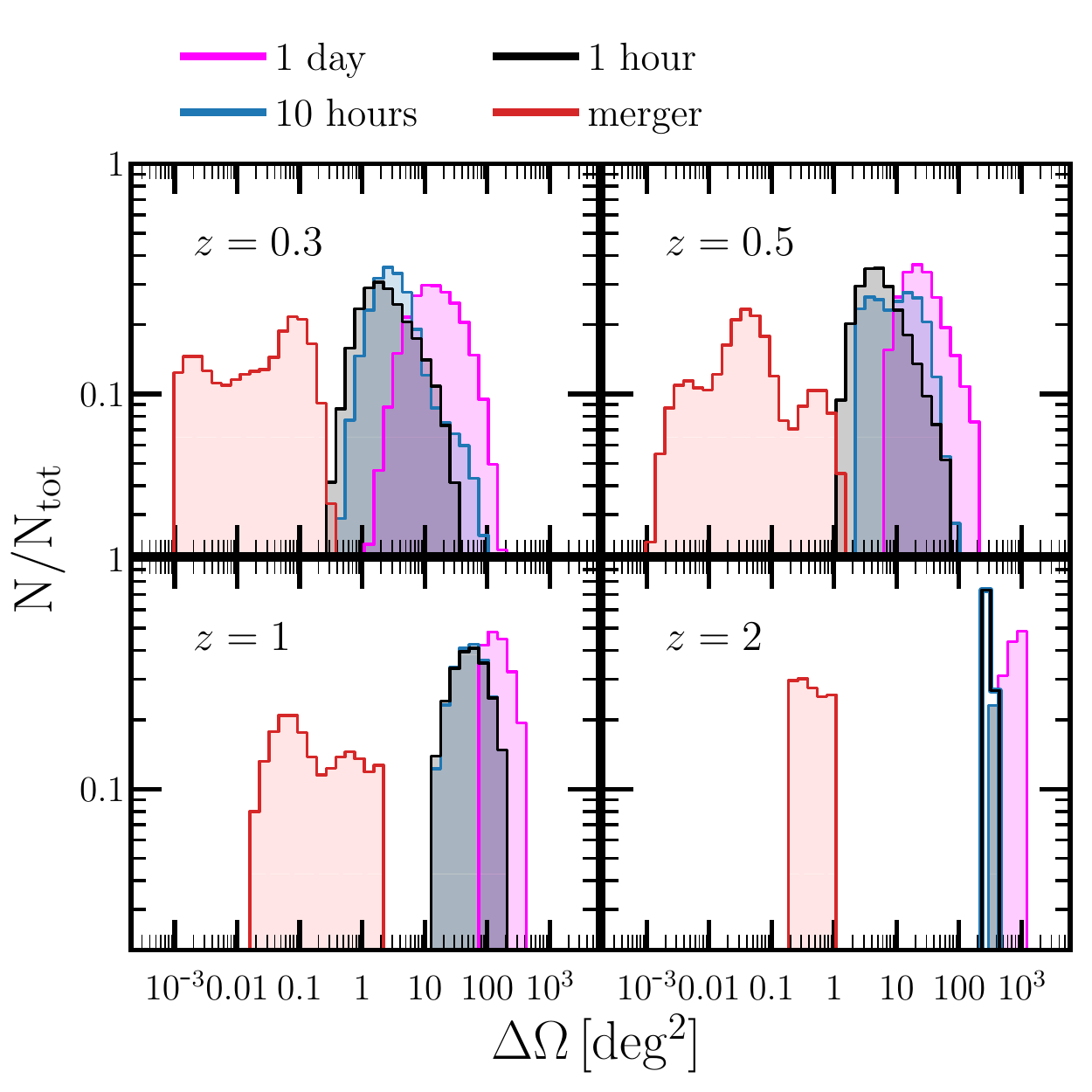}
    \caption[]{$\Delta \Omega$ associated to $200$ realizations per each binary of $\rm M_{\rm tot}\,{=}\, 3\,{\times}\,10^7 \msun$ at $z\,{=}\, 0.3, 0.5, 1, 2$. Different colors represent different times: $1$ day (magenta), $10$ hours (blue) and $1$ hour (black) before merger and the time of the final coalescence (red).}
    \label{fig:DeltaOmega_M3e7}
    \end{figure}

In Fig.~\ref{fig:DeltaOmega_M3e5} and Fig.~\ref{fig:DeltaOmega_M3e7} we present for the true hosts, the value of $\Delta \Omega$ associated to binaries with mass $3\,{\times}\,10^5\, \rm \msun$ and $3\,{\times}\,10^7\, \rm \msun$. As we can see, the larger is the time before the merger, the larger are the $\Delta \Omega$ values associated to the two binary mass. Interestingly, in Fig.~\ref{fig:DeltaOmega_M3e7} we can see that at $z\,{\geq}\,1$ the $\Delta \Omega$ distributions of 10 hours and 1 hour display the same values. Thus, the LISA sky-localization capabilities for a MBHB of $3\,{\times}\,10^7\, \rm \msun$ at $z\,{>}\,1$ do not vary between 10 hours and 1 hour before the merger.
    
\section{Number of cycles for the inspiralling MBHB} \label{appendix:Ncycles}

A plausible mechanism to identify the presence of a MBHB is through variability analysis in AGN lightcurves \citep[see e.g,][]{Graham2015,Charisi2016,Liu2016}. For instance, the theoretical work of \cite{Kelley2019} showed that the number of cycles of a binary system can imprint distinctive variations in the luminosity emitted by the active MBHB. Therefore, in this appendix we compute the number of cycles associated to the MBHBs ($\rm M_{\rm tot}\,{=}\,3\,{\times}\,10^{5-6-7}\, \msun$) as a function of the time before merger. To this end, we follow \cite{ColpiANDSesana2017} which showed that the number of cycles, $\mathcal{N}_{\rm cycles}$, covered by an inspiralling binary can be written as:
\begin{equation}\label{eq:NCycles}
\mathcal{N}_{\rm cycles} = \dfrac{6^{5/2}}{32 \pi} \dfrac{1}{\nu} \tilde{f}^{-5/3},
\end{equation}
\noindent where $\nu$ is the so-called \textit{symmetric mass ratio}, which can be expressed as a function of the binary mass ratio, $q$:
\begin{equation}
\nu \,{=}\, \frac{q}{(1+q)^2},
\end{equation}
%depends on the masses of the primary ($M_{\rm 1}$) and secondary ($M_{\rm 2}$) black holes through the following expression: 
%    \begin{equation}
%        \nu = \dfrac{M_{\rm 1} \, M_{\rm 2}}{(M_{\rm 1} + M_{\rm 2})^2}
%    \end{equation}
and $\tilde{f}$ is to the ratio between twice the Keplerian frequency, $f_{\rm K}$, and the frequency of the GW signal at the time of the final coalescence, $f_{\rm coal}$. These two quantities are computed, as:
\begin{equation}
f_{\rm K} \,{=}\,\dfrac{1}{2 \pi} \sqrt{\dfrac{G \, M_{\rm tot}}{a^3}},  %f_{\rm K} = \dfrac{1}{2 \pi} \sqrt{\dfrac{G \, (M_{\rm 1} + M_{\rm 2})}{a^3}}
\end{equation}
\begin{equation}
f_{\rm coal} \,{=}\, \dfrac{1}{\pi 6^{3/2}} \dfrac{c^3}{G \, M_{\rm tot}},%\dfrac{1}{\pi 6^{3/2}} \dfrac{c^3}{G \, (M_{\rm 1} + M_{\rm 2})}
\end{equation}
being $G$ the gravitational constant, $c$ the speed of light and $a$ the semi-major axis of the MBHB. Given that GW emission governs the shrinking process of the MBHBs during the inspiralling phase, we assume that the evolution of the binary semi-major axis can be determined as:
\begin{equation}
    \dfrac{da}{dt} \Big|_{\rm GW} \,{=}\, -\dfrac{B}{a^3},
\label{eq:SemiMajorAxis_Evolution}
\end{equation}
where $B$ is a coefficient that depends on the masses of the two black holes and the eccentricity, $e$, of the system:
\begin{equation}
    B \,{=}\, \dfrac{64 G^3 q \, M_{\rm tot}^3 }{5 c^5 (1+q)^2} (1-e)^{-7/2}\left[ 1 + \left( \frac{73}{24} \right)e^2  + \left( \frac{37}{96} \right)e^4 \right],  % \dfrac{64 G^3 q (M_{\rm 1} + M_{\rm 2})^3 F(e)}{5 c^5 (1+q)^2}
\end{equation}
Here, we assume $e = 0$ given that MBHBs at the evolutionary stage dominated by the GW emission undergo a circularization of the binary orbit. By solving Eq.~\ref{eq:SemiMajorAxis_Evolution} and assuming that at $t\,{=}\,0$ the system merges, we can derive the expression of the semi-major axis at a generic time, $t$, in the source frame as:
\begin{equation}
a(t) = (4Bt)^{1/4}.
\end{equation}

\begin{figure}
    \centering
    \includegraphics[width=0.9\columnwidth]{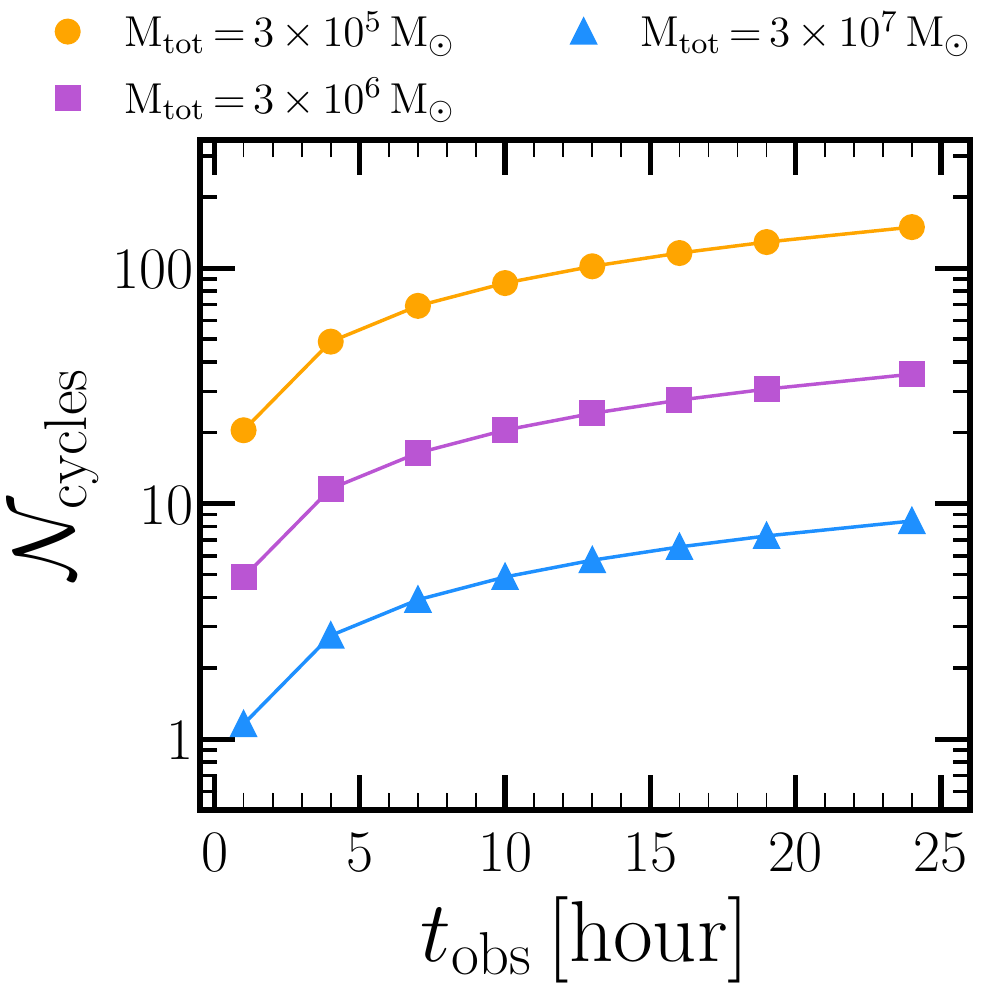}
    \caption[]{Median values of $\mathcal{N}_{\rm cycles}$ associated to the true hosts with mass $\rm M_{\rm tot} \,{=}\, 3 \,{\times}\, 10^5 \, \msun$ (orange), $\rm M_{\rm tot} \,{=}\, 3 \,{\times}\, 10^6 \,\msun$ (purple) and $\rm M_{\rm tot} \,{=}\, 3 \,{\times}\, 10^7 \, \msun$ (blue) at $z\,{=}\,0.5$.  }
    \label{fig:NCycles_z1}
\end{figure}

Taking into account the previous methodology, we can compute $\mathcal{N}_{\rm cycles}$ at the last hours preceeding the final plunge for binaries with $M_{\rm tot}\,{=}\,3\,{\times}\,10^{5}\,\msun$, $3\,{\times}\,10^{6}\,\msun$ and $3\,{\times}\,10^{7}\,\msun$. For simplicity, we have chosen the parameters associated to the true hosts at $z\,{=}\,0.5$. Similar results are expected at other redshifts given that mild evolution in the binary properties are seen. The results are presented in Fig.~\ref{fig:NCycles_z1} as a function of the observed time $t_{\rm obs} = t \, / (1+z)$, where $z$ is the redshift of the binary. As expected, the number of cycles decreases as the binary approaches to the final coalescence. Despite this trend is shared by all the three MBHBs, at fixed time the number of cycles depends on the total mass of the system, with higher values associated to lower-mass MBHBs. This behavior leads back to the fact that Eq.~\ref{eq:NCycles} is inversely proportional to the mass of the binary.

%\section{X-ray luminosity and observed flux of Eddington limited binaries} \label{appendix:Luminosity_and_Flux}
%In Table~\ref{tab:Lum_and_flux_Limit_AllBinaries} we present the luminosities in the soft ($\rm 0.5\,{-}\,2\, keV$) and hard ($\rm 2\,{-}\,10\, keV$) X-ray band associated to the MBHBs radiating at the Eddington-limit, and corrected by obscuration assuming an hydrogen column density equal to the median value of the $\rm N_{\rm H}$ distribution presented in Fig.~\ref{fig:NHDistribution} ($\rm N_{H}\,{=}\,10^{23}\, \rm cm^{-2}$). Besides, these luminosities have been transformed into observed fluxes at $z\,{=}\,0.3,0.5,1,2$ and $3$.

\section{High X-ray absorption in active host candidates}
\label{appendix:ComptonThinScenario}
In this appendix we re-analyze the results presented in Section~\ref{sec:DetectionInspirallingBinaries} and Section~\ref{sec:NAGNs} by using larger values of $\rm N_H$ for the population of X-ray AGNs.

\subsection{Exposure times in the high X-ray absorption scenario}  \label{appendixSubsection:Exposuretimes}

\begin{table}
\begin{adjustbox}{width=\columnwidth,center}
\begin{tabular}{lcccc|} \cline{2-5}
                       & \multicolumn{1}{|c|}{\diagbox[width=\dimexpr \textwidth/18+2\tabcolsep\relax, height=0.5cm]{ $z$ }{ $\rm Mass$}} & $3\,{\times}\,10^5\,[\msun]$   &  $3\,{\times}\,10^6\,[\msun]$  &  $3\,{\times}\,10^7\,[\msun]$   \\ \hline
\multicolumn{5}{|c|}{\cellcolor[HTML]{C0C0C0} \hspace{1.8cm}  $\mathbf{0.5\,{-}\,2}$ \textbf{keV}} \\
\multicolumn{1}{|l|}{$\rm L_{\rm X-ray}\,[erg\,s^{-1}]$}   &    -    &   $1.32\,{\times}\,10^{40}$    &   $7.79\,{\times}\,10^{40}$   &   $4.45\,{\times}\,10^{41}$  \\ \cline{1-5}
 \multicolumn{1}{|l|}{}                       & 0.3    &  $4.27\,{\times}\,10^{-17}$     &  $2.51\,{\times}\,10^{-16}$    &   $1.44\,{\times}\,10^{-15}$   \\
 \multicolumn{1}{|l|}{}                      & 0.5    &  $1.29\,{\times}\,10^{-17}$     &   $7.57\,{\times}\,10^{-17}$    &  $4.33\,{\times}\,10^{-16}$    \\
 \multicolumn{1}{|l|}{}                      & 1     &   $2.39\,{\times}\,10^{-18}$    &   $1.40\,{\times}\,10^{-17}$    &   $8.02\,{\times}\,10^{-17}$   \\
 \multicolumn{1}{|l|}{}                      & 2     & -     &  $2.55\,{\times}\,10^{-18}$   &  $1.46\,{\times}\,10^{-17}$   \\
\multicolumn{1}{|l|}{\multirow{-5}{*}{$f \, \rm [erg \,\, cm^{-2} \, s^{-1}]$}}    & 3      & -     &  $9.58\,{\times}\,10^{-19}$    &   $5.48 \,{\times}\,10^{-18}$  \\ \cline{1-5}
\multicolumn{5}{|c|}{\cellcolor[HTML]{C0C0C0} \hspace{1.8cm} $\mathbf{2\,{-}\,10}$ \textbf{keV}}  \\
\multicolumn{1}{|l|}{$\rm L_{\rm X-ray}\,[erg\,s^{-1}]$}  &    -    &   $1.53\,{\times}\,10^{42}$    &   $9.12\,{\times}\,10^{43}$   &   $5.21\,{\times}\,10^{43}$   \\ \cline{1-5}
 \multicolumn{1}{|l|}{}                       & 0.3    &   $4.93\,{\times}\,10^{-15}$    &  $2.94\,{\times}\,10^{-14}$    &    $1.68\,{\times}\,10^{-13}$  \\
 \multicolumn{1}{|l|}{}                      & 0.5    &    $1.49\,{\times}\,10^{-15}$   &   $8.87\,{\times}\,10^{-15}$   &    $5.06\,{\times}\,10^{-14}$  \\
 \multicolumn{1}{|l|}{}                      & 1      &    $2.75\,{\times}\,10^{-16}$   &   $1.64\,{\times}\,10^{-15}$   &   $9.37\,{\times}\,10^{-15}$   \\
 \multicolumn{1}{|l|}{}                      & 2      &   -   &     $2.99\,{\times}\,10^{-16}$  &  $1.71\,{\times}\,10^{-15}$    \\
\multicolumn{1}{|l|}{\multirow{-5}{*}{$f \, \rm [erg \,\, cm^{-2} \, s^{-1}]$}}    & 3      &  -     &  $1.12\,{\times}\,10^{-16}$     &  $6.40\,{\times}\,10^{-16}$ \\  \hline
\end{tabular}
\end{adjustbox}
\caption{Luminosities and fluxes in the $0.5\,{-}\,2\,\rm keV$ and $2\,{-}\,10\,\rm keV$ bands emitted by MBHBs accreting at the Eddington limit with an hydrogen column density of $\rm N_{\rm H} = 10^{23} \rm cm^{-2}$.}
\label{tab:Lum_and_flux_Limit_AllBinaries_ComptonThin}
\end{table}

In this section we explore the capabilities of the X-ray observatories to detect Eddington-limited AGNs associated to the MBHBs when accounting for an hydrogen column density of $\rm N_{\rm H} = 10^{23} \rm cm^{-2}$ (i.e Compton thin AGNs). The X-ray fluxes and luminosities associated to these Compton thin AGNs are shown in Table~\ref{tab:Lum_and_flux_Limit_AllBinaries_ComptonThin}. As we can see, these values are smaller than the ones presented in Table~\ref{tab:Lum_and_flux_Limit_AllBinaries}. This difference is particularly important in the soft X-ray band, where the sources are up to ${\sim}\,2\rm dex$ dimmer. As we will summarize in Table~\ref{tab:Exposure_times_ComptonThin}, this decrease in the brightness will have strong repercussion in the time required to detect the MBHBs. \\

\begin{figure}
\centering
\includegraphics[width=1\columnwidth]{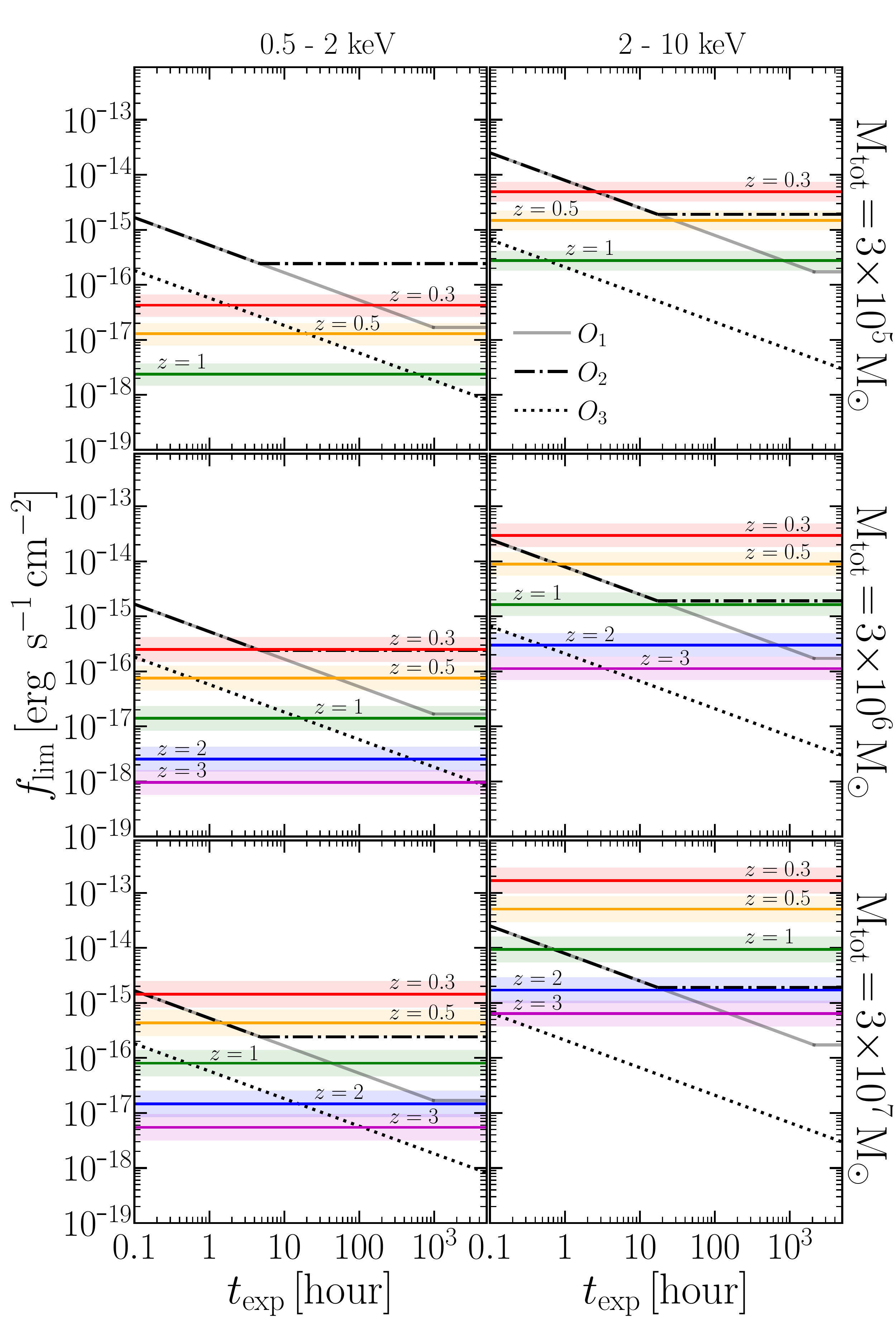}
\caption[]{Sensitivity curves of the three simulated X-ray observatories employed in this project to study a potential cooperation between LISA and an X-ray detector: $O_1$ (solid line), $O_2$ (dash-dotted line) and $O_3$ (dotted line). These functions have been compared with the fluxes at the Eddington limit (horizontal solid lines) associated to the MBHBs, which have been computed by using the bolometric corrections of \cite{Shen2020} and column density of $\rm N_{\rm H} = 10^{23} \rm cm^{-2}$. Shaded areas are delimited by the maximum and minimum fluxes corresponding to the maximum and minimum values of the bolometric corrections.}
\label{fig:SensitivityCurvesComptonThin}
\end{figure}

In Fig.~\ref{fig:SensitivityCurvesComptonThin} we present for the $0.5\,{-}\,2\,\rm keV$ and $2\,{-}\,10\,\rm keV$ bands a comparison between the sensitivity curves of the observatories and the fluxes associated to all the binary systems. As we can see, $O_3$ is able to detect in the hard X-ray band all the MBHBs studied in this work with exposure times $t_{\rm exp}\,{\sim}\,3$ hours, regardless of redshift. In the soft X-ray band these times increase rapidly, reaching values larger than $10^{3}$ hours. For instance, the detection of $3\,{\times}\,10^5\, \rm M_{\odot}$ and $3\,{\times}\,10^6\, \rm M_{\odot}$ binary at $z\,{=}\,1$ requires 580 and 17 hours, respectively. Such large difference between soft and hard X-ray band is due to the dust obscuration, which is very efficient in dimming the emission in the $0.5\,{-}\,2\,\rm keV$ band. With the value of $\rm N_{\rm H} = 10^{23}\, \rm cm^{-2}$, the fluxes associated to our sources in the soft X-ray band are reduced by a factor of $8\,{\times}\,10^{-3}$ while they only decrease by $0.6$ in the hard X-ray band. $O_1$ can also detect all the binaries in the hard X-rays with the exception of the MBHB of $3\times10^6 \msun$ at $z=3$, but the values of $t_{\rm exp}$ display important differences. For instance, in the $2\,{-}\,10\,\rm KeV$ band, $O_1$ needs ${\sim}\,28$ (${\sim}\,23$) hours to reach the minimum flux required to detect an Eddington-limited binary of $3\,{\times}\,10^5 \, \msun$ ($3\,{\times}\,10^6 \, \msun$) at $z\,{=}\,0.5$ ($z\,{=}\,1$). For the  $0.5\,{-}\,2\,\rm keV$ band, only binaries of $\rm M_{\rm tot} \geq 3\times10^6 \msun$ at $z\,{\lesssim}\,0.5$ are accessible with $t_{\rm exp}\,{\sim}\,4$ hours. Observatory $O_2$ behaves similarly to $O_1$, but its larger $f_c$ prevents the detection of any binary with redshift ${>}\,0.5$, with the exception of the MBHB of $3\times10^7 \msun$ at $z=1$. The values of $t_{\rm exp}$ are summarized in Table~\ref{tab:Exposure_times_ComptonThin}.\\      

\begin{table}
\begin{adjustbox}{width=\columnwidth,center}
\begin{tabular}{cccccccccc|}\cline{2-10}
                          & \multicolumn{3}{|c|}{$3\,{\times}\,10^5\,[\msun]$} & \multicolumn{3}{c|}{$3\,{\times}\,10^6\,[\msun]$} & \multicolumn{3}{c|}{$3\,{\times}\,10^7\,[\msun]$} \\ \cline{2-10}
\multicolumn{1}{c|}{}       & $O_1$     & $O_2$    & \multicolumn{1}{c|}{$O_3$}    & $O_1$     & $O_2$    & \multicolumn{1}{c|}{$O_3$}    & $O_1$     & $O_2$    & $O_3$    \\ \cline{1-10}
\rowcolor[HTML]{C0C0C0} 
\multicolumn{1}{|l|}{\diagbox[width=\dimexpr \textwidth/12+2\tabcolsep\relax, height=0.85cm]{ $z$ }{ $t_{\rm exp} \rm [h]$}} & \multicolumn{9}{c|}{\cellcolor[HTML]{C0C0C0}$\mathbf{0.5\,{-}\,2}$ \textbf{keV}}                    \\
\multicolumn{1}{|l|}{0.3}                       &  $152.3$    &  \xmark     &  \multicolumn{1}{l|}{$1.8$}     &    $4.4$    &   $4.4$    &   \multicolumn{1}{l|}{${<}\,0.1$}     &   $0.1$     &    $0.1$   &     ${<}\,0.1$  \\
\multicolumn{1}{|l|}{0.5}                      &    \xmark    &   \xmark    &   \multicolumn{1}{l|}{$20.0$}    &     $48.5$   &     \xmark  &   \multicolumn{1}{l|}{$0.6$}     &   $1.51$     &   $1.5$    &   ${<}\,0.1$    \\
\multicolumn{1}{|l|}{1}                       &    \xmark &    \xmark    &   \multicolumn{1}{l|}{$582.2$}    &   \xmark    &   \xmark   &  \multicolumn{1}{l|}{$17.0$}     &   $43.2$     &  \xmark     &   $0.5$    \\
\multicolumn{1}{|l|}{2}                       &    -        &   -        &   \multicolumn{1}{l|}{-}       &    \xmark    &    \xmark   &   \multicolumn{1}{l|}{$511.4$}    &   \xmark     &   \xmark     &   $15.6$    \\
\multicolumn{1}{|l|}{3}                       &    -        &    -       &    \multicolumn{1}{l|}{-}           &    \xmark    &  \xmark     &  \multicolumn{1}{l|}{3623.7}     &  \xmark     &    \xmark  &    $110.7$   \\ \cline{1-10}
 
\multicolumn{1}{c|}{}                          & $O_1$     & $O_2$    & \multicolumn{1}{c|}{$O_3$}    & $O_1$     & $O_2$    & \multicolumn{1}{c|}{$O_3$}    & $O_1$     & $O_2$    & $O_3$    \\ \cline{1-10}
\rowcolor[HTML]{C0C0C0} 
\multicolumn{1}{|l|}{\diagbox[width=\dimexpr \textwidth/12+2\tabcolsep\relax, height=0.85cm]{ $z$ }{ $t_{\rm exp} \rm [h]$}}& \multicolumn{9}{c|}{\cellcolor[HTML]{C0C0C0}$\mathbf{2\,{-}\,10}$ \textbf{keV}}    \\
\multicolumn{1}{|l|}{0.3}                       &  2.6    &  2.6    &  \multicolumn{1}{l|}{${<}\,0.1$}     &    ${<}\,0.1$    &   ${<}\,0.1$    &   \multicolumn{1}{l|}{${<}\,0.1$}     &   ${<}\,0.1$     &    ${<}\,0.1$   &     ${<}\,0.1$  \\
\multicolumn{1}{|l|}{0.5}                       &    28.2       &   \xmark         &   \multicolumn{1}{l|}{${<}\,0.1$}    &      0.8           &    0.8   &    \multicolumn{1}{l|}{${<}\,0.1$}    &   ${<}\,0.1$     &  ${<}\,0.1$     &   ${<}\,0.1$   \\
\multicolumn{1}{|l|}{1}                       &     826.4     &   \xmark        &   \multicolumn{1}{l|}{0.6}         &      23.2           &   \xmark   &   \multicolumn{1}{l|}{${<}\,0.1$}     &  0.7     &   0.7    &   ${<}\,0.1$    \\
\multicolumn{1}{|l|}{2}                       &     -          &    -            &   \multicolumn{1}{l|}{-}             &   699.1          &    \xmark   &     \multicolumn{1}{l|}{0.5}   &  21.4      &   \xmark    &   ${<}\,0.1$    \\
\multicolumn{1}{|l|}{3}                       &    -           &     -           &    \multicolumn{1}{l|}{-}            &       \xmark           &     \xmark  &   \multicolumn{1}{l|}{3.5}     &   152.6     &   \xmark    &   $0.1$   \\ \hline
\end{tabular}
\end{adjustbox}
\caption{Average minimum exposure times, $t_{\rm exp}$ (in hours) required to detect the binaries emitting at the Eddington limit with $\rm N_{H}\,{=}\,10^{23}\, \rm cm^{-2}$. The values of $t_{\rm exp}$ are computed for the three different X-rays observatories ($O_1$, $O_2$ and $O_3$) at five different redshifts ($z \,{=}\,0.3$, $0.5$, $1$, $2$ and $3$). The values have been presented for two different bands: $\rm 0.5\,{-}\,2\, keV$ (soft X-rays) and $\rm 2\,{-}\,10\, keV$ (hard X-rays). Crosses are drawn when the detection of a MBHB is not possible. These values correspond to the intersection between the horizontal solid lines in Fig~\ref{fig:SensitivityCurvesComptonThin} and the sensitivity curves of the three observatories.}
\label{tab:Exposure_times_ComptonThin}
\end{table}

\label{tab:Pointings_ComptonThin}
%\end{table}

%\gaia{In Table~\ref{tab:Pointings_ComptonThin} we report the average value of pointings,  $N_{\rm p}$, required by each observatory to cover the projected LISA error-box. The values are the same as in Table~\ref{tab:Pointings}, since they do not depend on the level of absorption assumed. Interestingly, the X-ray observatories turn out to be able to cover the full LISA projected error-box in the same cases discussed in Section~\ref{sec:DetectionInspirallingBinaries}, even though the hydrogen column density used to compute the exposure times of Table~\ref{tab:Exposure_times_ComptonThin} differs of $\sim 2$ orders of magnitude with respect to the scenario discussed in the main text. This result boils down to the fact that the $2-10$ keV band is much less affected by absorption than the $0.5-2$ keV energy window. Indeed, the feasibility of detecting Eddington-limited MBHBs severely affected by absorption by means of X-ray observatories is particularly hampered only in the soft X-ray band, as the corresponding exposure times become prohibitive. Therefore, limiting our anaylis at the hard x-ray band, we can conclude that a complete covering of the FOV of LISA within the inspiralling time will be \textit{only} feasible for binary systems of ${\leq}\,3\,{\times}\,10^7 \, \msun$ at $z\,{\lesssim}\,1$ regardless of the value chosen for $N_{\rm H}$ in the interval $10^{21-5} - 10^{23} \rm cm^{-2}$. }

\subsection{Number of host candidates detected as AGNs when a large absorption is assumed}
 
\begin{figure}
\centering  
\includegraphics[width=1\columnwidth]{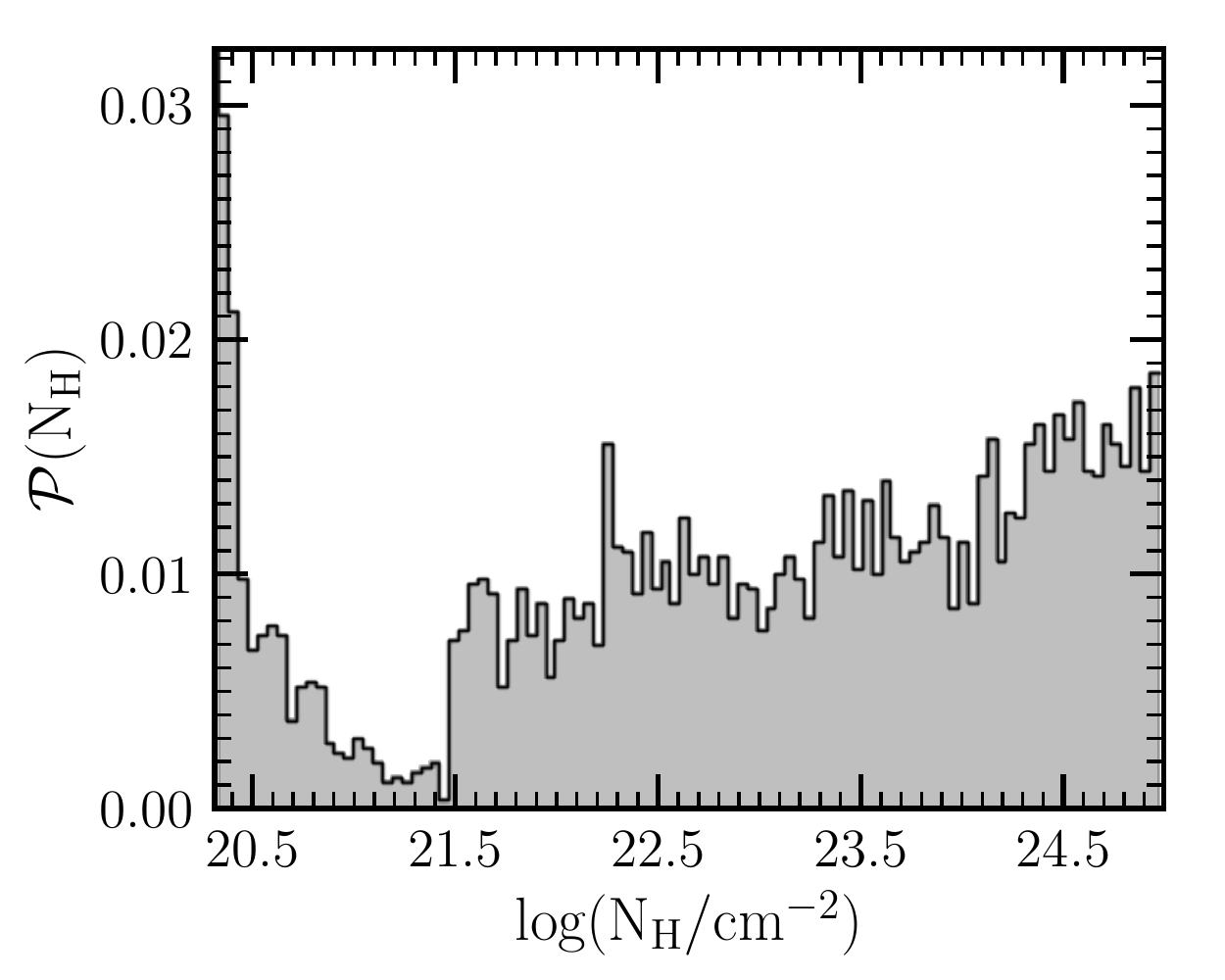}
\caption[]{Probability distribution of $\rm N_{H}$ used in this work. While the first peak corresponds to type-1 AGNs, the second one represents the population of type-2 AGNs.}
\label{fig:NHDistribution}
\end{figure}

In this section we perform the same analysis of Section~\ref{sec:NAGNs} but for hydrogen column density values larger than the ones used in the fiducial approach presented in Section~\ref{sec:BHPopulation}. For that, we assign to each AGN placed inside the lightcone a random value of $\rm N_{H}$ according to the distribution presented in Fig.~\ref{fig:NHDistribution}. This distribution is based on the works of \cite{Masoura2021} and \cite{Dwelly2005}, presenting the $\rm N_{H}$ values characterizing type-1 and type-2 AGNs, respectively. Specifically, starting from these works and assuming that type-1 (type-2) AGNs contribute with $1/4$ ($3/4$) of the total population of active MBHs \citep[see e.g.][]{Gilli2007,Lusso2011}, the final $\rm N_{H}$ distribution presented in Fig.~\ref{fig:NHDistribution} was computed.\\

\begin{figure}
        %\hspace{-0.7cm}
        \centering
        \includegraphics[width=1\columnwidth]{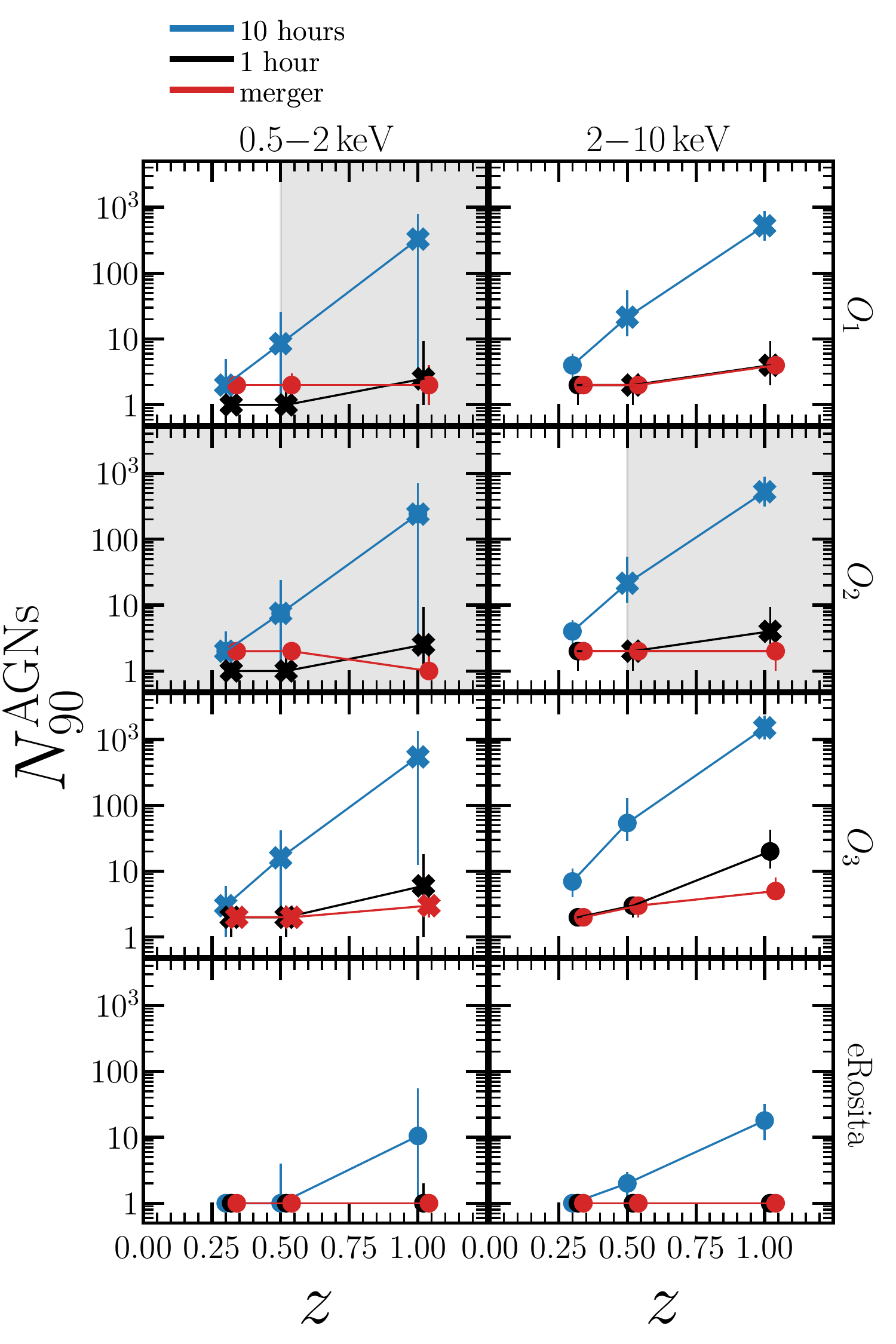}
        \caption{Number of X-ray AGNs (with $\rm N_{H}$ extracted from Fig.~\ref{fig:NHDistribution}) in soft (left panel) and hard bands (right panel) detected inside the LISA error-box associated to the true hosts of $\rm M_{\rm tot}\,{=}\, 3\,{\times}\,10^5 \msun$ at $z\,{=}\,0.3,0.5$ and $1$. The error bars correspond to the $32^{\rm th}\,{-}\,68^{\rm th}$ percentiles. Crosses are drawn when the number of pointings needed by the X-ray observatories to cover the full LISA error-box require exposure times larger than the inspiral time of the binary. Different colors represent different times during the evolution of the systems: $10$ hours before merger (blue), $1$ hour before merger (black) and merger (red). Finally, shaded gray areas highlight the redshifts at which the observatories are not able to detect any AGN associated to the MBHBs.}
        \label{fig:NAGN_M3e5_ComptonThin}
    \end{figure}

\begin{figure}
        %\hspace{-0.7cm}
        \centering
        \includegraphics[width=1\columnwidth]{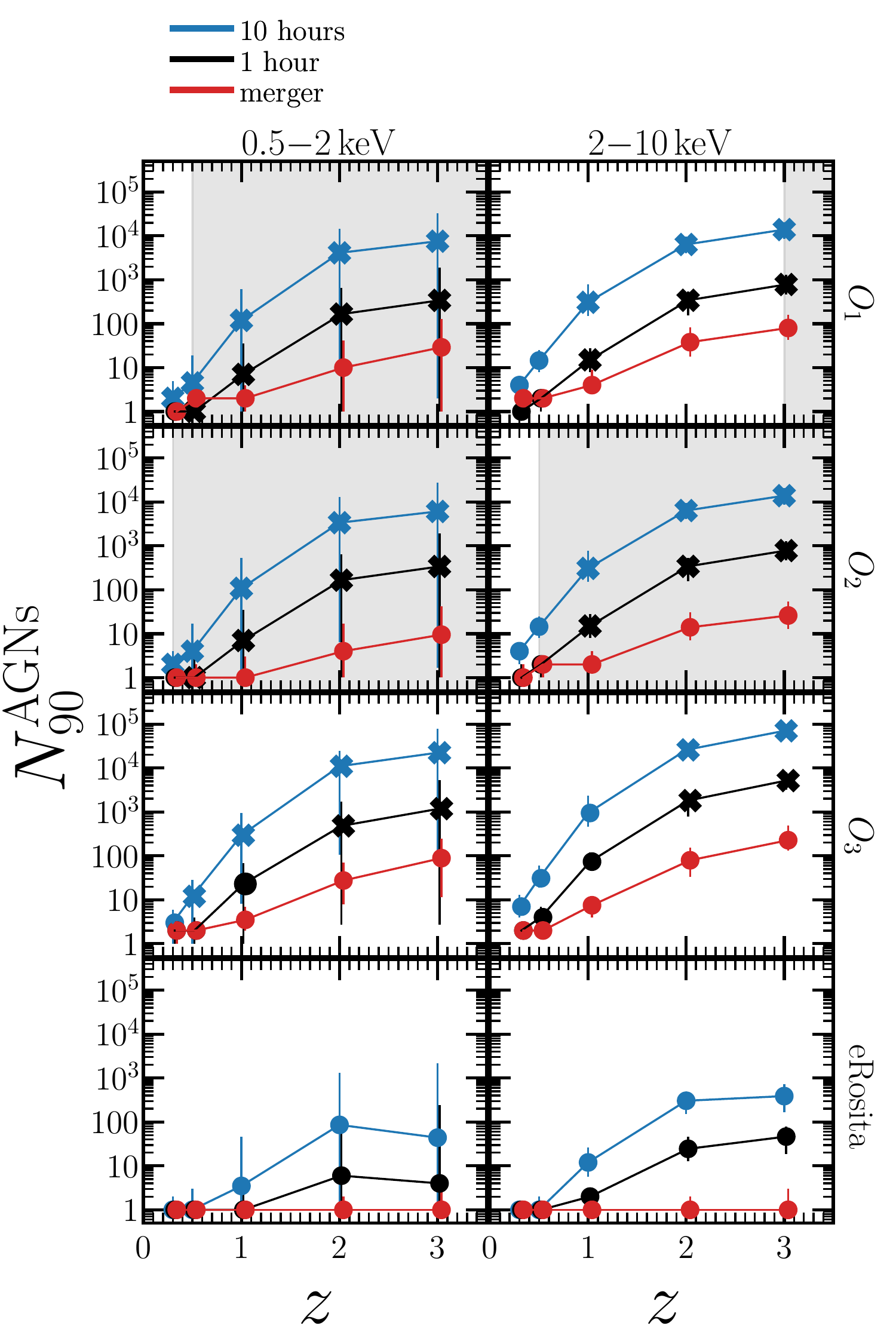}
        \caption{Number of X-ray AGNs (with $\rm N_{H}$ extracted from Fig.~\ref{fig:NHDistribution}) in soft (left panel) and hard bands (right panel) detected inside the LISA error-box associated to the true hosts of $\rm M_{\rm tot}\,{=}\, 3\,{\times}\,10^6 \msun$ at $z\,{=}\,0.3,0.5,1,2$ and $3$. The error bars correspond to the $32^{\rm th}\,{-}\,68^{\rm th}$ percentiles. Crosses are drawn when the number of pointings needed by the X-ray observatories to cover the full LISA error-box require exposure times larger than the inspiral time of the binary. Different colors represent different times during the evolution of the systems: $10$ hours before merger (blue), $1$ hour before merger (black) and merger (red). Finally, shaded gray areas highlight the redshifts at which the observatories are not able to detect any AGN associated to the MBHBs.}
        \label{fig:NAGN_M3e6_ComptonThin}
    \end{figure}

\begin{figure}
        %\hspace{-0.7cm}
        \centering
        \includegraphics[width=1\columnwidth]{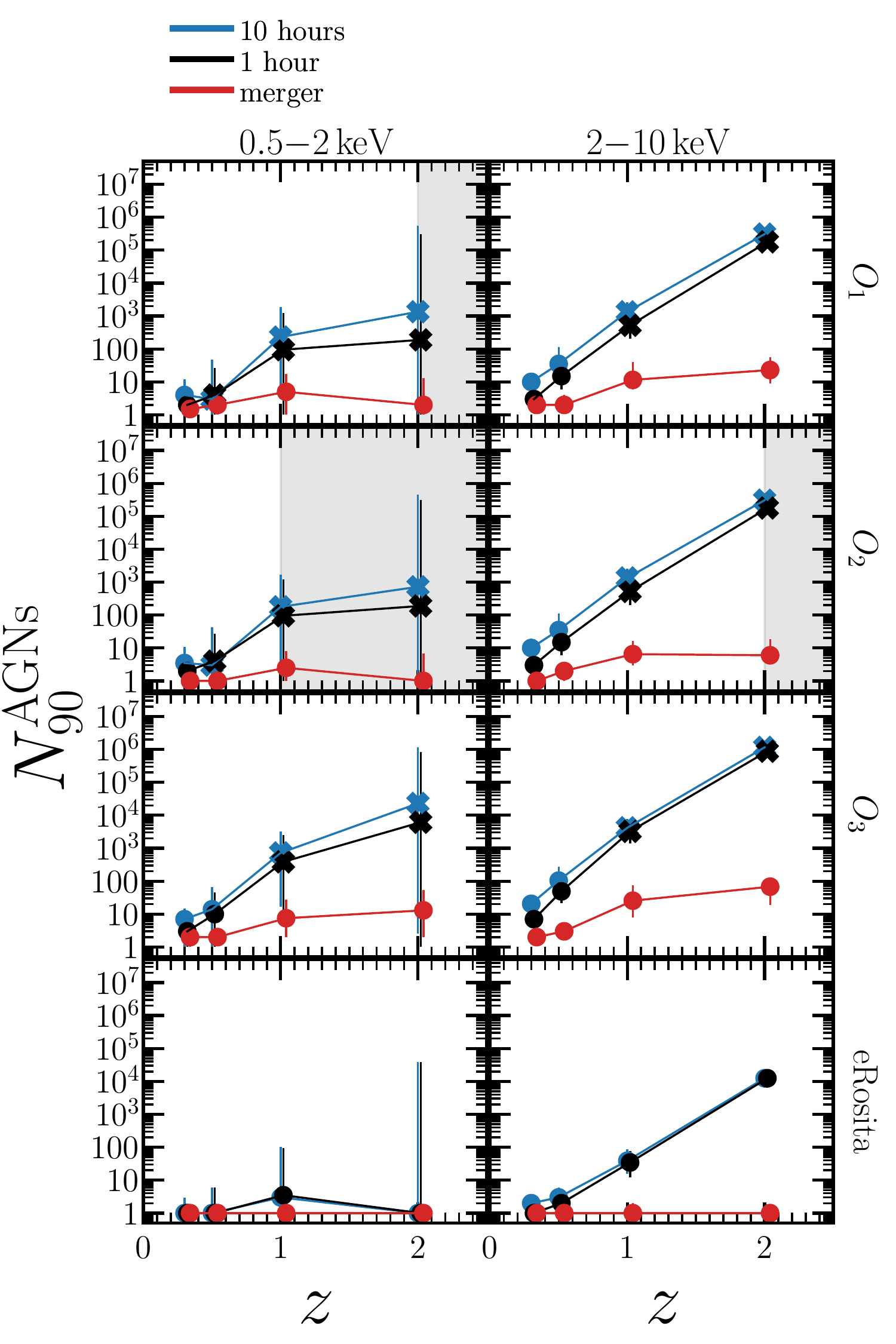}
        \caption{Number of X-ray AGNs (with $\rm N_{H}$ extracted from Fig.~\ref{fig:NHDistribution}) in soft (left panel) and hard bands (right panel) detected inside the LISA error-box associated to the true hosts of $\rm M_{\rm tot}\,{=}\, 3\,{\times}\,10^7 \msun$ at $z\,{=}\,0.3,0.5,1$ and $2$. The error bars correspond to the $32^{\rm th}\,{-}\,68^{\rm th}$ percentiles. Crosses are drawn when the number of pointings needed by the X-ray observatories to cover the full LISA error-box require exposure times larger than the inspiral time of the binary. Different colors represent different times during the evolution of the systems: $10$ hours before merger (blue), $1$ hour before merger (black) and merger (red). Finally, shaded gray areas highlight the redshifts at which the observatories are not able to detect any AGN associated to the MBHBs.}
        \label{fig:NAGN_M3e7_ComptonThin}
    \end{figure}

To determine the $N_{90}$ galaxies that are detected by $O_1$, $O_2$ and $O_3$ as AGNs with the new $\rm N_{H}$ distribution ($N_{90}^{\rm AGN}$), we proceeded as in Section~\ref{sec:NAGNs}. In brief, by using Eq.~\eqref{eq:BolometricCorrection_hard}, Eq.~\eqref{eq:BolometricCorrection_soft} and Eq.~\eqref{eq:ColumDensityAttenuation} we transformed the bolometric luminosity into X-ray luminosity for all the MBHs/MBHBs hosted by the $N_{90}$ galaxies. Then, we checked their observability by imposing that the X-ray flux must be above the flux limit of a given observatory. In Fig.~\ref{fig:NAGN_M3e5_ComptonThin}, Fig.~\ref{fig:NAGN_M3e6_ComptonThin} and Fig.~\ref{fig:NAGN_M3e7_ComptonThin} we show the values of $N_{90}^{\rm AGN}$ for the MBHBs of $3\,{\times}\,10^{5}\, \rm M_{\odot}$, $3\,{\times}\,10^{6}\, \rm M_{\odot}$ and $3\,{\times}\,10^{7}\, \rm M_{\odot}$, respectively. As one could expect, assuming an higher level of absorption implies a decrease in the value of $N_{90}^{\rm AGN}$, since more sources are characterized by fluxes falling below the detection limit of the X-ray observatories. At any redshift, we see a drop of $\sim 1$ dex ($\sim 3$ dex) for systems of $<3\times10^6 \msun$ ($3\times10^7 \msun$). This difference between high and low mass binaries is probably due to the different environments inhabited by the AGNs inside the LISA error-box (see Fig.~\ref{fig:Mass_ratio_betwee_AGNS_TrueHosts})

% Don't change these lines
\bsp	% typesetting comment
\label{lastpage}
\end{document}